\documentstyle[11pt]{article}
\begin{document}
\newtheorem{thm}{Theorem}[section]
\newtheorem{defin}[thm]{Definition}
\newtheorem{lemma}[thm]{Lemma}
\newtheorem{propo}[thm]{Proposition}
\newtheorem{cor}[thm]{Corollary}

\centerline{\huge \bf State-Sum Invariants of 4-Manifolds, I}
\bigskip

\centerline{\bf Louis Crane}
\smallskip

\centerline{\parbox{2.4in}{\small \em  Department of Mathematics \\
Kansas State
University \\ Manhattan, KS 66506}}
\bigskip

\centerline{\bf Louis H. Kauffman}

\centerline{\parbox{2.4in}{\small \em Department of Mathematics, Statistics
\vspace*{.1in} and Computer Science \\ University of Illinois at Chicago \\
851 S. Morgan Street \\
Chicago, IL 60607-7045}}
\bigskip

\centerline{\bf David N. Yetter}
\smallskip

\centerline{\parbox{2.4in}{\small \em  Department of Mathematics \\
Kansas State
University \\ Manhattan, KS 66506}}
\bigskip

\noindent{\small {\bf Abstract:} We provide, with proofs,
a complete description
of the authors' construction of state-sum invariants announced in [CY], and
its generalization to an arbitrary (artinian) semisimple tortile category.
We also discuss the relationship of these invariants to generalizations of
Broda's surgery invariants [Br1,Br2] using techniques developed in the case of
the semi-simple sub-quotient of $Rep(U_q(sl_2))$ ($q$ a principal $4r^{th}$
root of unity) by Roberts [Ro1].
We briefly discuss the generalizations to invariants of 4-manifolds
equipped with 2-dimensional (co)homology classes introduced by Yetter [Y6] and
Roberts [Ro2], which are the subject of the sequel. }
\bigskip

\clearpage \section{Introduction}

Ever since the seminal papers [A] and [W], mathematicians and
physicists have been interested in the problem of construction of
topological quantum field theories. From the beginning of the subject,
the indication has been that the most important example would be in
dimension four, namely, Donaldson-Floer theory. The belief, which has
never been rigorously substantiated in general, is that 4D-TQFTs can
be constructed, which would generalize the line of development
initiated by Donaldson [D] (cf. also [DK]),
which has led to the recent advances in our
knowledge of the smooth structures on 4 manifolds.

If we examine the current progress of the efforts at construction of
TQFTs, we see that there is a very deep gap between the situation in
dimensions 2 and 3, and that in dimension 4. The picture in D=2,3 is
that the TQFTs can be directly constructed in various ways by
connecting structures from abstract algebra (including categorical
algebra) to various decompositions of manifolds. For example, in D=2,
we can construct TQFTs either by connecting a handlebody decomposition
to a commutative Frobenius algebra, or by connecting a triangulation
to a semisimple algebra [CFS]. In D=3, we can either produce a theory by
relating a Heegaard splitting (or more generally, a handlebody
decomposition) or a triangulation to a Hopf algebra or
modular tensor category [TV,Cr1,Ku]. More recently,
it has also been shown how to get a 3D theory by connecting a
surgery presentation to a Hopf algebra [RT]. (cf. also [KaLi])

In contrast, the picture in D=4 is much less clear. Aside from trivial
examples involving finite groups, the 4D theories under active study
have not been constructed in general. Special cases are computed,
either by extremely difficult methods involving analysis on moduli
spaces of instantons [D,DK], or, on a nonrigorous level, by means of
techniques from quantum field theory [W1].

The purpose of this paper is to begin to bridge the gap between the 3D
and 4D situations. In [CY], the authors produced a new 4D-TQFT by using
methods analogous to the lower dimensional constructions. More
specifically, we showed how to produce a 4D-TQFT from the category of
representations of the quantum group $SU(2)_q$ with q a root of unity.
The present paper provides formal proofs and a general setting for the
construction. We prove the theorem that the analog of the construction
in [CY] for an artinian semisimple tortile category gives rise to a 4D-TQFT.
We also present an approach to proving that a state sum is
topological, the blob property, which may have applications in other settings.

In our original case, it has been demonstrated in [CKY1] and [Ro1], that the
invariant which we obtain for a closed 4 manifold is a combination of
the Euler character and the signature. In particular, our formula
gives a new solution to the classical problem, first solved by
Gelfand, of finding a combinatorial formula for the signature of a 4
manifold.

There remains the question of how closely our theory is related to
Donaldson-Floer theory. One might naively think, that since the
invariants we attach
to closed manifolds are topological, i.e., not dependent on smooth
structure, that there would be no hope of any significant
relationship.

The actual situation is somewhat more complex, and turns on the issue of
insertions. It is quite a general phenomenon that constructions of
TQFTs can be extended to give invariants of manifolds with labelled
imbedded submanifolds. For example, the CSW 3D-TQFT can be easily
extended to give invariants of framed labelled graphs, which are, in
fact, generalizations of the Jones polynomial [W2, Cr1, Wa RT].

If we could construct DF theory as a TQFT,its topological
significance would depend, in an essential manner, on two types of insertions,
one on surfaces, and one on points. The insertion on points
corresponds to a twist in the bundle, ie to changing the second
Chern class of the bundle in which the DF theory takes place. The
insertion on surfaces is represented in the picture of DF theory from
moduli spaces, by restricting to submanifolds of moduli space
corresponding to connections which have nonvanishing index when
restricted to the surfaces. In [W3], Witten formally reproduces DF
theory, in the special case of Kahler manifolds, as a TQFT with just
such insertions, although in a nonrigorous approach.

It is then natural to ask if we can find topologically invariant ways
of modifying our state sum to include insertions, and if so, whether they
produce results
related to DF theory.

Our investigation of this question is so far incomplete, but the
results are interesting. In [Y2] and [Ro2], two closely related procedures
have been found for modifying our formula to include insertions on
surfaces. These prescriptions are invariant under homotopy of the
surfaces. In [Ro2], it was demonstrated that the invariant of a manifold
with embedded surfaces counts the intersection numbers of the
surfaces.

This is an intriguing result, since there has recently appeared some
new information about DF theory, deduced rigorously in [KM] and
nonrigorously (but very beautifully) in [W3]. The formula these
sources derive shows that the generating function for the DF
invariants
can be expressed as an exponential involving the Euler character and
signature, times a quadratic exponential involving the intersection
form, times a sum of ``subdominant'' exponentials which are sensitive
to the smooth structure of the manifold. It is not possible to see the
effect of the subdominant exponentials without looking at terms
corresponding to twisted bundles.

It follows that the question of how much information our theory can
detect is closely connected to the question of whether we can find a
natural way to modify it to include twists in the bundle, and what
effect they have on the sum. We have not solved this problem as of this
writing, but we see several natural approaches.
Unfortunately, the dimension of the vector space which our TQFT
assigns to $S^3$ is 1, so it is hard to see how an ``instanton'' could
contribute anything more than a multiplier. Thus, it is
still not completely clear whether a natural modification of our
expression will
make it sensitive to smooth structure, but it is most probable that we
have reconstructed the fictitious cousin of DF theory discussed in [W3].

Our formula is not the only possible approach to a 4D state sum. In
[CF], another approach is outlined, making use of a more subtle piece
of algebraic structure, a Hopf category. There is reason to
hope that the entire picture in 4D can be rendered as algebraic as the
lower dimensional cases.

\smallskip

{\bf Physical Applications}

\smallskip

Let us also mention the possibility that 4D topological state sums may
play a role in the problem of quantizing gravity. The first piece of
evidence we can cite is the work of Regge and Ponzano [RP] on spin
networks. They reinterpreted Penrose's spin network [P] approach tn
quantum gravity by using the techniques of the graphical calculus to
rewrite the evaluation of a spin network as a  sort of discretized
path integral. The form of their expression is identical to the TQFT
of Turaev and Viro [TV], except that they use a Lie group instead of
a quantum group.
Their state sum gives an interesting approach to quantum theory
for 3D gravity. It is then natural to wonder if a 4D state sum
model could play a similar role. This leads into great complexities of
interpretation, but see [Cr2] for a possible approach.

A topological state sum has many attractive features as a tool to
describe a quantum theory of gravity. It occupies a position
intermediate between a path integral for a continuum theory and a
lattice approximation to the theory,
as a sort of magic lattice theory which is invariant under any
change of the lattice. This resonates nicely with the old idea that it
is not possible to measure the distance between two physical points and get a
value less than the Planck scale.

In summation, topological 4D state sums are a very new construction,
whose possibilities have not been explored fully, which may have many
applications.
\smallskip

Throughout, all manifolds are assumed to be piecewise-linear (equivalently
smooth) oriented, and unless stated to the contrary to have empty boundary.

\clearpage \section{Tortile Categories}

This section has two subsections. In the first, we review the properties
of two versions of the recoupling theory associated with the quantum group
$U_q(sl_2)$. These recoupling theories and their properties can be used
directly to build the simplest cases of the 4-manifold invariants discussed
in this paper. This is the example which has been most fully understood.
The reader who is interested primarily in this case of the construction
can read the first subsection, then proceed directly to Section 3.

The second subsection describes the general setting for our construction
in terms of semi-simple tortile categories. This general setting for
the invariants is of great potential value since it gives a framework
in which future applications to quantum groups or other
categories can be cradled.

\subsection{ $U_q(sl_2)$ Recoupling Theory}

In this section we give a quick resum\'{e} of two versions of $U_q(sl_2)$
recoupling theory and the relationships between them.  Both formulations
are useful in studying our 4-manifold invariants, and in the following
sections we shall express the invariants in terms of both.

We begin with a review of the knot-theoretic and combinatorial Temperley-Lieb
recoupling theory.  The principal reference for this version is [KaLi]
We shall refer to this as the {\em TL theory}.

The TL theory is based on the bracket polynomial model for the original
Jones polynomial [J].  Recall that the bracket polynomial $<K>$ is a
Laurent polynomial in $A$ satisfying the relations in Figure \ref{brax}

\begin{figure}[htb]
\begin{centering}
\setlength{\unitlength}{0.0125in}%
\begin{picture}(440,180)(45,620)
\thinlines
\put( 60,800){\line(-1,-2){ 15}}
\put( 60,740){\line(-1, 2){ 15}}
\put(120,800){\line( 1,-2){ 15}}
\put(120,740){\line( 1, 2){ 15}}
\put(230,800){\line(-1,-2){ 15}}
\put(230,740){\line(-1, 2){ 15}}
\put(290,800){\line( 1,-2){ 15}}
\put(290,740){\line( 1, 2){ 15}}
\put(410,800){\line(-1,-2){ 15}}
\put(410,740){\line(-1, 2){ 15}}
\put(470,800){\line( 1,-2){ 15}}
\put(470,740){\line( 1, 2){ 15}}
\put( 70,790){\line( 1,-1){ 40}}
\put(110,790){\line(-1,-1){ 15}}
\put( 85,765){\line(-1,-1){ 15}}
\multiput(280,790)(-0.25000,-0.50000){21}{\makebox(0.4444,0.6667){\sevrm .}}
\multiput(275,780)(-0.50000,-0.25000){21}{\makebox(0.4444,0.6667){\sevrm .}}
\put(265,775){\line(-1, 0){  5}}
\multiput(240,790)(0.25000,-0.50000){21}{\makebox(0.4444,0.6667){\sevrm .}}
\multiput(245,780)(0.50000,-0.25000){21}{\makebox(0.4444,0.6667){\sevrm .}}
\put(255,775){\line( 1, 0){  5}}
\multiput(240,750)(0.25000,0.50000){21}{\makebox(0.4444,0.6667){\sevrm .}}
\multiput(245,760)(0.50000,0.25000){21}{\makebox(0.4444,0.6667){\sevrm .}}
\put(255,765){\line( 1, 0){  5}}
\multiput(280,750)(-0.25000,0.50000){21}{\makebox(0.4444,0.6667){\sevrm .}}
\multiput(275,760)(-0.50000,0.25000){21}{\makebox(0.4444,0.6667){\sevrm .}}
\put(265,765){\line(-1, 0){  5}}
\multiput(420,790)(0.50000,-0.25000){21}{\makebox(0.4444,0.6667){\sevrm .}}
\multiput(430,785)(0.25000,-0.50000){21}{\makebox(0.4444,0.6667){\sevrm .}}
\put(435,775){\line( 0,-1){  5}}
\multiput(420,750)(0.50000,0.25000){21}{\makebox(0.4444,0.6667){\sevrm .}}
\multiput(430,755)(0.25000,0.50000){21}{\makebox(0.4444,0.6667){\sevrm .}}
\put(435,765){\line( 0, 1){  5}}
\multiput(460,750)(-0.50000,0.25000){21}{\makebox(0.4444,0.6667){\sevrm .}}
\multiput(450,755)(-0.25000,0.50000){21}{\makebox(0.4444,0.6667){\sevrm .}}
\put(445,765){\line( 0, 1){  5}}
\multiput(460,790)(-0.50000,-0.25000){21}{\makebox(0.4444,0.6667){\sevrm .}}
\multiput(450,785)(-0.25000,-0.50000){21}{\makebox(0.4444,0.6667){\sevrm .}}
\put(445,775){\line( 0,-1){  5}}
\put( 60,680){\line(-1,-2){ 15}}
\put( 60,620){\line(-1, 2){ 15}}
\put(120,680){\line( 1,-2){ 15}}
\put(120,620){\line( 1, 2){ 15}}
\put(280,680){\line(-1,-2){ 15}}
\put(280,620){\line(-1, 2){ 15}}
\put(340,680){\line( 1,-2){ 15}}
\put(340,620){\line( 1, 2){ 15}}
\put( 75,650){\circle{30}}
\put(160,760){\makebox(0,0)[lb]{\raisebox{0pt}[0pt][0pt]{\twlrm $=$}}}
\put(200,760){\makebox(0,0)[lb]{\raisebox{0pt}[0pt][0pt]{\twlrm $A$}}}
\put(335,760){\makebox(0,0)[lb]{\raisebox{0pt}[0pt][0pt]{\twlrm $+$}}}
\put(370,760){\makebox(0,0)[lb]{\raisebox{0pt}[0pt][0pt]{\twlrm $A^-1$}}}
\put(100,640){\makebox(0,0)[lb]{\raisebox{0pt}[0pt][0pt]{\twlrm $K$}}}
\put(160,640){\makebox(0,0)[lb]{\raisebox{0pt}[0pt][0pt]{\twlrm $=$}}}
\put(195,640){\makebox(0,0)[lb]{\raisebox{0pt}[0pt][0pt]
{\twlrm $(-A^2-A^{-2})$}}}
\put(305,640){\makebox(0,0)[lb]{\raisebox{0pt}[0pt][0pt]{\twlrm $K$}}}
\end{picture}
\end{centering}
\caption{\label{brax} Axioms for the Bracket Polynomial}
\end{figure}

In the first equation,
 the small diagrams stand for larger link diagrams differing only
at the site shown. In the second, the circle stands for an extra
component disjoint from the rest of the diagram. We let $d = -A^2-A^{-2}$.

The bracket evaluation is an invariant of regular isotopy---the equivalence
relation generated by the Reidemeister moves II and III, shown in Figure
\ref{regiso}.

\begin{figure}[htb]
\begin{centering}
\setlength{\unitlength}{0.0125in}%
\begin{picture}(300,240)(40,540)
\thinlines
\multiput(140,780)(-0.50000,-0.25000){21}{\makebox(0.4444,0.6667){\sevrm .}}
\put(130,775){\line(-2,-3){ 10}}
\multiput(120,760)(-0.25000,-0.50000){21}{\makebox(0.4444,0.6667){\sevrm .}}
\put(115,750){\line( 0,-1){ 10}}
\multiput(140,700)(-0.50000,0.25000){21}{\makebox(0.4444,0.6667){\sevrm .}}
\put(130,705){\line(-2, 3){ 10}}
\multiput(120,720)(-0.25000,0.50000){21}{\makebox(0.4444,0.6667){\sevrm .}}
\put(115,730){\line( 0, 1){ 10}}
\multiput(120,620)(-0.50000,-0.25000){21}{\makebox(0.4444,0.6667){\sevrm .}}
\multiput(110,615)(-0.25000,-0.50000){21}{\makebox(0.4444,0.6667){\sevrm .}}
\multiput(100,595)(-0.25000,-0.50000){21}{\makebox(0.4444,0.6667){\sevrm .}}
\put( 95,585){\line( 0,-1){  5}}
\multiput(120,540)(-0.50000,0.25000){21}{\makebox(0.4444,0.6667){\sevrm .}}
\multiput(110,545)(-0.25000,0.50000){21}{\makebox(0.4444,0.6667){\sevrm .}}
\multiput(100,565)(-0.25000,0.50000){21}{\makebox(0.4444,0.6667){\sevrm .}}
\put( 95,575){\line( 0, 1){  5}}
\put( 80,620){\line( 1,-1){ 80}}
\put(160,620){\line(-1,-1){ 35}}
\put(115,575){\line(-1,-1){ 35}}
\multiput(300,540)(0.50000,0.25000){21}{\makebox(0.4444,0.6667){\sevrm .}}
\multiput(310,545)(0.25000,0.50000){21}{\makebox(0.4444,0.6667){\sevrm .}}
\multiput(320,565)(0.25000,0.50000){21}{\makebox(0.4444,0.6667){\sevrm .}}
\put(325,575){\line( 0, 1){  5}}
\multiput(300,620)(0.50000,-0.25000){21}{\makebox(0.4444,0.6667){\sevrm .}}
\multiput(310,615)(0.25000,-0.50000){21}{\makebox(0.4444,0.6667){\sevrm .}}
\multiput(320,595)(0.25000,-0.50000){21}{\makebox(0.4444,0.6667){\sevrm .}}
\put(325,585){\line( 0,-1){  5}}
\put(340,540){\line(-1, 1){ 80}}
\put(260,540){\line( 1, 1){ 35}}
\put(305,585){\line( 1, 1){ 35}}
\multiput( 80,780)(0.50000,-0.25000){21}{\makebox(0.4444,0.6667){\sevrm .}}
\put( 90,775){\line( 2,-3){ 10}}
\multiput(100,760)(0.25000,-0.50000){21}{\makebox(0.4444,0.6667){\sevrm .}}
\put(105,750){\line( 0,-1){ 10}}
\multiput( 80,700)(0.50000,0.25000){21}{\makebox(0.4444,0.6667){\sevrm .}}
\put( 90,705){\line( 2, 3){ 10}}
\multiput(100,720)(0.25000,0.50000){21}{\makebox(0.4444,0.6667){\sevrm .}}
\put(105,730){\line( 0, 1){ 10}}
\put(260,700){\line( 5, 1){ 25}}
\put(285,705){\line( 4, 3){ 20}}
\multiput(305,720)(0.25000,0.50000){21}{\makebox(0.4444,0.6667){\sevrm .}}
\put(310,730){\line( 0, 1){ 10}}
\put(260,780){\line( 5,-1){ 25}}
\put(285,775){\line( 4,-3){ 20}}
\multiput(305,760)(0.25000,-0.50000){21}{\makebox(0.4444,0.6667){\sevrm .}}
\put(310,750){\line( 0,-1){ 10}}
\put(320,780){\line(-4,-1){ 20}}
\put(320,700){\line(-4, 1){ 20}}
\put(285,765){\line(-2,-3){ 10}}
\put(275,750){\line( 0,-1){ 10}}
\put(285,715){\line(-2, 3){ 10}}
\put(275,730){\line( 0, 1){ 10}}
\put(180,740){\vector(-1, 0){  0}}
\put(180,740){\vector( 1, 0){ 40}}
\put(190,580){\vector(-1, 0){  0}}
\put(190,580){\vector( 1, 0){ 40}}
\put( 40,730){\makebox(0,0)[lb]{\raisebox{0pt}[0pt][0pt]{\twlrm {\bf II.}}}}
\put( 40,580){\makebox(0,0)[lb]{\raisebox{0pt}[0pt][0pt]{\twlrm {\bf III.}}}}
\end{picture}
\end{centering}
\caption{\label{regiso} Regular Isotopy---Reidemeister Moves II and III}
\end{figure}

\begin{figure}[htb]
\begin{centering}
\setlength{\unitlength}{0.0125in}%
\begin{picture}(285,180)(105,620)
\thinlines
\put(120,800){\line(-1,-2){ 15}}
\put(120,740){\line(-1, 2){ 15}}
\put(180,800){\line( 1,-2){ 15}}
\put(180,740){\line( 1, 2){ 15}}
\put(315,800){\line(-1,-2){ 15}}
\put(315,740){\line(-1, 2){ 15}}
\put(375,800){\line( 1,-2){ 15}}
\put(375,740){\line( 1, 2){ 15}}
\put(315,680){\line(-1,-2){ 15}}
\put(315,620){\line(-1, 2){ 15}}
\put(375,680){\line( 1,-2){ 15}}
\put(375,620){\line( 1, 2){ 15}}
\put(120,680){\line(-1,-2){ 15}}
\put(120,620){\line(-1, 2){ 15}}
\put(180,680){\line( 1,-2){ 15}}
\put(180,620){\line( 1, 2){ 15}}
\put(120,780){\line( 1, 0){ 10}}
\put(130,780){\line( 4,-1){ 20}}
\multiput(150,775)(0.41667,-0.41667){13}{\makebox(0.4444,0.6667){\sevrm .}}
\put(155,770){\line( 0,-1){  5}}
\multiput(155,765)(-0.25000,-0.50000){21}{\makebox(0.4444,0.6667){\sevrm .}}
\put(150,755){\line(-1, 0){  5}}
\multiput(145,755)(-0.41667,0.41667){13}{\makebox(0.4444,0.6667){\sevrm .}}
\put(140,760){\line( 0, 1){  5}}
\multiput(140,765)(0.41667,0.41667){13}{\makebox(0.4444,0.6667){\sevrm .}}
\multiput(155,775)(0.50000,0.25000){21}{\makebox(0.4444,0.6667){\sevrm .}}
\put(165,780){\line( 1, 0){ 15}}
\put(180,660){\line(-1, 0){ 10}}
\put(170,660){\line(-4,-1){ 20}}
\multiput(150,655)(-0.41667,-0.41667){13}{\makebox(0.4444,0.6667){\sevrm .}}
\put(145,650){\line( 0,-1){  5}}
\multiput(145,645)(0.25000,-0.50000){21}{\makebox(0.4444,0.6667){\sevrm .}}
\put(150,635){\line( 1, 0){  5}}
\multiput(155,635)(0.41667,0.41667){13}{\makebox(0.4444,0.6667){\sevrm .}}
\put(160,640){\line( 0, 1){  5}}
\multiput(160,645)(-0.41667,0.41667){13}{\makebox(0.4444,0.6667){\sevrm .}}
\multiput(145,655)(-0.50000,0.25000){21}{\makebox(0.4444,0.6667){\sevrm .}}
\put(135,660){\line(-1, 0){ 15}}
\put(320,780){\line( 1, 0){ 50}}
\put(320,660){\line( 1, 0){ 50}}
\put(230,760){\makebox(0,0)[lb]{\raisebox{0pt}[0pt][0pt]{\twlrm $=$}}}
\put(265,760){\makebox(0,0)[lb]{\raisebox{0pt}[0pt][0pt]{\twlrm $-A^3$}}}
\put(225,640){\makebox(0,0)[lb]{\raisebox{0pt}[0pt][0pt]{\twlrm $=$}}}
\put(265,640){\makebox(0,0)[lb]{\raisebox{0pt}[0pt][0pt]{\twlrm $-A^{-3}$}}}
\end{picture}
\end{centering}
\caption{\label{brRI} The Bracket under Reidemeister I}
\end{figure}

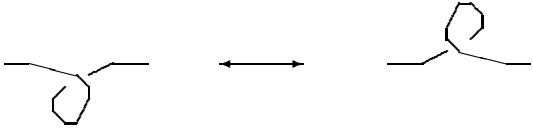
\begin{figure}[htb]
\begin{centering}
\setlength{\unitlength}{0.0125in}%
\begin{picture}(220,50)(120,755)
\thinlines
\put(120,780){\line( 1, 0){ 10}}
\put(130,780){\line( 4,-1){ 20}}
\multiput(150,775)(0.41667,-0.41667){13}{\makebox(0.4444,0.6667){\sevrm .}}
\put(155,770){\line( 0,-1){  5}}
\multiput(155,765)(-0.25000,-0.50000){21}{\makebox(0.4444,0.6667){\sevrm .}}
\put(150,755){\line(-1, 0){  5}}
\multiput(145,755)(-0.41667,0.41667){13}{\makebox(0.4444,0.6667){\sevrm .}}
\put(140,760){\line( 0, 1){  5}}
\multiput(140,765)(0.41667,0.41667){13}{\makebox(0.4444,0.6667){\sevrm .}}
\multiput(155,775)(0.50000,0.25000){21}{\makebox(0.4444,0.6667){\sevrm .}}
\put(165,780){\line( 1, 0){ 15}}
\put(340,780){\line(-1, 0){ 10}}
\put(330,780){\line(-4, 1){ 20}}
\multiput(310,785)(-0.41667,0.41667){13}{\makebox(0.4444,0.6667){\sevrm .}}
\put(305,790){\line( 0, 1){  5}}
\multiput(305,795)(0.25000,0.50000){21}{\makebox(0.4444,0.6667){\sevrm .}}
\put(310,805){\line( 1, 0){  5}}
\multiput(315,805)(0.41667,-0.41667){13}{\makebox(0.4444,0.6667){\sevrm .}}
\put(320,800){\line( 0,-1){  5}}
\multiput(320,795)(-0.41667,-0.41667){13}{\makebox(0.4444,0.6667){\sevrm .}}
\multiput(305,785)(-0.50000,-0.25000){21}{\makebox(0.4444,0.6667){\sevrm .}}
\put(295,780){\line(-1, 0){ 15}}
\put(210,780){\vector(-1, 0){  0}}
\put(210,780){\vector( 1, 0){ 35}}
\end{picture}
\end{centering}
\caption{\label{ffRI} The Framed First Reidemeister Move}
\end{figure}

Under the first Reidemeister move, we have the equations of Figure \ref{brRI}
 from which it is readily seen that the bracket is, moreover,
invariant under framed isotopy---the equivalence generated by Reidemeister
Moves II and III, and the ``framed first Reidemeister move'' of Figure
\ref{ffRI}---an observation which is important to the applications
of the bracket to surgery descriptions of manifolds, and to the categorical
formulation of the second subsection.

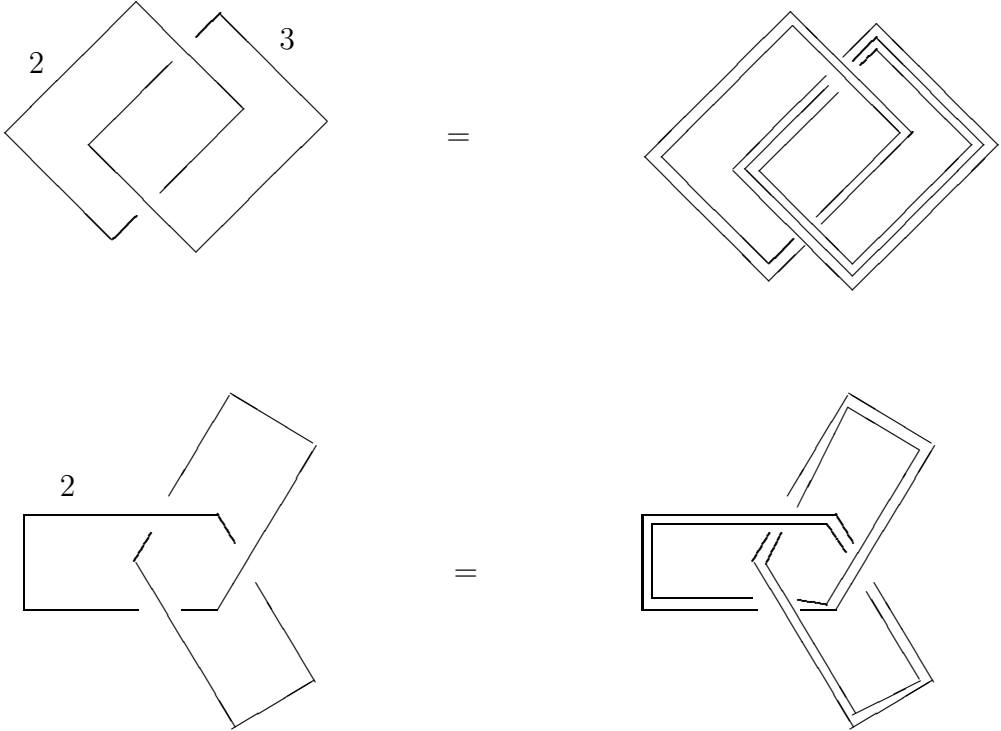
\begin{figure}[htb]
\begin{centering}
\setlength{\unitlength}{0.0125in}%
\begin{picture}(416,305)(45,440)
\thinlines
\put(100,745){\line( 1,-1){ 45}}
\put(145,700){\line(-1,-1){ 35}}
\put(125,640){\line(-1, 1){ 45}}
\put( 80,685){\line( 1, 1){ 35}}
\put(100,745){\line(-1,-1){ 55}}
\put( 45,690){\line( 1,-1){ 45}}
\multiput( 90,645)(0.40000,0.40000){26}{\makebox(0.4444,0.6667){\sevrm .}}
\put(125,640){\line( 1, 1){ 55}}
\put(180,695){\line(-1, 1){ 45}}
\multiput(135,740)(-0.40000,-0.40000){26}{\makebox(0.4444,0.6667){\sevrm .}}
\put(375,735){\line( 1,-1){ 45}}
\put(420,690){\line(-1,-1){ 35}}
\put(400,630){\line(-1, 1){ 45}}
\put(355,675){\line( 1, 1){ 35}}
\put(375,735){\line(-1,-1){ 55}}
\put(320,680){\line( 1,-1){ 45}}
\multiput(365,635)(0.40000,0.40000){26}{\makebox(0.4444,0.6667){\sevrm .}}
\put(400,630){\line( 1, 1){ 55}}
\put(455,685){\line(-1, 1){ 45}}
\multiput(410,730)(-0.40000,-0.40000){26}{\makebox(0.4444,0.6667){\sevrm .}}
\put(134,530){\line(-1, 0){ 81}}
\put( 53,530){\line( 0,-1){ 40}}
\put( 53,490){\line( 1, 0){ 48}}
\put(119,490){\line( 1, 0){ 15}}
\put(100,510){\line( 3,-5){ 41.471}}
\put(140,440){\line( 5, 3){ 34.559}}
\put(175,460){\line(-3, 5){ 24.882}}
\multiput(141,518)(-0.28588,0.47647){26}{\makebox(0.4444,0.6667){\sevrm .}}
\put(134,490){\line( 3, 5){ 41.471}}
\put(174,560){\line(-5, 3){ 34.559}}
\put(139,580){\line(-3,-5){ 25.323}}
\multiput(106,522)(-0.28588,-0.47647){26}{\makebox(0.4444,0.6667){\sevrm .}}
\put(393,530){\line(-1, 0){ 81}}
\put(312,530){\line( 0,-1){ 40}}
\put(312,490){\line( 1, 0){ 48}}
\put(378,490){\line( 1, 0){ 15}}
\put(359,510){\line( 3,-5){ 41.471}}
\put(399,440){\line( 5, 3){ 34.559}}
\put(434,460){\line(-3, 5){ 24.882}}
\multiput(400,518)(-0.28588,0.47647){26}{\makebox(0.4444,0.6667){\sevrm .}}
\put(393,490){\line( 3, 5){ 41.471}}
\put(433,560){\line(-5, 3){ 34.559}}
\put(398,580){\line(-3,-5){ 25.323}}
\multiput(365,522)(-0.28588,-0.47647){26}{\makebox(0.4444,0.6667){\sevrm .}}
\put(387,652){\line( 1, 1){ 36}}
\multiput(423,688)(0.40000,0.40000){6}{\makebox(0.4444,0.6667){\sevrm .}}
\put(425,690){\line(-1, 1){ 51}}
\put(374,741){\line(-1,-1){ 61}}
\put(313,680){\line( 1,-1){ 52}}
\put(365,628){\line( 1, 1){ 15}}
\put(350,674){\line( 1,-1){ 50}}
\put(400,624){\line( 1, 1){ 61}}
\put(461,685){\line(-1, 1){ 51}}
\put(410,736){\line(-1,-1){ 14}}
\put(389,714){\line(-1,-1){ 39}}
\put(394,707){\line(-1,-1){ 33}}
\put(361,674){\line( 1,-1){ 39}}
\put(400,635){\line( 1, 1){ 50}}
\put(450,685){\line(-1, 1){ 40}}
\multiput(410,725)(-0.41176,-0.41176){18}{\makebox(0.4444,0.6667){\sevrm .}}
\multiput(397,514)(-0.32000,0.48000){26}{\makebox(0.4444,0.6667){\sevrm .}}
\put(389,526){\line(-1, 0){ 73}}
\put(316,526){\line( 0,-1){ 31}}
\put(316,495){\line( 1, 0){ 42}}
\multiput(370,522)(-0.25600,-0.51200){26}{\makebox(0.4444,0.6667){\sevrm .}}
\put(364,509){\line( 3,-5){ 37.324}}
\put(400,446){\line( 2, 1){ 28.400}}
\put(428,461){\line(-3, 5){ 22.147}}
\multiput(377,494)(0.57143,-0.09524){22}{\makebox(0.4444,0.6667){\sevrm .}}
\put(389,492){\line( 3, 5){ 39}}
\put(428,557){\line(-5, 3){ 30}}
\put(398,575){\line(-1,-2){ 21}}
\put( 55,715){\makebox(0,0)[lb]{\raisebox{0pt}[0pt][0pt]{\twlrm 2}}}
\put(160,725){\makebox(0,0)[lb]{\raisebox{0pt}[0pt][0pt]{\twlrm 3}}}
\put(230,685){\makebox(0,0)[lb]{\raisebox{0pt}[0pt][0pt]{\twlrm =}}}
\put( 68,538){\makebox(0,0)[lb]{\raisebox{0pt}[0pt][0pt]{\twlrm 2}}}
\put(233,502){\makebox(0,0)[lb]{\raisebox{0pt}[0pt][0pt]{\twlrm =}}}
\end{picture}
\end{centering}
\caption{\label{cabex} Examples of cabling}
\end{figure}

The recoupling formulation based on the bracket polynomial
involves applying the
bracket polynomial to links with parallel cables.  Parallel cabling is
indicated by labelling a link component with an integer $a\geq 0$. That
component is then replaced by $a$ parallel components. \footnote{The reader
must remember that we are dealing with {\em diagrams}, if one insists on
thinking of the link in 3-dimensional space, we are using the ``blackboard
framing''.} An example is given in Figure \ref{cabex}

\begin{figure}[htb]
\begin{centering}
\setlength{\unitlength}{0.0125in}%
\begin{picture}(326,95)(90,605)
\thinlines
\put(356,644){\framebox(60,15){}}
\put(387,660){\line( 0, 1){ 40}}
\put(387,643){\line( 0,-1){ 35}}
\put(383,647){\makebox(0,0)[lb]{\raisebox{0pt}[0pt][0pt]
{\twlrm $\hat{\sigma}$}}}
\put(283,664){\line( 0, 1){  4}}
\put(283,668){\line(-1, 0){ 30}}
\put(253,668){\line( 1,-1){ 18}}
\put(283,636){\line( 0,-1){  4}}
\put(283,632){\line(-1, 0){ 30}}
\put(253,632){\line( 1, 1){ 18}}
\put(217,650){\line( 1, 0){ 17}}
\put(226,657){\makebox(0,0)[lb]{\raisebox{0pt}[0pt][0pt]{\twlrm $1$}}}
\put(221,632){\makebox(0,0)[lb]{\raisebox{0pt}[0pt][0pt]{\twlrm $\{a\}!$}}}
\put( 90,640){\framebox(60,15){}}
\put(120,655){\line( 0, 1){ 40}}
\put(120,640){\line( 0,-1){ 35}}
\put(105,670){\makebox(0,0)[lb]{\raisebox{0pt}[0pt][0pt]{\twlrm $a$}}}
\put(255,615){\makebox(0,0)[lb]{\raisebox{0pt}[0pt][0pt]
{\twlrm $\sigma \in {\cal S}_n$}}}
\put(300,647){\makebox(0,0)[lb]{\raisebox{0pt}[0pt][0pt]
{\twlrm $(A^{-3})^{t(\sigma)}$}}}
\put(181,644){\makebox(0,0)[lb]{\raisebox{0pt}[0pt][0pt]{\twlrm $=$}}}
\end{picture}
\end{centering}
\caption{\label{qsym} q-symmetrizers}
\end{figure}

We then define q-symmetrizers (here $A = \sqrt{q}$) by the formula of
Figure \ref{qsym}
 in which $\{a\}! = \sum_{\sigma \in {\cal S}_n} (A^{-4})^{t(\sigma)}$,
${\cal S}_n$ is the permutation group on n letters, $\hat \sigma$ is the
usual lift of $\sigma$ to positive braids, and $t(\sigma)$ is the least
number of transpositions needed to express $\sigma$.

\begin{figure}[htb]
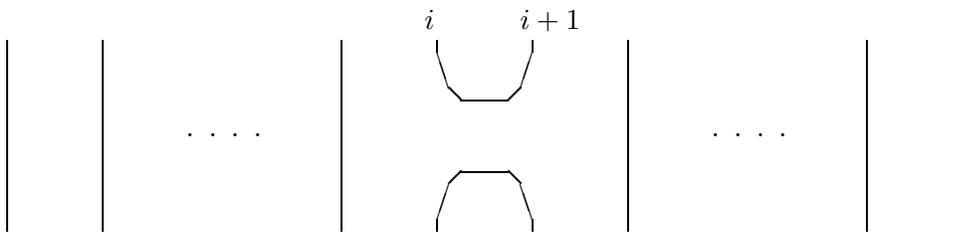

\begin{centering}
\setlength{\unitlength}{0.0125in}%
%
\end{centering}
\caption{\label{TLgen} The $i^{th}$ Temperley-Lieb Generator}
\end{figure}

One then has that the q-symmetrizers are projectors in the sense given
in Figure \ref{qproj} in which $i$ ranges between $1$ and $a-1$,
and $U_i$ is the standard
generator for the Temperley-Lieb algebra shown in Figure \ref{TLgen}.

\begin{figure}[htb]
\begin{centering}
\setlength{\unitlength}{0.0125in}%

\end{centering}
\caption{\label{qproj} }
\end{figure}

\begin{figure}[htb]
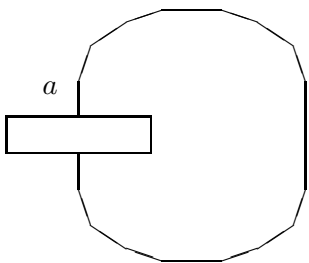

\begin{centering}
\setlength{\unitlength}{0.0125in}%
%
\end{centering}
\caption{\label{nobrack} Brackets Hidden}
\end{figure}

\begin{figure}[htb]
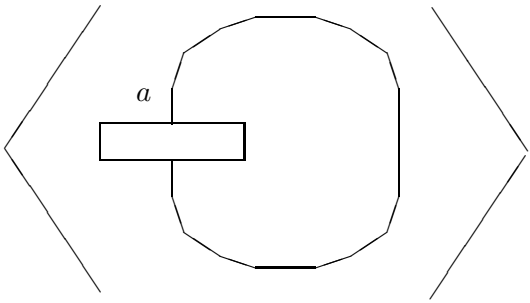

\begin{centering}
\setlength{\unitlength}{0.0125in}%
%
\end{centering}
\caption{\label{brack} Brackets Shown}
\end{figure}

Equations of the sort in Figure \ref{qproj} mean identities for bracket
evaluations of closed diagrams. Hereafter, we omit the brackets when
writing labelled link diagrams in algebraic contexts. Thus, for
example, we use the
expression in Figure \ref{nobrack} instead of that in Figure \ref{brack}

Other useful formulae about the symmetrizers are given in Figure \ref{useful}.

\begin{figure}[htb]
\begin{centering}
\setlength{\unitlength}{0.0125in}%
%
\end{centering}
\caption{\label{useful} Useful Formulae}
\end{figure}

\begin{figure}[htb]
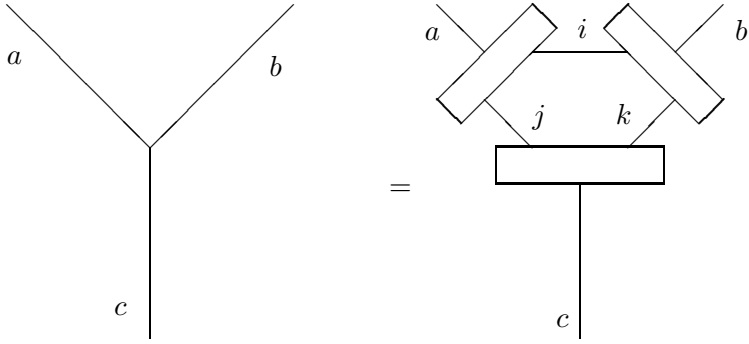

\begin{centering}
\setlength{\unitlength}{0.0125in}%
%
\end{centering}
\caption{\label{TL3} Definition of 3-vertices}
\end{figure}

The $q$-symmetrizers are used to build {\em 3-vertices} and these are the
core of the recoupling theory. They are defined in Figure \ref{TL3}
in which $i+j = a, \ i+k = b, \ j+k = c$.

The $(a,b,c)$ vertex exists exactly when $a+b+c$ is even and the sum
of any two elements of the set ${a,b,c}$ is greater than or equal to
the third.  (We will occasionally write vertices where these conditions
fail, the evaluation of any diagram with such a non-existent vertex is
$0$.)

One can check that crossings and 3-vertices are related as in Figure
\ref{br3} in which  $x^\prime = x(x+2)$.

\begin{figure}[htb]
\begin{centering}
\setlength{\unitlength}{0.0125in}%
%
\end{centering}
\caption{\label{br3} Braiding at a 3-vertex}
\end{figure}

Other properties, such as that shown in Figure \ref{brov}
follow directly from the genesis of the 3-vertex in terms of
bracket evaluation.

\begin{figure}[htb]
\begin{centering}
\setlength{\unitlength}{0.0125in}%
%
\end{centering}
\caption{\label{brov} Braiding past a 3-vertex}
\end{figure}

\begin{figure}[htb]
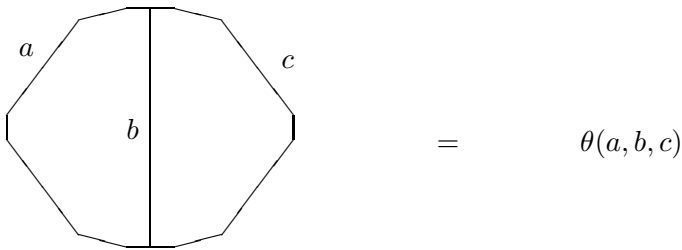

\begin{centering}
\setlength{\unitlength}{0.0125in}%
%
\end{centering}
\caption{\label{thetanet} The Theta Net $\theta (a,b,c)$}
\end{figure}

\begin{figure}[htb]
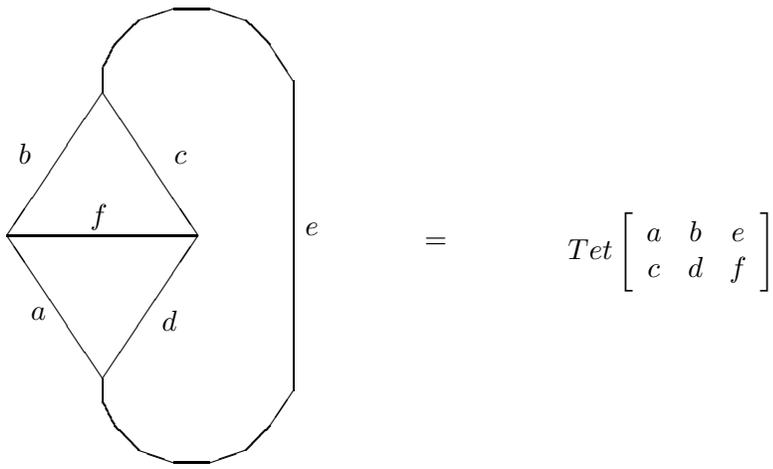

\begin{centering}
\setlength{\unitlength}{0.0125in}%
%
\end{centering}
\caption{\label{tetnet} The Tetrahedral Net  }
\end{figure}

Two important network evaluations have explicit formulas that we omit here
(see Kauffman and Lins [KaLi]).
These are the {\em theta net}, $\theta (a,b,c)$,
 shown in
Figure \ref{thetanet} and the {\em tetrahedral net},
\[ Tet
\left[ \begin{array}{ccc}
a & b & e \\
c & d & f \\
\end{array} \right], \]

\noindent shown in Figure \ref{tetnet}.

\begin{figure}[htb]
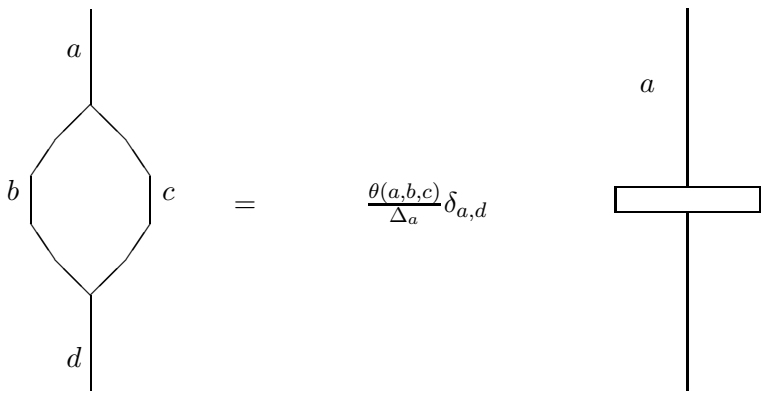

\begin{centering}
\setlength{\unitlength}{0.0125in}%
%
\end{centering}
\caption{\label{TLSchur} Schur's lemma in TL theory}
\end{figure}

A typical instance of Schur's lemma
(see the second subsection) in the TL theory is the
formula of Figure \ref{TLSchur} in which
$\delta_{a,d}$ is the Kronecker delta.

\begin{figure}[h]
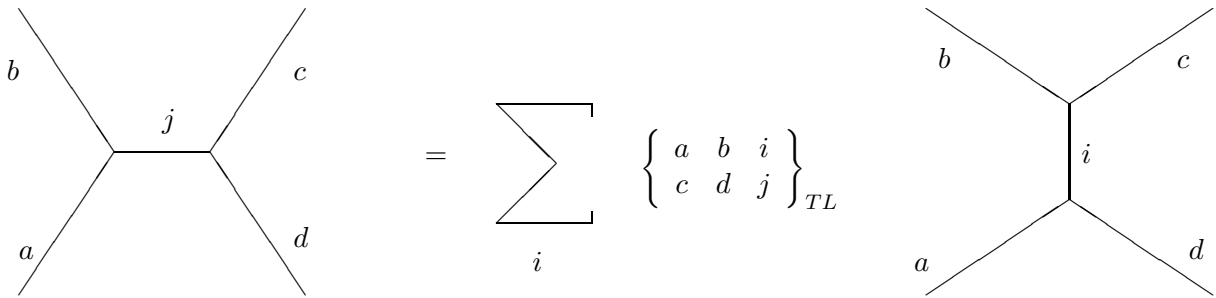

\begin{centering}
\setlength{\unitlength}{0.0125in}%
%
\end{centering}
\caption{\label{fusion} Recoupling via q-6j symbols}
\end{figure}

Recoupling of 3-vertices is given in terms of q-6j symbols, which we
write with a subscript TL to indicate the Temperley-Lieb algebra
formulation as in Figure \ref{fusion}.

These q-6j symbols have an explicit expression in terms of net-evaluations
given by

\[ \left\{ \begin{array}{ccc}
a & b & i \\
c & d & j \\
\end{array} \right\} = \frac{Tet\left[ \begin{array}{ccc}
a & b & i \\
c & d & j \\
\end{array} \right] \Delta_i}{\theta (a,d,i) \theta (b,c,i)}. \]

The q-6j symbols satisfy orthogonality and Biedenharn-Elliot identities. They
can be used to construct 3-manifold invariants, and, as we shall see,
4-manifold invariants.

The TL theory is rooted in the combinatorics of link diagrams, and it is a
direct generalization (q-deformation) of Penrose spin networks (cf. Penrose
[P]). Its advantage to us here is that there is no dependence in the
diagrammatics of the TL theory on maxima or minima or on the orientation
of diagrams with respect to a direction in the plane. Thus TL networks
can be freely embedded in manifolds, a feature that we shall use in later
sections.

	The Kirillov-Reshetikhin formulation of $U_q(sl_2)$ recoupling
theory (cf. Kirillov and Reshetikhin [KR]) is based directly on the
representation theory of the quantum group $U_q(sl_2)$. Although it lacks
the geometric naturalness of the TL theory observed in the last paragraph,
the KR theory has good categorical propeties:  the vertices (in two
flavors---two-in-one-out and one-in-two-out---reading from top to bottom)
are projections and inclusions from a tensor product of irreducible
representations to its irreducible direct summands. It is the
KR theory which is generalized to arbitary semisimple tortile categories
in the next subsection.

	The basic information needed to transform KR nets into TL nets is
the relationship between their 3-vertices shown in Figure \ref{KRtoTL},
in which $\theta (a,b,c)$ is the TL evaluation of a theta-net and we
indicate the theory by
a label on the vertex.

\begin{figure}
\begin{centering}
\setlength{\unitlength}{0.0125in}%
\begin{picture}(460,141)(60,620)
\thinlines
\put(400,761){\line( 1,-1){ 60}}
\put(520,761){\line(-1,-1){ 60}}
\put(460,701){\line( 0,-1){ 80}}
\put(400,736){\makebox(0,0)[lb]{\raisebox{0pt}[0pt][0pt]{\twlrm $a$}}}
\put(510,731){\makebox(0,0)[lb]{\raisebox{0pt}[0pt][0pt]{\twlrm $b$}}}
\put(445,631){\makebox(0,0)[lb]{\raisebox{0pt}[0pt][0pt]{\twlrm $c$}}}
\put( 60,760){\line( 1,-1){ 60}}
\put(180,760){\line(-1,-1){ 60}}
\put(120,700){\line( 0,-1){ 80}}
\put( 60,735){\makebox(0,0)[lb]{\raisebox{0pt}[0pt][0pt]{\twlrm $a$}}}
\put(170,730){\makebox(0,0)[lb]{\raisebox{0pt}[0pt][0pt]{\twlrm $b$}}}
\put(105,630){\makebox(0,0)[lb]{\raisebox{0pt}[0pt][0pt]{\twlrm $c$}}}
\put(132,685){\makebox(0,0)[lb]{\raisebox{0pt}[0pt][0pt]{\twlrm \bf KR}}}
\put(470,687){\makebox(0,0)[lb]{\raisebox{0pt}[0pt][0pt]{\twlrm \bf TL}}}
\put(219,680){\makebox(0,0)[lb]{\raisebox{0pt}[0pt][0pt]{\twlrm $=$}}}
\put(261,679){\makebox(0,0)[lb]{\raisebox{0pt}[0pt][0pt]
{\twlrm $\frac{\sqrt{\Delta_c}}{\sqrt{\theta (a,b,c)}}$}}}
\end{picture}
\end{centering}
\caption{\label{KRtoTL} KR 3-vertices in terms of TL 3-vertices}
\end{figure}

Note that the KR
vertex now has a distinguished direction (since the label on the
downward leg is distinguished from the others), while the TL vertex
does not have any distinguished direction dependent on leg placement.
The value of a closed loop $\Delta_i$ is the same in both formulations.
\footnote{The reader should note that KR nets are often labelled with
half-integral
``spins''. In that convention, the legs of the KR vertex would be
labelled by $\frac{a}{2}, \ \frac{b}{2}, \ \frac{c}{2}$. To avoid
complicating notation, we shall label both KR nets and TL nets with
integers (twice spin, number of cables, or (non-quantum) dimension$-1$).  }

\clearpage

The formula relating the KR vertex and the TL vertex lets us derive the
identities of KR theory involving cups and caps directly.  In the
next subsection, we will see that similar identities are a general
phenomenon for projection and inclusion maps
for direct summands of tensor products in semi-simple tortile categories.

For example, we have

\begin{propo} \label{cupprop}
?
\end{propo}

\begin{figure}[h]
\begin{centering}
\setlength{\unitlength}{0.0125in}%
\begin{picture}(325,105)(105,665)
\thinlines
\put(365,770){\line( 2,-3){ 30}}
\put(395,725){\line( 0,-1){ 50}}
\put(395,725){\line( 0,-1){ 50}}
\put(395,725){\line( 0,-1){ 50}}
\put(425,770){\line(-2,-3){ 30}}
\put(155,665){\line( 2, 3){ 30}}
\put(185,710){\line( 0, 1){ 50}}
\put(185,710){\line( 0, 1){ 50}}
\put(185,710){\line( 0, 1){ 50}}
\put(215,665){\line(-2, 3){ 30}}
\put(155,665){\line(-1, 0){ 15}}
\put(140,665){\line(-1, 2){ 20}}
\put(120,705){\line( 0, 1){ 55}}
\put(105,750){\makebox(0,0)[lb]{\raisebox{0pt}[0pt][0pt]{\twlrm $a$}}}
\put(170,750){\makebox(0,0)[lb]{\raisebox{0pt}[0pt][0pt]{\twlrm $b$}}}
\put(195,665){\makebox(0,0)[lb]{\raisebox{0pt}[0pt][0pt]{\twlrm $c$}}}
\put(190,710){\makebox(0,0)[lb]{\raisebox{0pt}[0pt][0pt]{\twlrm {\bf KR}}}}
\put(255,710){\makebox(0,0)[lb]{\raisebox{0pt}[0pt][0pt]{\twlrm $=$}}}
\put(285,705){\makebox(0,0)[lb]{\raisebox{0pt}[0pt][0pt]
{\twlrm $\frac{\sqrt{\Delta_b}}{\sqrt{\Delta_c}}$}}}
\put(350,760){\makebox(0,0)[lb]{\raisebox{0pt}[0pt][0pt]{\twlrm $a$}}}
\put(430,755){\makebox(0,0)[lb]{\raisebox{0pt}[0pt][0pt]{\twlrm $b$}}}
\put(405,670){\makebox(0,0)[lb]{\raisebox{0pt}[0pt][0pt]{\twlrm $c$}}}
\put(405,710){\makebox(0,0)[lb]{\raisebox{0pt}[0pt][0pt]{\twlrm {\bf KR}}}}
\end{picture}
\end{centering}
\caption{Rotational properties of KR vertices as derived from TL theory}
\end{figure}

\noindent{\bf Proof:} Given in Figure \ref{cupKR}.

\begin{figure}[htb]
\begin{centering}
\setlength{\unitlength}{0.0125in}%
\begin{picture}(430,560)(105,200)
\thinlines
\put(155,665){\line( 2, 3){ 30}}
\put(185,710){\line( 0, 1){ 50}}
\put(185,710){\line( 0, 1){ 50}}
\put(185,710){\line( 0, 1){ 50}}
\put(215,665){\line(-2, 3){ 30}}
\put(155,665){\line(-1, 0){ 15}}
\put(140,665){\line(-1, 2){ 20}}
\put(120,705){\line( 0, 1){ 55}}
\put(105,750){\makebox(0,0)[lb]{\raisebox{0pt}[0pt][0pt]{\twlrm $a$}}}
\put(170,750){\makebox(0,0)[lb]{\raisebox{0pt}[0pt][0pt]{\twlrm $b$}}}
\put(195,665){\makebox(0,0)[lb]{\raisebox{0pt}[0pt][0pt]{\twlrm $c$}}}
\put(475,665){\line( 2, 3){ 30}}
\put(505,710){\line( 0, 1){ 50}}
\put(505,710){\line( 0, 1){ 50}}
\put(505,710){\line( 0, 1){ 50}}
\put(535,665){\line(-2, 3){ 30}}
\put(430,605){\line( 2,-3){ 30}}
\put(460,560){\line( 0,-1){ 50}}
\put(460,560){\line( 0,-1){ 50}}
\put(460,560){\line( 0,-1){ 50}}
\put(490,605){\line(-2,-3){ 30}}
\put(420,590){\makebox(0,0)[lb]{\raisebox{0pt}[0pt][0pt]{\twlrm $a$}}}
\put(490,585){\makebox(0,0)[lb]{\raisebox{0pt}[0pt][0pt]{\twlrm $b$}}}
\put(465,510){\makebox(0,0)[lb]{\raisebox{0pt}[0pt][0pt]{\twlrm $c$}}}
\put(425,440){\line( 2,-3){ 30}}
\put(455,395){\line( 0,-1){ 50}}
\put(455,395){\line( 0,-1){ 50}}
\put(455,395){\line( 0,-1){ 50}}
\put(485,440){\line(-2,-3){ 30}}
\put(415,425){\makebox(0,0)[lb]{\raisebox{0pt}[0pt][0pt]{\twlrm $a$}}}
\put(485,420){\makebox(0,0)[lb]{\raisebox{0pt}[0pt][0pt]{\twlrm $b$}}}
\put(460,345){\makebox(0,0)[lb]{\raisebox{0pt}[0pt][0pt]{\twlrm $c$}}}
\put(425,295){\line( 2,-3){ 30}}
\put(455,250){\line( 0,-1){ 50}}
\put(455,250){\line( 0,-1){ 50}}
\put(455,250){\line( 0,-1){ 50}}
\put(485,295){\line(-2,-3){ 30}}
\put(415,280){\makebox(0,0)[lb]{\raisebox{0pt}[0pt][0pt]{\twlrm $a$}}}
\put(485,275){\makebox(0,0)[lb]{\raisebox{0pt}[0pt][0pt]{\twlrm $b$}}}
\put(460,200){\makebox(0,0)[lb]{\raisebox{0pt}[0pt][0pt]{\twlrm $c$}}}
\put(475,665){\line(-1, 0){ 15}}
\put(460,665){\line(-1, 2){ 20}}
\put(440,705){\line( 0, 1){ 55}}
\put(190,710){\makebox(0,0)[lb]{\raisebox{0pt}[0pt][0pt]{\twlrm {\bf KR}}}}
\put(255,710){\makebox(0,0)[lb]{\raisebox{0pt}[0pt][0pt]{\twlrm $=$}}}
\put(515,710){\makebox(0,0)[lb]{\raisebox{0pt}[0pt][0pt]{\twlrm {\bf TL}}}}
\put(300,700){\makebox(0,0)[lb]{\raisebox{0pt}[0pt][0pt]
{\twlrm $\frac{\sqrt{\Delta_b}}{\sqrt{\theta(a,b,c)}}$}}}
\put(245,565){\makebox(0,0)[lb]{\raisebox{0pt}[0pt][0pt]{\twlrm $=$}}}
\put(295,560){\makebox(0,0)[lb]{\raisebox{0pt}[0pt][0pt]
{\twlrm $\frac{\sqrt{\Delta_b}}{\sqrt{\theta(a,b,c)}}$}}}
\put(425,750){\makebox(0,0)[lb]{\raisebox{0pt}[0pt][0pt]{\twlrm $a$}}}
\put(490,750){\makebox(0,0)[lb]{\raisebox{0pt}[0pt][0pt]{\twlrm $b$}}}
\put(515,665){\makebox(0,0)[lb]{\raisebox{0pt}[0pt][0pt]{\twlrm $c$}}}
\put(245,400){\makebox(0,0)[lb]{\raisebox{0pt}[0pt][0pt]{\twlrm $=$}}}
\put(295,395){\makebox(0,0)[lb]{\raisebox{0pt}[0pt][0pt]
{\twlrm
$\frac{\sqrt{\Delta_b}}{\sqrt{\theta(a,b,c)}}\frac{\sqrt{\theta(a,b,c)}}{\sqrt{\Delta_c}}$}}}
\put(250,230){\makebox(0,0)[lb]{\raisebox{0pt}[0pt][0pt]{\twlrm $=$}}}
\put(295,220){\makebox(0,0)[lb]{\raisebox{0pt}[0pt][0pt]
{\twlrm $\frac{\sqrt{\Delta_b}}{\sqrt{\Delta_c}}$}}}
\put(470,550){\makebox(0,0)[lb]{\raisebox{0pt}[0pt][0pt]{\twlrm {\bf TL}}}}
\put(465,380){\makebox(0,0)[lb]{\raisebox{0pt}[0pt][0pt]{\twlrm {\bf KR}}}}
\put(465,235){\makebox(0,0)[lb]{\raisebox{0pt}[0pt][0pt]{\twlrm {\bf KR}}}}
\end{picture}
\end{centering}
\caption{\label{cupKR} Proof of Proposition \protect\ref{cupprop} }
\end{figure}

In this comparison, the cups and caps of KR theory are the same as
those of TL theory, and thus an $a$ labelled cup (resp. cap) is
the same as a TL $(a, a, 0)$ 3-vertex in the appropriate
orientation or  $\sqrt{\Delta_a}$ times a KR 3-vertex with
two upward legs labelled $a$, and a downward one labelled $0$ (resp.
times a KR 3-vertex with two downward legs labelled $a$ and an upward
one labelled $0$).

In the next subsection, we develop the general form of the categorical
data needed for our constructions. KR and TL recoupling networks provide
two formulations of the most fundamental example: the representation theory
of $U_q(sl_2)$.

\clearpage
\subsection{\label{generalsstc} General semi-simple tortile categories}

The initial data required for our construction is the same as that
required for the constructions of Yetter [Y3]: a semisimple
tortile category over a field $K$. Non-degeneracy conditions as in
Turaev [T] will be required only for the surgical versions given in
Section 4.  We review the necessary axiomatics
and categorical results.

The axioms fall into two types: those dealing with the linearity structure
over a field $K$, and those dealing with the monoidal and duality structure
of the
category. We begin with the latter. We assume familiarity with the basic
notions of monoidal category theory and abelian category theory (cf.
Mac Lane [CWM]) and with
basic notions associated with categories of tangles (cf. Freyd/Yetter
[FY1,FY2], Joyal/Street[JS1,JS2], Resetikhin/Turaev [RT], Shum [S], Yetter
[Y1]).

	Our categories will all be $K$-linear abelian monoidal categories
with $\otimes $ exact in both variables, but will be equipped with additional
structure. One
piece of structure we will require is the presence of dual objects:

\begin{defin} A right (resp. left) dual to an object $X$ in a monoidal category
 $\cal C$ is an object $X^\ast $ (resp. $^\ast X$) equipped with maps
$\epsilon :X\otimes X^\ast \rightarrow I$ and
$\eta :I\rightarrow X^\ast \otimes X$ (resp.
$e:^\ast X\otimes X\rightarrow I$ and
$h:I\rightarrow X\otimes ^\ast X$) such that
the composites

\[ X \stackrel{\rho ^{-1}}{\longrightarrow } X\otimes I
\stackrel{X\otimes \eta}{\longrightarrow } X\otimes(X^\ast \otimes X)
\stackrel{\alpha ^{-1}}{\longrightarrow }
	(X\otimes X^\ast )\otimes X
\stackrel{\epsilon \otimes X}{\longrightarrow } I\otimes X
\stackrel{\lambda }{\longrightarrow } X \]

\noindent and

\[ X^\ast \stackrel{\lambda ^{-1}}{\longrightarrow } I\otimes X^\ast
\stackrel{\eta \otimes X^\ast}{\longrightarrow }
(X^\ast \otimes X)\otimes X^\ast \stackrel{\alpha }{\longrightarrow }
	X^\ast \otimes (X\otimes X^\ast )
\stackrel{X^\ast \otimes \epsilon }{\longrightarrow } X^\ast \otimes I
\stackrel{\rho }{\longrightarrow } X^\ast \]

\noindent (resp.

\[ X \stackrel{\lambda ^{-1}}{\longrightarrow } I\otimes X
\stackrel{h\otimes X}{\longrightarrow } (X\otimes ^\ast X)\otimes X
\stackrel{\alpha }{\longrightarrow }
	X\otimes (^\ast X\otimes X)
\stackrel{X \otimes e}{\longrightarrow } X\otimes I
\stackrel{\rho }{\longrightarrow } X \]

\noindent and

\[ ^\ast X \stackrel{\rho ^{-1}}{\longrightarrow } ^\ast X\otimes I
\stackrel{^\ast X\otimes h}{\longrightarrow }
^\ast X\otimes(X\otimes ^\ast X) \stackrel{\alpha ^{-1}}{\longrightarrow }
	(^\ast X\otimes X)\otimes ^\ast X
\stackrel{e \otimes ^\ast X}{\longrightarrow } I\otimes ^\ast X
\stackrel{\lambda }{\longrightarrow } ^\ast
X {\rm )} \]

\noindent are both identity maps.
\end{defin}

	Observe that a choice of right (resp. left) dual object for each
object of a (small) monoidal category $\cal C$ extends to a contravariant
monoidal functor from
$\cal C$ to its opposite category with $\otimes $ reversed, and that there are
canonical natural isomorphisms $k:^\ast(X^\ast )\rightarrow X$ and
$\kappa :(^\ast X)^\ast \rightarrow X$.

	The categories we consider will have {\em two-sided duals}. To make
sense of this in the non-symmetric setting, we need

\begin{defin} \label{sovereign}
{\rm [Freyd/Yetter [FY2]]} A monoidal category is sovereign if it is equipped
with a choice of left and right duals for all objects, and a (chosen)
monoidal natural isomorphism $\phi:X^\ast \rightarrow ^\ast X$ such that

\[ \phi k = (\phi ^{-1})^\ast \kappa \]

\noindent and $\phi _I = 1_I$.
\end{defin}

\begin{defin} \label{tortile}
{\rm [Shum [S]]} A tortile (tensor) category is a monoidal category
${\cal C} = ({\sl C},\otimes ,I, \alpha, \rho, \lambda)$ in which every object
 has a right dual,
equipped moreover with  natural isomorphisms
$\sigma _{A,B}:A\otimes B\rightarrow B\otimes A$ (the braiding) and
$\theta _{A}:A\rightarrow A$ (the twist) satisfying
\bigskip

\[
\begin{array}{lcl}
 	\mbox{\bf the hexagons} &
		(\tau _{A,B}\otimes C)\alpha _{B,A,C}(B\otimes
\tau _{A,C}) = \alpha _{A,B,C}\tau _{A,B\otimes C}\alpha _{B,C,A}  &
\mbox{for $\tau\ = \sigma ^{\pm 1}$} \\[8pt]
	\mbox{\bf balance} & \theta _{A\otimes B} =
\sigma _{A,B}\sigma _{B,A}(\theta _A\otimes \theta _B) & \\[8pt]
	\mbox{\bf $\theta $ self-dual} & \theta _{A^\ast } =
\theta _A^\ast & \\
\end{array}  \]
\smallskip

\end{defin}

	Of course, the presence of a braiding makes right duals into left
 duals by $e = \sigma ^{-1}\epsilon$ and $h = \eta \sigma $. However,
these are not two-sided duals in the sense of Definition \ref{sovereign}
unless the category satisfies the balance axiom. Indeed, we have

\begin{propo}  {\rm [Deligne [D]] (cf. Yetter [Y1])} A braided monoidal
category with right duals for all objects is balanced if and only if the
category is sovereign
with the chosen right duals and the left duals of the previous paragraph. More
 precisely, a choice of twist is equivalent to a choice of the natural
isomorphism in
Definition \ref{sovereign}.
\end{propo}

	Since our categories will be tortile, we will consider chosen right
duals as two-sided duals under the structure of the previous Proposition.

	In what follows, we will use a diagrammatic notation, similar to
Penrose's [P] notation for tensors, for maps in our categories (see
{\bf Appendix on Diagrammatic
Notation} below). Its use is justified by the following
theorem of Shum [S] (cf. also Freyd/Yetter [FY2], Joyal/Street [JS2],
Reshetikhin/Turaev [RT], Yetter[Y1]):

\begin{thm} {\rm [Shum [S]]} The tortile category freely generated by a single
 object is monoidally equivalent to the category of framed tangles.
\end{thm}

	The second sort of structure we require involves the linear and abelian
 structure on the category.

\begin{defin} \label{simple.obj}
An object $X$ in a $K$-linear category $\cal C$ is {\em simple}
if ${\cal C}[X,X]$ is
 1-dimensional.
\end{defin}

\begin{defin} \label{semisimple}
A $K$-linear abelian category $\cal C$ is {\em completely reducible}
if every object is
isomorphic to a direct sum of simple objects, and a completely reducible
 category is
semisimple if there
are only finitely many
isomorphism classes of simple objects. If $\cal C$ is equipped with an exact
monoidal structure, we also require that the monoidal identity object $I$ be
simple.
\end{defin}

	In what follows we will be concerned with $K$-linear semisimple
abelian categories equipped with an exact tortile structure (for $K$ a field).
For brevity we refer to these
as semisimple tortile categories over $K$.

\begin{lemma}
If $S$ is a simple object in any category $\cal C$ over $K$, then for any
object $X$ ${\cal C}[X,S]$ and ${\cal C}[S,X]$ are canonically
dual as vector-spaces.
\end{lemma}

\noindent {\bf proof:} The composition map $\circ :{\cal C}[S,X]\otimes
{\cal C}[X,S]\longrightarrow {\cal C}[S,S]\cong K$ defines
a non-degenerate bilinear pairing. $\Box$

\begin{lemma}  \label{sumlemma}
Let $\cal C$ be any semisimple category over $K$ with $\cal S$ a family of
representatives for the isomorphism classes of
simple objects in $\cal C$. If $X$ is any object of $\cal C$, then a choice of
bases ${b_{1,S}, \ldots ,b_{d_S,S}} \subset {\cal C}[X,S]$
for each $S \in {\cal S}$ (where $d_S = dim_K{\cal C}[X,S]$) determines a
direct sum decomposition

\[ X = \bigoplus_{S \in {\cal S}} \bigoplus_{i=1}^{d_S} S. \]

\noindent in which the $b_{i,S}$'s are the projections onto the direct
summands, and there are splittings $\overline{b_{i,S}}$ satisfying

\[ \sum_{S \in {\cal S}} \sum_{i=1}^{d_S} b_{i,S}\overline{b_{i,S}} = 1_X \]

\noindent  and $\overline{b_{i,S}}b_{j,T}$ is the zero map from $S$ to $T$,
unless $(i,S) = (j,T)$, in which case it is $1_S$.
\end{lemma}

\noindent {\bf proof:} Now, by the definition of semisimplicity, $X$ admits a
direct sum decomposition of the form given, though not
{\em a priori} having the $b_{i,S}$'s as projections. Let $p_{i,S}$ (resp.
$\overline{p_{i,S}}$) denote the projections (resp. inclusions)
of the summands in this direct sum decomposition. Now for each
$S \in {\cal S}$, $\{p_{i,S}\}$ form a basis for ${\cal C}[X,S]$, while
$\{\overline{p_{i,S}}\}$ form a basis for ${\cal C}[S,X]$ which is the dual
basis under the identification of the previous lemma.
But if $B$ is the change of basis matrix transforming the $p_{i,S}$'s to the
$b_{i,S}$'s then $B^{-1}$ transforms the $\overline{p_{i,S}}$'s
to the $\overline{b_{i,S}}$'s, and thus the $b_{i,S}$'s are the projections
for a (generally different) direct sum decomposition. $\Box$

	We adopt the convention that if we have a basis of ${\cal C}[X,S]$
{\it or of} ${\cal C}[S,X]$ for $S$ any simple object, then for
any basis element, $x$, $\overline{x}$ is the corresponding element of the
dual basis, and thus $\overline{\overline{x}} = x$.

\begin{defin} \label{def:trace}
If $f:X\longrightarrow X$ is any endomorphism in a tortile
category $\cal C$ over $K$ then the trace of $f$, denoted
$tr(f)$, is the map $h(f\otimes 1_{X^\ast })\varepsilon $, or equivalently,
the corresponding element of $K$ under the identification of
${\cal C}[I,I]$ with $K$. The dimension of an object $X$ is $dim(X) = tr(1_X)$.
\end{defin}

The name trace follows from the following, originally proved in the symmetric
case by Kelly and Laplaza [KL]:

\begin{propo}
In any tortile category over $K$, if $f:X\longrightarrow Y$ and
$g:Y\longrightarrow X$ then

\[ tr(fg) = tr(gf). \]
\end{propo}

The following lemma is an immediate consequence of the coherence theorem of
Shum [S]:

\begin{lemma} \label{dual.dim}
If $X$ is any object in a tortile category over $K$, and $X^\ast $ is its dual
object, then

\[ dim(X) = dim(X^\ast ). \]
\end{lemma}

On the other hand in the presence of the additive and linear structure on a
semisimple category over $K$ we have

\begin{lemma} \label{dim.sum}
If $X$ and $Y$ are objects in a semisimple tortile category over $K$ then

\[ dim(X\oplus Y) = dim(X) + dim(Y). \]
\end{lemma}
\smallskip

\noindent {\bf proof:} It follows from the exactness of $( )^\ast$ that
$h_{X\oplus Y} = \delta_I(h_X\oplus h_Y)$ and
$\varepsilon _{X\oplus Y} = (\varepsilon _X\oplus \varepsilon _Y)+$. $\Box$
\smallskip

\begin{lemma} \label{dim.prod}
If $X$ and $Y$ are objects in a semisimple tortile category over $K$ then

\[ dim(X\otimes Y) = dim(X)dim(Y). \]
\end{lemma}
\smallskip

\noindent {\bf proof:} It is immediate from the definition of dimension and
the coherence theorem of Shum [S] that
$dim(X\otimes Y) = dim(X)\otimes dim(Y)$
when the dimensions are regarded as endomorphisms of $I$. But for endomorphisms
 of $I$, $\otimes $, composition, and multiplication
of coefficients of $1_I$ all coincide. $\Box$
\smallskip

The following trivial lemma will be used throughout our construction:

\begin{lemma} \label{schur} {\bf (``Schur's Lemma'')}
If $X$ is any simple object in a semisimple tortile category over $K$ with
$dim(X) \neq 0$ and
$f:X\longrightarrow X$ is any map, then $f$ is
$\frac{tr(f)}{dim(X)} 1_X$.
\end{lemma}

\noindent {\bf proof:} Since $f$ is a scalar multiple of $1_X$ by simplicity,
it suffices to observe that $tr(f)$ must be the multiple
of $dim(X)$ by the same scalar. $\Box $
\bigskip

	We shall assume in what follows that all of our semisimple tortile
categories are non-degenerate in the sense that all simple objects $X$ have
$dim(X) \neq 0$.

An important, though easy consequence of Lemma \ref{schur} concerns the
behaviour of bases ${\cal B} = \{b_1,...,b_n\}$ and dual bases
$\overline{\cal B} = \{\overline{b_1},...,\overline{b_n}\}$ for
${\cal C}[A\otimes B, C]$ and
${\cal C}[C,A\otimes B]$ under dualization of one or more objects.

Even without Lemma \ref{schur} it is clear that $(\, )^\ast$ carries
$\overline{\cal B}$ to a basis for ${\cal C}[B^\ast \otimes A^\ast , C^\ast ]$
and $\cal B$ to the dual basis for ${\cal C}[C^\ast , B^\ast \otimes A^\ast ]$.
\footnote{Throughout this section, we are using the coherence theorem
of Freyd/Yetter [FY2] for sovereign categories
to suppress mention of certain canonical isomorphisms,
for instance, between $(A \otimes B)^\ast $ and $B^\ast \otimes A^\ast$.}

\begin{figure}
\begin{centering}
\setlength{\unitlength}{0.0125in}%
\begin{picture}(491,337)(28,468)
\thinlines
\put( 90,745){\circle{42}}
\put(102,762){\line( 2, 3){ 26}}
\multiput(128,801)(-0.50000,-0.50000){3}{\makebox(0.4444,0.6667){\sevrm .}}
\put( 91,722){\line( 0,-1){ 16}}
\multiput( 91,706)(0.30000,-0.50000){21}{\makebox(0.4444,0.6667){\sevrm .}}
\multiput( 97,696)(0.52632,-0.21053){20}{\makebox(0.4444,0.6667){\sevrm .}}
\put(107,692){\line( 1, 0){ 19}}
\put(126,692){\line( 6, 5){ 12}}
\put(138,702){\line( 3, 5){ 15}}
\put(153,727){\line( 1, 3){ 12}}
\put(165,763){\line( 0, 1){ 42}}
\put( 80,763){\line(-2, 5){ 12}}
\multiput( 68,793)(-0.30000,0.50000){11}{\makebox(0.4444,0.6667){\sevrm .}}
\multiput( 65,798)(-0.55556,0.11111){19}{\makebox(0.4444,0.6667){\sevrm .}}
\put( 55,800){\line(-1, 0){ 11}}
\multiput( 44,800)(-0.50000,-0.33333){13}{\makebox(0.4444,0.6667){\sevrm .}}
\multiput( 38,796)(-0.33333,-0.44444){19}{\makebox(0.4444,0.6667){\sevrm .}}
\put( 32,788){\line(-1,-5){  4}}
\put( 28,768){\line( 0,-1){105}}
\put(421,746){\circle{42}}
\put(407,760){\line(-1, 2){ 22}}
\put(422,724){\line( 0,-1){ 61}}
\put(434,762){\line( 2, 3){ 16}}
\multiput(450,786)(0.44444,0.33333){19}{\makebox(0.4444,0.6667){\sevrm .}}
\put(458,792){\line( 6, 1){ 18}}
\put(476,795){\line( 1, 0){ 14}}
\multiput(490,795)(0.50000,-0.30000){21}{\makebox(0.4444,0.6667){\sevrm .}}
\put(500,789){\line( 2,-5){  6}}
\put(506,774){\line( 0,-1){109}}
\put(420,526){\circle{42}}
\put(406,512){\line(-1,-2){ 22}}
\put(421,548){\line( 0, 1){ 61}}
\put(433,510){\line( 2,-3){ 16}}
\multiput(449,486)(0.44444,-0.33333){19}{\makebox(0.4444,0.6667){\sevrm .}}
\put(457,480){\line( 6,-1){ 18}}
\put(475,477){\line( 1, 0){ 14}}
\multiput(489,477)(0.50000,0.30000){21}{\makebox(0.4444,0.6667){\sevrm .}}
\put(499,483){\line( 2, 5){  6}}
\put(505,498){\line( 0, 1){109}}
\put( 91,528){\circle{42}}
\put(103,511){\line( 2,-3){ 26}}
\multiput(129,472)(-0.50000,0.50000){3}{\makebox(0.4444,0.6667){\sevrm .}}
\put( 92,551){\line( 0, 1){ 16}}
\multiput( 92,567)(0.30000,0.50000){21}{\makebox(0.4444,0.6667){\sevrm .}}
\multiput( 98,577)(0.52632,0.21053){20}{\makebox(0.4444,0.6667){\sevrm .}}
\put(108,581){\line( 1, 0){ 19}}
\put(127,581){\line( 6,-5){ 12}}
\put(139,571){\line( 3,-5){ 15}}
\put(154,546){\line( 1,-3){ 12}}
\put(166,510){\line( 0,-1){ 42}}
\put( 81,510){\line(-2,-5){ 12}}
\multiput( 69,480)(-0.30000,-0.50000){11}{\makebox(0.4444,0.6667){\sevrm .}}
\multiput( 66,475)(-0.55556,-0.11111){19}{\makebox(0.4444,0.6667){\sevrm .}}
\put( 56,473){\line(-1, 0){ 11}}
\multiput( 45,473)(-0.50000,0.33333){13}{\makebox(0.4444,0.6667){\sevrm .}}
\multiput( 39,477)(-0.33333,0.44444){19}{\makebox(0.4444,0.6667){\sevrm .}}
\put( 33,485){\line(-1, 5){  4}}
\put( 29,505){\line( 0, 1){105}}
\put( 40,665){\makebox(0,0)[lb]{\raisebox{0pt}[0pt][0pt]{\twlrm $A^\ast$}}}
\put( 79,782){\makebox(0,0)[lb]{\raisebox{0pt}[0pt][0pt]{\twlrm $A$}}}
\put(128,776){\makebox(0,0)[lb]{\raisebox{0pt}[0pt][0pt]{\twlrm $B$}}}
\put(102,705){\makebox(0,0)[lb]{\raisebox{0pt}[0pt][0pt]{\twlrm $C$}}}
\put(175,787){\makebox(0,0)[lb]{\raisebox{0pt}[0pt][0pt]{\twlrm $C^\ast$}}}
\put(519,671){\makebox(0,0)[lb]{\raisebox{0pt}[0pt][0pt]{\twlrm $B^\ast$}}}
\put( 38,590){\makebox(0,0)[lb]{\raisebox{0pt}[0pt][0pt]{\twlrm $A^\ast$}}}
\put( 59,498){\makebox(0,0)[lb]{\raisebox{0pt}[0pt][0pt]{\twlrm $A$}}}
\put(121,495){\makebox(0,0)[lb]{\raisebox{0pt}[0pt][0pt]{\twlrm $B$}}}
\put( 75,557){\makebox(0,0)[lb]{\raisebox{0pt}[0pt][0pt]{\twlrm $C$}}}
\put(178,472){\makebox(0,0)[lb]{\raisebox{0pt}[0pt][0pt]{\twlrm $C^\ast$}}}
\put(446,761){\makebox(0,0)[lb]{\raisebox{0pt}[0pt][0pt]{\twlrm $B$}}}
\put(450,497){\makebox(0,0)[lb]{\raisebox{0pt}[0pt][0pt]{\twlrm $B$}}}
\put(514,591){\makebox(0,0)[lb]{\raisebox{0pt}[0pt][0pt]{\twlrm $B^\ast$}}}
\put(385,766){\makebox(0,0)[lb]{\raisebox{0pt}[0pt][0pt]{\twlrm $A$}}}
\put(384,500){\makebox(0,0)[lb]{\raisebox{0pt}[0pt][0pt]{\twlrm $A$}}}
\put(432,591){\makebox(0,0)[lb]{\raisebox{0pt}[0pt][0pt]{\twlrm $C$}}}
\put(432,666){\makebox(0,0)[lb]{\raisebox{0pt}[0pt][0pt]{\twlrm $C$}}}
\put( 88,738){\makebox(0,0)[lb]{\raisebox{0pt}[0pt][0pt]{\twlrm $b$}}}
\put( 88,522){\makebox(0,0)[lb]{\raisebox{0pt}[0pt][0pt]{\twlrm
	$\overline{b}$}}}
\put(417,738){\makebox(0,0)[lb]{\raisebox{0pt}[0pt][0pt]{\twlrm $b$}}}
\put(417,522){\makebox(0,0)[lb]{\raisebox{0pt}[0pt][0pt]{\twlrm
	$\overline{b}$}}}
\end{picture}
\end{centering}

\caption{Transformation of bases under partial dualization \label{partdual}}
\end{figure}
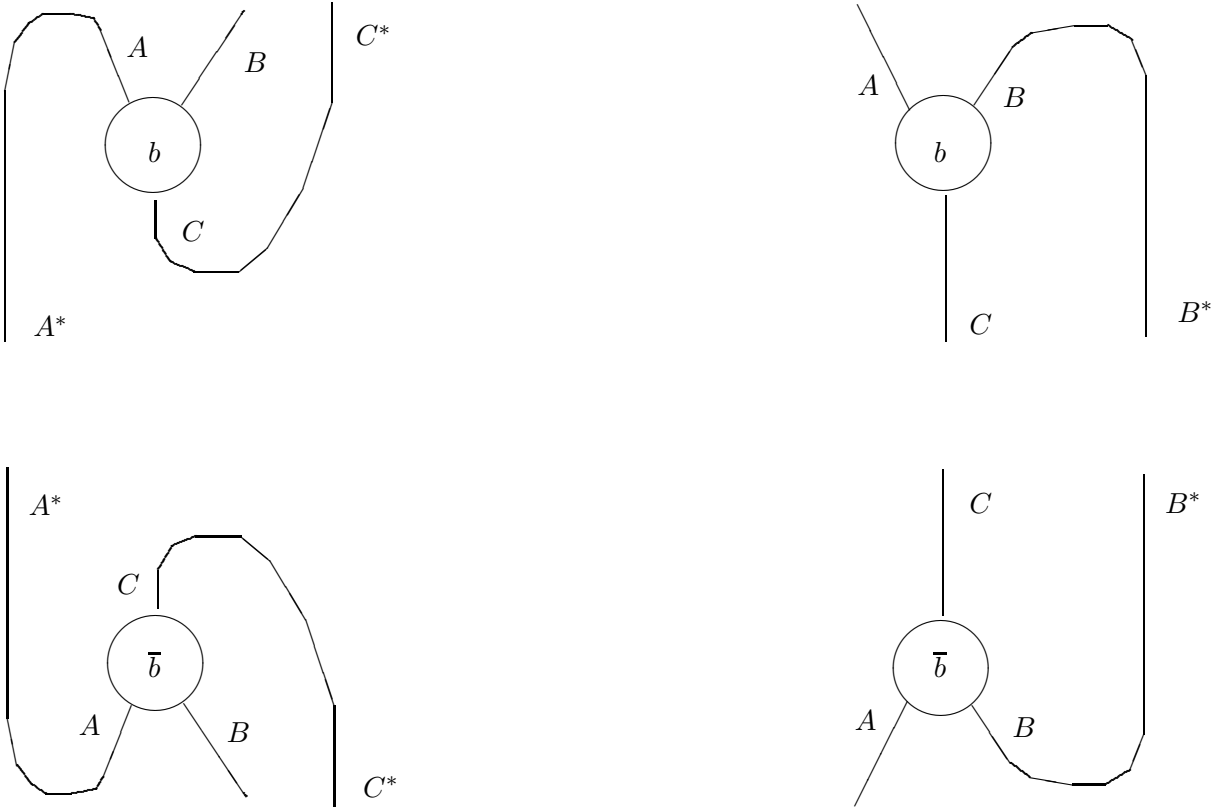

Similarly, without resort to Lemma \ref{schur} we can see that $\cal B$ gives
rise to bases for ${\cal C}[B\otimes C^\ast, A^\ast ]$ (resp.
${\cal C}[A^\ast , B\otimes C^\ast ]$; ${\cal C}[A, C\otimes B^\ast ]$;
and ${\cal C}[C\otimes B^\ast , A]$ ) by  applying
$A^\ast \otimes - \otimes C^\ast $ to the basis elements then
precomposing with $\eta_A \otimes 1_{B\otimes C^\ast}$ and postcomposing with
$1_{A^\ast}\otimes \epsilon_C$
(resp. applying $A^\ast \otimes - \otimes C^\ast$ to the splitting
then precomposing
with $1_{A^\ast }\otimes h_C$ and postcomposing with
$e_A\otimes 1_{B\otimes C^\ast }$; applying $-\otimes B^\ast $ to the
basis elements then
precomposing with $1_A\otimes h_B$; applying $-\otimes B^\ast $ to the
splittings then postcomposing with
$1_A \otimes \epsilon_B$ ). (These are represented graphically in
Figure \ref{partdual}.) [go through and put in $\phi$'s]

What is not immediately clear is the relationship between the first
and second (resp. third and fourth) of the transformed bases in the
previous paragraph.  In fact, a calculation using Lemma \ref{schur}
provides the following generalization of the rules in the Kirillov-Reshetikhin
[KR] formulation of $U_q(sl_2)$-recoupling theory:

\begin{lemma} \label{pdb}
The bases of ${\cal C}[B\otimes C^\ast, A^\ast ]$ and
${\cal C}[A^\ast , B\otimes C^\ast ]$ (resp. ${\cal C}[C\otimes B^\ast , A]$
and ${\cal C}[A, C\otimes B^\ast ]$ ) obtained by multiplying the
transformed bases described two paragraphs ago by $\sqrt{dim(A)}/\sqrt{dim(C)}$
are splittings of each other, moreover giving the projection and inclusions of
a direct sum decomposition for the given tensor product.
\end{lemma}

\noindent {\bf proof:} By application of Lemma \ref{schur} it suffices to
show that the trace of composition of an element in the second basis
of each pair with one in the first is $dim(A^\ast) = dim(A)$ (resp. $dim(A)$)
if the
elements are the multiple of the tranformation of an element of $\cal B$ and
its splitting, and 0 otherwise.  But this follows directly from
Lemma \ref{sumlemma}.

We conclude with a categorical notion, introduced in [CKY2]
which will be important when we consider
the interpretation of the invariants constructed:

\begin{defin} \label{center}
The {\em center} $Z({\cal C})$ of a braided monoidal category $\cal C$
is the full-subcategory
of all objects $B$ with the property that

\[ \forall X \in Ob({\cal C}) \; \sigma_{X,B}\sigma_{B,X} = 1_{X\otimes B}. \]

A braided monoidal category has {\em trivial center} if the center is
the full-subcategory of objects isomorphic to a finite (possibly empty) direct
sum of copies of $I$.
\end{defin}

\clearpage \section{ Coloring Triangulations and State-Sum Invariants}

Throughout this section we let $\cal C$ be a fixed semisimple tortile
category, let $\cal S$ be a chosen set of representatives for the
isomorphism classes of simple objects including, as a representative of its
class, the chosen monoidal identity object $I$, and let $\cal B$ be a choice
for each triple of elements of $a,b,c \in {\cal S}$ of a basis
${\cal B}^{ab}_{c}$ for the hom-space ${\cal C}[a\otimes b,c]$ and by
abuse of notation the disjoint union of these bases. Assume without
loss of generality that the choice of dual objects has been made so that
$()^\ast$ induces an involution on $\cal S$. For the reader who has skipped
Subsection \ref{generalsstc},
the specific example of ${\cal C} = Rep_!(U_q(sl_2))$,
${\cal S} = \{0,...,r-2\}$ and ${\cal B}^{ab}_c$ given by choosing
the KR vertex with upward legs labelled $a$ and $b$ and downward leg labelled
$c$ should be considered.  Notes directed to readers interested in this
level of generality will occur from time to time. The translation from KR to
TL recoupling theory is given in Subsection \ref{translation}.

If {\bf T} is a triangulation of a 4-manifold $M$, we let ${\bf T}_{(i)}$
denote the set of (non-degenerate) $i$-simplices of the triangulation. In what
follows we will be concerned with ordered triangulations, that is
triangulations equipped with total orderings of their vertices.

We are now in a position to define the colorings which index our state-sums.

\begin{defin}
A $\cal CSB$-coloring $\lambda $ (or simply a coloring if no confusion is
possible)
of an ordered triangulation of a 4-manifold is a triple of maps

\[ (\lambda :{\bf T}_{(2)} \cup {\bf T}_{(3)} \longrightarrow {\cal S}, \,
  \lambda^+ :{\bf T}_{(3)}\longrightarrow {\cal B}, \,
\lambda^- :{\bf T}_{(3)}\longrightarrow {\cal B}) \]

\noindent such that $\lambda^+ \{a,b,c,d\} \in {\cal B}^{\lambda \{b,c,d\},
\lambda \{a,b,d\}}_{\lambda \{a,b,c,d\}}$ and
$\lambda^- \{a,b,c,d\} \in {\cal B}^{\lambda \{a,c,d\},
\lambda \{a,b,c\}}_{\lambda \{a,b,c,d\}}$, where $a < b < c < d$ in the
ordering on the vertices. We denote the set of $\cal CSB$-colorings of an
ordered triangulation {\bf T} by
$\Lambda _{\cal CSB}({\bf T})$.
\end{defin}

Readers interested in the $U_q(sl_2)$ case only should note that in that
case, the content of a coloring is given entirely by a choice of
$\lambda :{\bf T}_{(2)} \cup {\bf T}_{(3)} \longrightarrow {\cal S} $
for which the labels on the positive (resp. negative) part of the boundary of
each 3-simplex couple to the label on the 3-simplex.

In what follows, we let $N = \sum_{A \in \cal S} dim(A)^2$ and $n_i = |{\bf
T_{(i)}}|$

Now, given an ordered triangulation {\bf T}  of a 4-manifold, we can assign to
each coloring $\lambda $ a number $\ll \lambda \gg$ defined by

\[ \ll \lambda \gg = N^{n_0 - n_1} \prod_{\parbox{.6in}{\small faces
$\sigma$}}
dim(\lambda(\sigma)) \prod_{\parbox{.7in}{\small tetrahedra $\tau $}}
dim(\lambda
(\tau))^{-1} \prod_{\parbox{.8in}{\small 4-simplices $\xi$}}
\| \lambda, \xi \| \]

\noindent where $\| \lambda, \xi \|$ is given by the endomorphism of
$I$ graphically
in Figure \ref{gen15j} if the orientation of $\xi $ induced by the
ordering of the vertices is the same as the ambient orientation, and by
the endomorphism of $I$ represented graphically by the network obtained
from that in Figure \ref{gen15j} by mirror-imaging the network, applying
$(-)^\ast $ to all object labels and $\overline{(-)}^\ast $ to all
labels by maps in $\cal B$.

For readers interested in the $U_q(sl_2)$ case, the picture is simpler:
the nodes are KR vertices with legs as shown in Figure \ref{gen15j} and
the dualizing of labels is unnecessary (since all representations of
$U_q(sl_2)$ are self-dual).

\begin{centering}
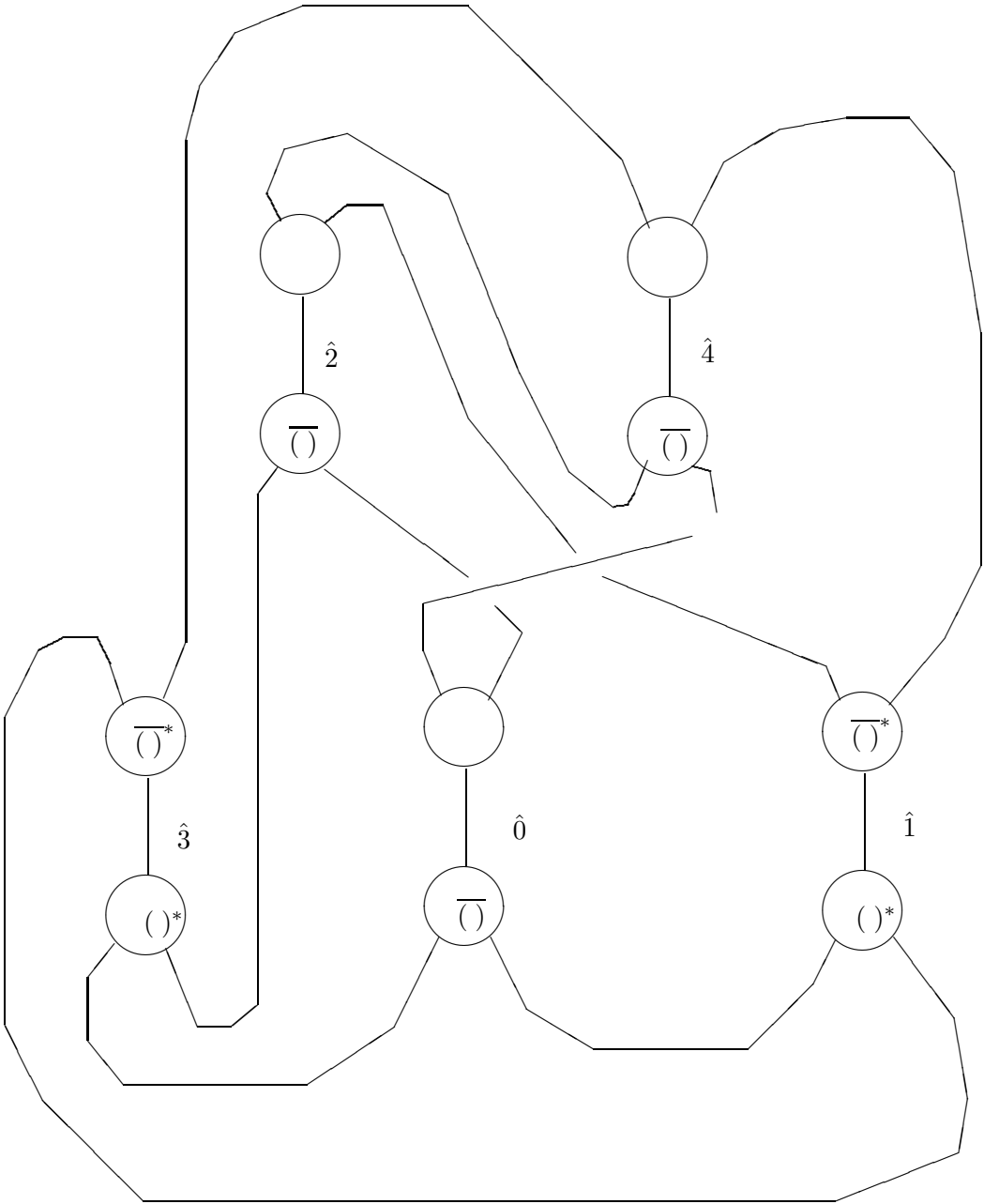
\begin{figure}

\setlength{\unitlength}{0.0125in}%
\begin{picture}(436,534)(13,257)
\thinlines
\put(145,680){\circle{36}}
\put(145,600){\circle{36}}
\put(146,661){\line( 0,-1){ 43}}
\put(309,679){\circle{36}}
\put(309,599){\circle{36}}
\put(310,660){\line( 0,-1){ 43}}
\put(218,469){\circle{36}}
\put(218,389){\circle{36}}
\put(219,450){\line( 0,-1){ 43}}
\put(396,467){\circle{36}}
\put(396,387){\circle{36}}
\put(397,448){\line( 0,-1){ 43}}
\put( 76,465){\circle{36}}
\put( 76,385){\circle{36}}
\put( 77,446){\line( 0,-1){ 43}}
\put(208,483){\line(-2, 5){  8}}
\put(200,503){\line( 0, 1){ 21}}
\put(200,524){\line( 4, 1){120}}
\put(320,554){\line( 1, 1){ 11}}
\put(331,565){\line(-1, 6){  3}}
\multiput(328,583)(-0.57143,0.14286){15}{\makebox(0.4444,0.6667){\sevrm .}}
\put(156,584){\line( 4,-3){ 64}}
\put(232,523){\line( 1,-1){ 12}}
\put(244,511){\line(-1,-2){ 15}}
\put(135,585){\line(-3,-4){  9}}
\put(126,573){\line( 0,-1){228}}
\put(126,345){\line(-6,-5){ 12}}
\put(114,335){\line(-1, 0){ 15}}
\put( 99,335){\line(-2, 5){ 14}}
\put( 62,372){\line(-4,-5){ 12}}
\put( 50,357){\line( 0,-1){ 28}}
\put( 50,329){\line( 4,-5){ 16}}
\put( 66,309){\line( 1, 0){ 82}}
\put(148,309){\line( 3, 2){ 39}}
\put(187,335){\line( 1, 2){ 20}}
\put(230,375){\line( 1,-2){ 16}}
\put(246,343){\line( 5,-3){ 30}}
\put(276,325){\line( 1, 0){ 69}}
\put(345,325){\line( 1, 1){ 29}}
\put(374,354){\line( 1, 2){ 10}}
\put(410,375){\line( 3,-4){ 27}}
\put(437,339){\line( 1,-6){  6}}
\put(443,303){\line(-1,-6){  4}}
\put(439,279){\line(-5,-2){ 55}}
\put(384,257){\line(-1, 0){309}}
\put( 75,257){\line(-1, 1){ 45}}
\put( 30,302){\line(-1, 2){ 17}}
\put( 13,336){\line( 0, 1){137}}
\put( 13,473){\line( 1, 2){ 15}}
\multiput( 28,503)(0.50000,0.25000){25}{\makebox(0.4444,0.6667){\sevrm .}}
\put( 40,509){\line( 1, 0){ 14}}
\multiput( 54,509)(0.25000,-0.50000){25}{\makebox(0.4444,0.6667){\sevrm .}}
\put( 60,497){\line( 1,-3){  6}}
\multiput(136,695)(-0.25000,0.50000){25}{\makebox(0.4444,0.6667){\sevrm .}}
\put(130,707){\line( 2, 5){  8}}
\put(138,727){\line( 4, 1){ 28}}
\put(166,734){\line( 5,-3){ 45}}
\put(211,707){\line( 2,-5){ 32}}
\put(243,627){\line( 1,-2){ 22}}
\put(265,583){\line( 5,-4){ 20}}
\multiput(285,567)(0.60000,0.10000){11}{\makebox(0.4444,0.6667){\sevrm .}}
\multiput(291,568)(0.30000,0.50000){11}{\makebox(0.4444,0.6667){\sevrm .}}
\put(294,573){\line( 2, 5){  6}}
\multiput(156,694)(0.43478,0.34783){24}{\makebox(0.4444,0.6667){\sevrm .}}
\put(166,702){\line( 1, 0){ 16}}
\put(182,702){\line( 2,-5){ 38}}
\put(220,607){\line( 4,-5){ 48}}
\put(280,536){\line( 5,-2){ 85}}
\put(365,502){\line( 5,-2){ 15}}
\put(380,496){\line( 2,-5){  6}}
\put(320,693){\line( 1, 2){ 14}}
\put(334,721){\line( 5, 3){ 25}}
\put(359,736){\line( 6, 1){ 30}}
\put(389,741){\line( 1, 0){ 28}}
\put(417,741){\line( 5,-6){ 20}}
\put(437,717){\line( 1,-6){ 12}}
\put(449,645){\line( 0,-1){104}}
\put(449,541){\line(-1,-2){ 16}}
\put(433,509){\line(-5,-6){ 25}}
\put( 84,482){\line( 2, 5){ 10}}
\put( 94,507){\line( 0, 1){224}}
\put( 94,731){\line( 1, 4){  6}}
\put(100,755){\line( 2, 3){ 16}}
\put(116,779){\line( 5, 2){ 30}}
\put(146,791){\line( 1, 0){ 74}}
\put(220,791){\line( 1,-1){ 35}}
\put(255,756){\line( 1,-1){ 34}}
\put(289,722){\line( 2,-5){ 12}}
\put(240,419){\makebox(0,0)[lb]{\raisebox{0pt}[0pt][0pt]{\twlrm $\hat{0}$}}}
\put(156,630){\makebox(0,0)[lb]{\raisebox{0pt}[0pt][0pt]{\twlrm $\hat{2}$}}}
\put(324,632){\makebox(0,0)[lb]{\raisebox{0pt}[0pt][0pt]{\twlrm $\hat{4}$}}}
\put(414,420){\makebox(0,0)[lb]{\raisebox{0pt}[0pt][0pt]{\twlrm $\hat{1}$}}}
\put( 90,415){\makebox(0,0)[lb]{\raisebox{0pt}[0pt][0pt]{\twlrm $\hat{3}$}}}
\put(140,592){\makebox(0,0)[lb]{\raisebox{0pt}[0pt][0pt]{\twlrm
$\overline{( \; )}$}}}
\put(306,591){\makebox(0,0)[lb]{\raisebox{0pt}[0pt][0pt]{\twlrm
$\overline{( \; )}$}}}
\put(215,381){\makebox(0,0)[lb]{\raisebox{0pt}[0pt][0pt]{\twlrm
$\overline{( \; )}$}}}
\put( 75,378){\makebox(0,0)[lb]{\raisebox{0pt}[0pt][0pt]{\twlrm
$( \; )^\ast$}}}
\put(393,380){\makebox(0,0)[lb]{\raisebox{0pt}[0pt][0pt]{\twlrm
$( \; )^\ast$}}}
\put( 71,458){\makebox(0,0)[lb]{\raisebox{0pt}[0pt][0pt]{\twlrm
$\overline{( \; )}^\ast$}}}
\put(391,461){\makebox(0,0)[lb]{\raisebox{0pt}[0pt][0pt]{\twlrm
$\overline{( \; )}^\ast$}}}
\end{picture}

\caption{The generalized 15-j symbol associated to a correctly oriented
vertex-ordered 4-simplex \label{gen15j}}
\end{figure}
\end{centering}

The main result of this section is then

\begin{thm} \label{main}
The state-sum

\[ CY_{\cal C}(M) = \sum_{\lambda \in \Lambda_{\cal CSB}({\bf T})}
 \ll \lambda \gg \]

\noindent is independent of the choice of ordered triangulation {\bf T}, of
representative simple objects $\cal S$ and of bases $\cal B$.  Thus for
any semisimple tortile category $\cal C$,
 $CY_{\cal C}(-)$ is an invariant
of piecewise-linear 4-manifolds. In the case of ${\cal C} = Rep_!(U_q(sl_2))$,
the invariant is the original Crane-Yetter invariant of [CY].
\end{thm}

To prove this we use an auxiliary notion

\begin{defin}
A {\em $d$-blob} is a $d$-cell equipped with an ordered triangulation
of its boundary.

\end{defin}

Although the initial verification of the invariance of the Crane-Yetter
state-sum [CY] was carried out using Pachner's moves, the notion of blobs
provides an alternative method for verifying that state-sums on triangulations
give rise to PL-manifold invariants:  in general, one must show that
the state-sum associated to any blob with an arbitrary extension of the
triangulation to the interior and fixed initial data on the boundary is
independent of the extension of the triangulation to the interior.

Observe that Pachner's moves in any dimension are of this form.  At first,
it might appear that the method suggested above is worse than verifying
Pachner's moves.  However, in our case (and potentially in the case of
more refined state-sums) the use of blobs allows a uniform inductive
proof. Indeed, it follows from general principles enunciated in Yetter [Y6]
that any state-summation on ordered triangulations in which weights are
assigned
to simplices (or simplicial flags with suitable compatibilities imposed
for shared simplices) must satisfy a ``blob lemma'' of this sort if it is
to be a PL-invariant. It will
depend on the exact circumstances whether it will be easier to verify this
``blob lemma'' inductively or to check Pachner's moves.

It should also be noted, that the use of blobs provides an immediate check
for feasibility of finding normalization factors on lower-dimensional
simplices (or simplicial flags) to make a proposed state-sum topologically
invariant:  one must be able to find a prescription for a weight on a labelled
blob (for example as a product or ratio of recombination diagrams) which
restricts to the proposed weight on an ordered 4-simplex.

In the present case, we set up the induction (carried out
in the proof of Lemma \ref{blob}) as follows:  first, we describe a
network naming an endomorphism of $I$, and hence a number, for any
$4$-blob in an oriented 4-manifold, then show that the state-sum can
be rewritten in terms of a decomposition into blobs (as observed above,
without regard to
the triangulation of their interiors).

This will complete the proof of invariance, and independence from the
triangulation, while independence from the ordering of the vertices
will follow from the
simple expedient of observing that the star of a vertex is a $4$-blob,
rewriting with the vertex missing, then reversing the process to insert the
vertex somewhere else in the ordering.

To construct our networks associated to $4$-blobs, first notice
that a vertex ordering on a tetrahedron in the boundary of a $4$-blob
(or $4$-simplex as a special case) together with the orientation
gives rise to a chosen side of the tetrahedron.

(Specifically,
in a local coordinate
system identified with a ball in ${\bf R^4}$ the orientation gives a way
to choose a fourth vector orthogonal to any given ordered triple of vectors
---here the vectors
are the tangents to the edges from the lowest numbered vertex to the
others in order. We only care about the side the fourth vector lies
on, inside the blob or outside, so the result doesn't depend on what
orientation preserving map we used to identify the chart with the ball.)

Now, place the highest numbered vertex at $\infty$, identity the rest
of the bounding ${\bf S^3}$ with ${\bf R^3}$, and choose a plane to project
on and vertical and horizontal directions in the plane.
In each 3-simplex place a vertical ``dumbbell'' (as in
the network for the generalized $15j$ symbol in Figure \ref{gen15j}
with ends representing places to be colored with maps, and a bar representing
a place to be colored with a simple object) in a plane parallel
to the plane of projection. Add arcs connecting the ``dumbbells'' so that
for tetrahedra with inward normals the bottom right (resp. top right,
bottom left, top left) is connected through the face obtained by omitting
the lowest (resp. second, third, highest) numbered vertex of the tetrahedron,
while for tetrahedra with outward normals, proceed as above, but reversing
left and right. Finally,
in tetrahedra with inward normal vectors
 place  an overline in the lower end of the ``dumbbell'' (to indicate
that the map here will be the splitting of the color from $\cal B$), while
in those with outward normal vectors, place an overline and a $^\ast$
in the upper end of the ``dumbbell'' and a $^\ast$ in the lower end (to
indicate that the map here will be the dual of the splitting, respectively
the dual, of the color from $\cal B$). For brevity it will be
convenient to refer to the ``dumbbells'' as ``inward'' or ``outward dumbbells''
according to the structure of connections and labellings.

Observe

\begin{lemma}
The generalized $15j$ symbol of Figure \ref{gen15j} is precisely the
network associated to a 4-simplex whose order-orientation agrees with
the induced orientation by the procedure just outlined.
\end{lemma}

\noindent {\bf proof:} It suffices to observe that the normal vector
induced on $\hat{0},\hat{2},\hat{4}$ are opposite to those induced
on $\hat{1},\hat{3}$, and that the agreement of orientations implies
that the normal vector on $\hat{0}$ is on the same side of $\hat{0}$
as 4, and thus inward.
\smallskip

Finally we are in a position to state the key lemma in the proof.
In its proof, we will regard the state-sum as an evaluation of linear
combinations of colored embedded trivalent (ribbon) graphs interpreted
in the now standard way (cf. Reshetikhin/Turaev [RT]).

\begin{lemma} \label{blob}
{\bf ``The Blob Lemma''}  If $M$ is a 4-manifold equipped with an ordered
triangulation {\bf T}, and $D$ is a 4-blob formed by the union of 4-simplices
in {\bf T}, then the state-sum

\[ \sum_{\lambda \in \Lambda_{\cal CSB}({\bf T})} N^{n_0 - n_1}
\prod_{\parbox{.6in}{\small faces
$\sigma$}}
dim(\lambda(\sigma)) \prod_{\parbox{.7in}{\small tetrahedra $\tau $}}
dim(\lambda
(\tau))^{-1} \prod_{\parbox{.8in}{\small 4-simplices $\xi$}}
\| \lambda, \xi \| \]

decomposes as

{\small
\[ \sum_{\parbox{.6in}{\tiny $\lambda $ a coloring of ${\bf T}|_{\partial D}$}}
\left[
\sum_{\parbox{.6in}{\tiny $\mu $ a coloring of ${\bf T}|_D$ extending
$\lambda$}}
N^{|{\bf T_{(0)}} \cup int(D)| - |{\bf T_{(0)}} \cup int(D)|}
\prod_{\parbox{.4in}{\tiny faces $\sigma \subset int(D)$} } dim(\mu(\sigma))
\prod_{\parbox{.4in}{\tiny tetrahedra $\tau \subset int(D)$}}
dim(\mu(\tau))^{-1}
\prod_{\parbox{.5in}{\tiny 4-simplices $\xi \subset D$}}
\| \mu, \xi \| \right] \]

\[
\times
\left[  \sum_{\parbox{.6in}{\tiny $\nu $ a coloring of ${\bf T}|_{M \setminus
int(D)}$
extending $\lambda$}}
N^{|{\bf T_{(0)}} \setminus int(D)| - |{\bf T_{(0)}} \setminus int(D)|}
\prod_{\parbox{.4in}{\tiny faces $\sigma \subset M \setminus int(D)$ }}
dim(\nu(\sigma))
\prod_{\parbox{.4in}{\tiny tetrahedra $\tau \subset M \setminus int(D)$}}
dim(\nu(\tau)^{-1}
\prod_{\parbox{.5in}{\tiny 4-simplices $\xi \subset M \setminus int(D)$}}
\| \nu, \xi \| \right] \]
}

\noindent and for each $\lambda$ the first factor (the sum on $\mu$)
is equal to the evaluation of the network associated to the boundary of
the 4-blob $D$.
\end{lemma}

\noindent {\bf proof:} We proceed by induction on the number $n$ of 4-simplices
in $D$. If $n = 1$ there is nothing to show, by the preceding lemma.

Now suppose $n > 1$, and we have shown for the lemma for all 4-blobs with
$n-1$ 4-simplices.  Select a 4-simplex $\sigma$
in $D$ which intersect the boundary
in a cell of dimension 3. Then $D^\prime =
\overline{D \setminus \overline{\sigma}}$ is a 4-blob. Now observe that
$D^\prime \cap \overline{\sigma}$ is a union of closed 3-simplices of {\bf T},
and that those assigned ``inward dumbbells''  in one of $D^\prime $
or $\overline{\sigma}$ are assigned ``outward dumbbells''  in the
other and vice-versa. Similarly, the pattern of connections between
the ``dumbbells'' in shared tetrahedra (and from shared tetrahedra out
to unspecified ``dumbbells'') will be mirror images.

\begin{figure}
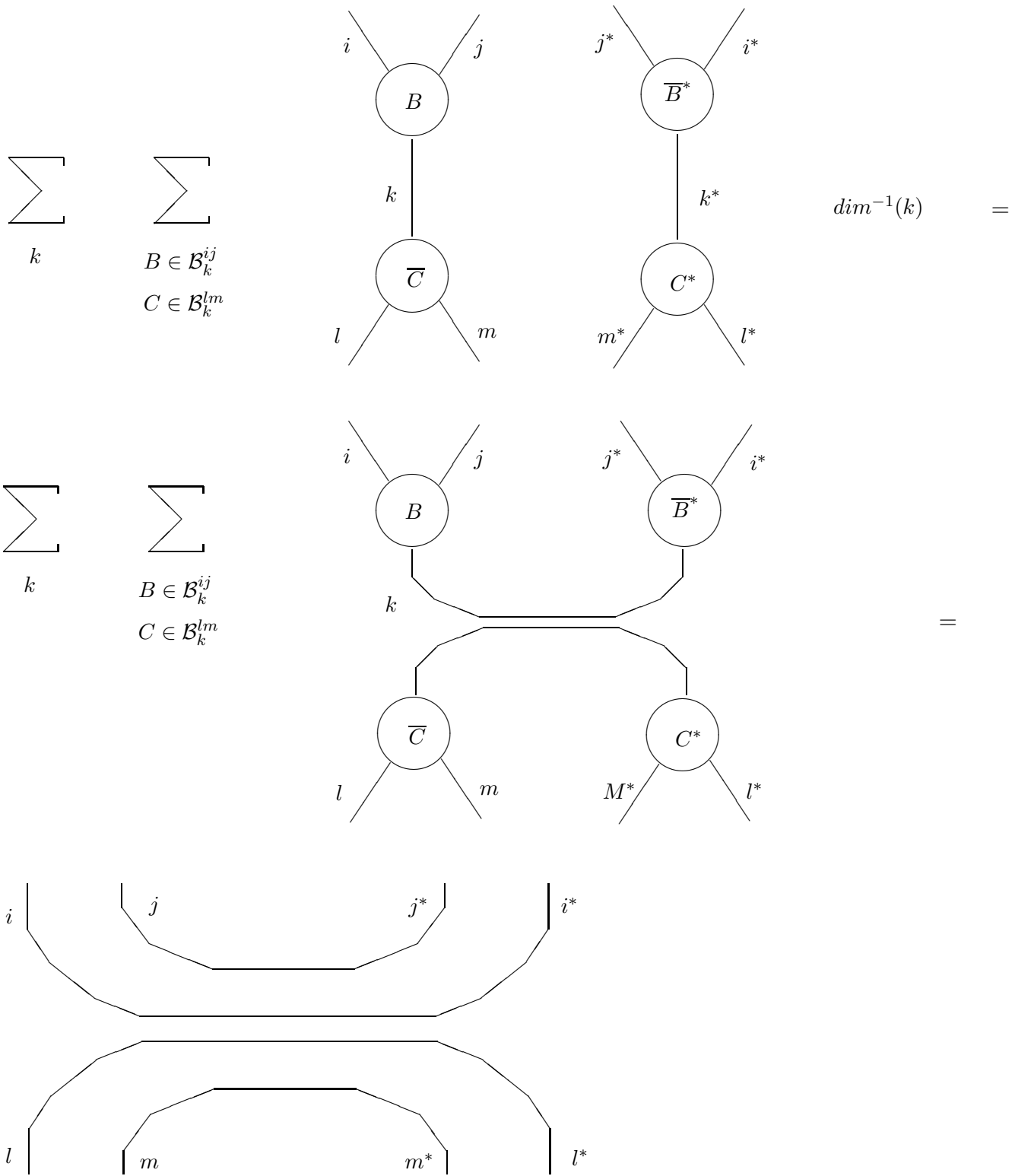

\setlength{\unitlength}{0.0125in}%


\caption{\label{blob1} Joining blobs which share one tetrahedron}
\end{figure}

\begin{figure}
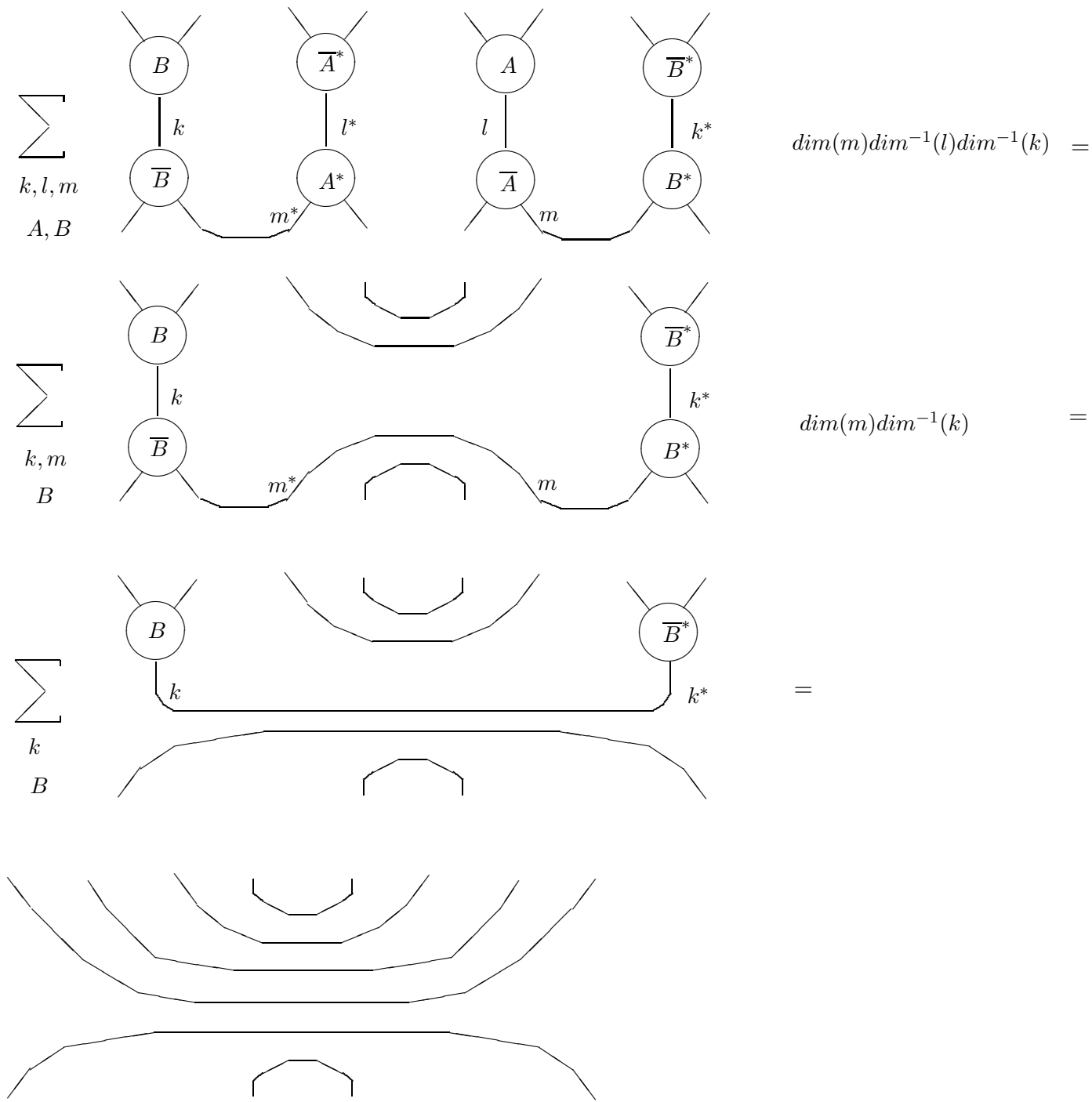

\setlength{\unitlength}{0.0125in}%
%

\caption{\label{blob2} One case of joining blobs which share two tetrahedra}
\end{figure}

\begin{figure}[b]
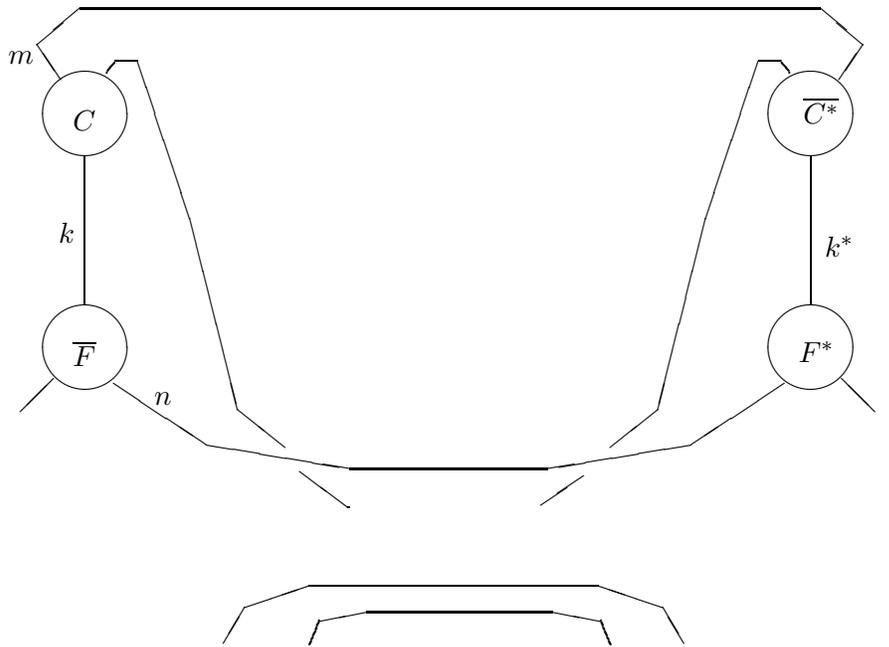

\setlength{\unitlength}{0.0125in}%
%
\caption{\label{blob3.1} One case of joining blobs that share three tetrahedra
(beginning)}
\end{figure}

\begin{figure}[t]
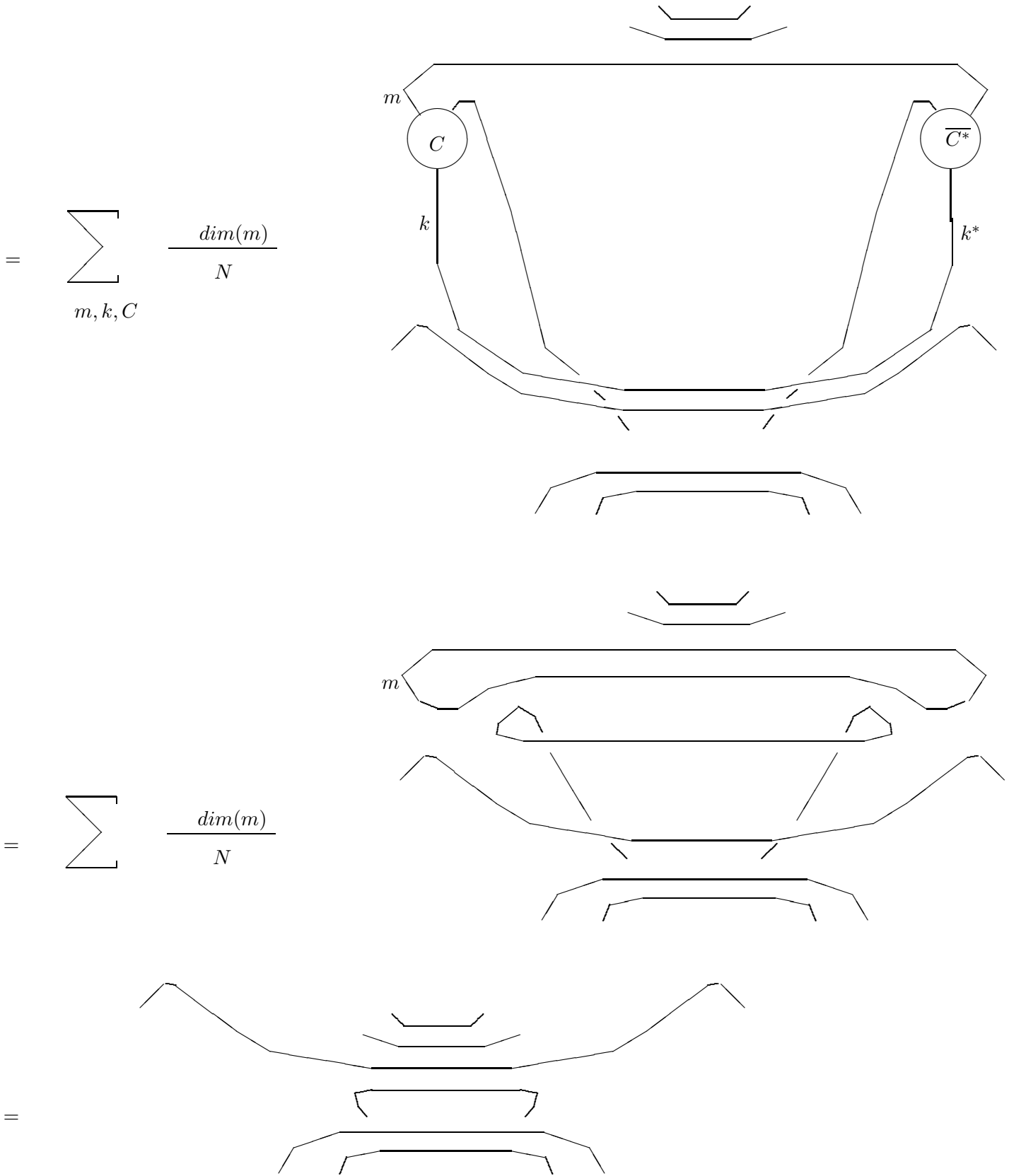

\setlength{\unitlength}{0.0125in}%
%

\caption{\label{blob3.2} One case of joining blobs which share three tetrahedra
(conclusion)}
\end{figure}

\begin{figure}
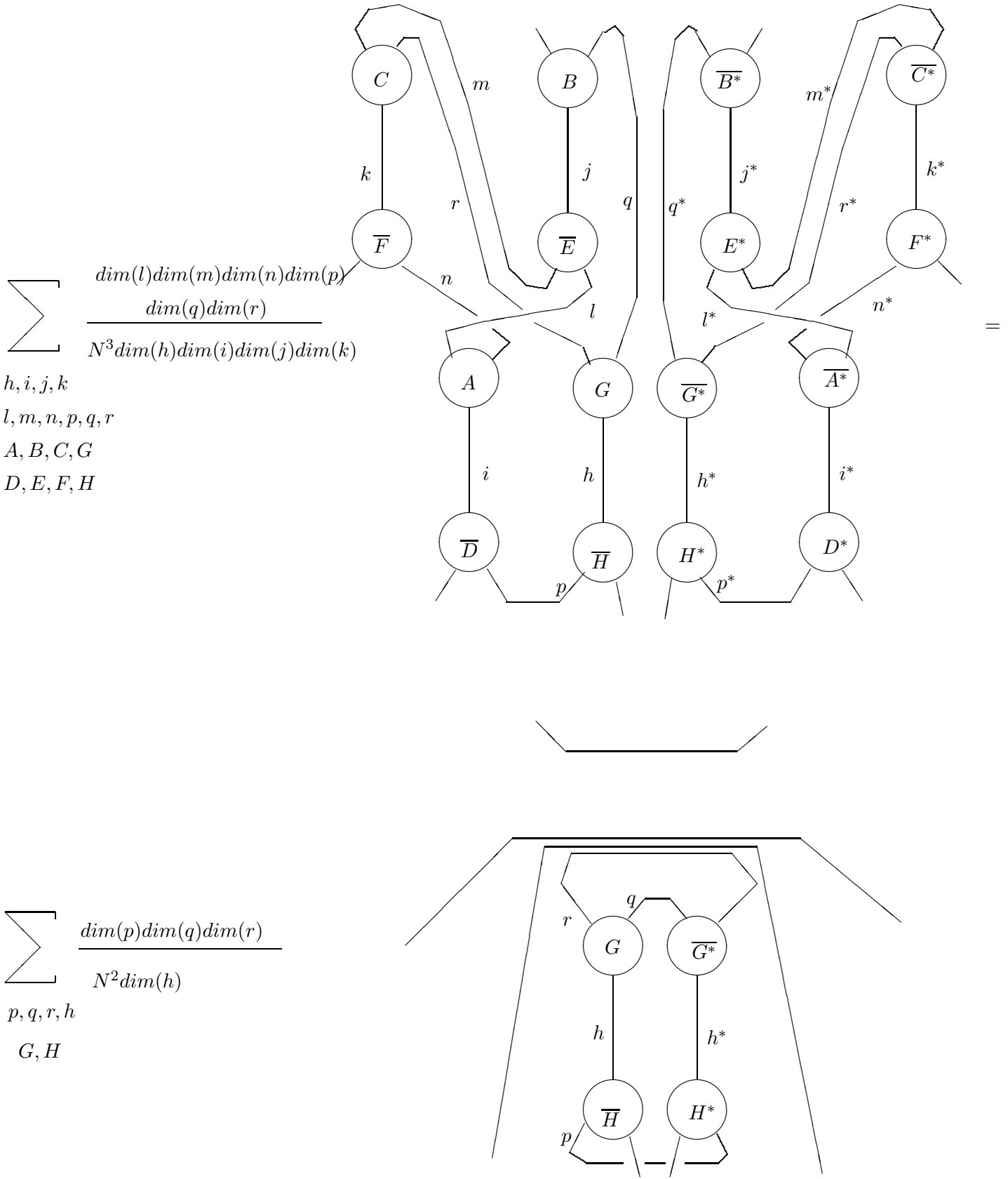

\setlength{\unitlength}{0.0125in}%
%

\caption{\label{blob4} One case of joining blobs which share four tetrahedra
(beginning)}
\end{figure}

To complete the proof, it suffices to show that a local evaluation of
the part of the diagram describing the state-sum which includes the
contributions of $\sigma$, and all simplices in
$D^\prime \cap \sigma$ is equal to the local evalutation of the
remaining faces of $\sigma$. A series of diagrammatic calculations
verifies this.  A sample of the calculations are given in Figures \ref{blob1},
\ref{blob2}, \ref{blob3.1}, \ref{blob3.2},  \ref{blob4} and \ref{blob4.2}.
The others (in which other portions of the 15-j symbol are involved) are
completely analogous, and are left to the reader. Only for the first (Figure
\ref{blob1})
do we
give a detailed description of where labels are drawn from.

\begin{figure}
\setlength{\unitlength}{0.0125in}%
%

\caption{\label{blob4.2} One case of joining blobs which share four tetrahedra
(conclusion)}
\end{figure}
Independence of the choice of representative simple objects will follow
by inserting isomorphisms between the chosen representative objects around
each of the nodes in the generalized 15-j symbol, and the following
lemma, which shows independence from the choice of bases.

\begin{lemma}
The state-sum is invariant under the change of bases from which the
map components of the labels are chosen on any one tetrehedron (and hence
on all).

\end{lemma}

\noindent {\bf proof:} It suffices to show that a local evaluation of
the diagram describing the state-sum including all occurences of
the basis elements chosen on a particular tetrahedron is independent
of the choice of basis.  Observe that this follows immediately from
the first of the calculations in the proof of Lemma \ref{blob} (Figure
\ref{blob1}).

This completes the proof of Theorem \ref{main}.

\clearpage
\section{\label{3mfsec} Surgical Versions
and an Aside About 3-Manifold Invariants}

	At about the same time as the announcement of [CY] appeared, B. Broda
[B] announced the construction of a 4-manifold invariant calculated from
a surgery description of the 4-manifold (cf. Kirby [Ki]) by a framed link
with a distinguished unlink.

	In this section, we describe analogs of Broda's invariant for
arbitrary semisimple tortile categories and of the generalized
Reshetikhin/Turaev 3-manifold invariants of Turaev [T] (without the
``modularity'' assumption on the category). The detour through 3-manifold
invariants is necessary, as the generalization of Roberts' results
relating the surgical and state-sum invariants, and interpreting
the former in terms of signature requires the use of the 3-manifold
invariants.

	The key here is the idea that in addition to being able to label
components of a framed link diagram with objects of a k-linear
tortile category, and thereby (via the freeness theorem of Shum [S])
interpret the diagram as giving an endomorphism of $I$ (that is
a number, when $I$ is simple), we can also label them with
linear combinations of objects (again obtaining a number).
\footnote{The {\em very} categorically minded will recognize that we
are really labelling each strand with a direct sum of objects,
and placing on each strand a node with the map which multiplies each
direct summand by the coefficient. This view of the construction
will doubtless seem a bit stretched and confusing to most readers,
but may be essential to generalizations to more non-commutative
``Hopf categories''--cf. Crane/Frenkel [CF]}

	If the linear combination used to label the components is
carefully chosen, the resulting framed link invariant will be
invariant under handle-sliding, and thus (upon suitable normalization)
can be turned into an invariant of the 3- or 4-manifold
described by surgery on the link (a little care must be taken
in the 4-manifold case to correctly deal with the distingushed
unlink whose curves represent places to ``hollow out a 2-handle'',
equivalent to attaching a 1-handle, but we will deal with that
when the time comes.)

	The key here is a generalization of the elegant demonstration
given in Lickorish [L] (based on ideas of Roberts and Viro, cf. also
Kauffman/Lins [KaLi]) that the linear combination of the simple objects
in the semisimple subquotient category of $Rep(U_q(sl_2))$ at
a root of unity with
their (internal or quantum) dimensions as coefficents gives
rise to a framed link invariant which is invariant under handle-sliding.

	In fact the phenomenon is quite general:

\begin{propo} \label{handleslide}
If $\cal C$ is any semisimple $k$-linear tortile category with
$\cal S$ and $\cal B$ as in the previous section, then
\end{propo}

\begin{figure}[h]
\setlength{\unitlength}{0.0125in}%
\begin{picture}(511,103)(23,730)
\thinlines
\put( 50,800){\line(-1, 0){ 25}}
\put( 25,800){\line( 1,-1){ 15}}
\put( 50,770){\line(-1, 0){ 25}}
\put( 25,770){\line( 1, 1){ 15}}
\put(179,788){\line( 0,-1){ 20}}
\multiput(179,768)(0.25000,-0.50000){21}{\makebox(0.4444,0.6667){\sevrm .}}
\multiput(184,758)(0.50000,-0.25000){21}{\makebox(0.4444,0.6667){\sevrm .}}
\put(194,753){\line( 1, 0){ 10}}
\put(179,780){\line( 0, 1){ 20}}
\multiput(179,800)(0.25000,0.50000){21}{\makebox(0.4444,0.6667){\sevrm .}}
\multiput(184,810)(0.50000,0.25000){21}{\makebox(0.4444,0.6667){\sevrm .}}
\put(194,815){\line( 1, 0){ 10}}
\put(225,780){\line( 0, 1){ 20}}
\multiput(225,800)(-0.25000,0.50000){21}{\makebox(0.4444,0.6667){\sevrm .}}
\multiput(220,810)(-0.50000,0.25000){21}{\makebox(0.4444,0.6667){\sevrm .}}
\put(210,815){\line(-1, 0){ 10}}
\put(225,788){\line( 0,-1){ 20}}
\multiput(225,768)(-0.25000,-0.50000){21}{\makebox(0.4444,0.6667){\sevrm .}}
\multiput(220,758)(-0.50000,-0.25000){21}{\makebox(0.4444,0.6667){\sevrm .}}
\put(210,753){\line(-1, 0){ 10}}
\put(120,775){\framebox(35,25){}}
\put(130,800){\line( 0, 1){  5}}
\multiput(130,805)(-0.25000,0.50000){21}{\makebox(0.4444,0.6667){\sevrm .}}
\multiput(125,815)(-0.41667,0.41667){13}{\makebox(0.4444,0.6667){\sevrm .}}
\put(120,820){\line(-1, 0){ 10}}
\multiput(110,820)(-0.50000,-0.25000){21}{\makebox(0.4444,0.6667){\sevrm .}}
\multiput(100,815)(-0.41667,-0.41667){13}{\makebox(0.4444,0.6667){\sevrm .}}
\put( 95,810){\line( 0,-1){ 25}}
\put(130,775){\line( 0,-1){  5}}
\multiput(130,770)(-0.25000,-0.50000){21}{\makebox(0.4444,0.6667){\sevrm .}}
\multiput(125,760)(-0.41667,-0.41667){13}{\makebox(0.4444,0.6667){\sevrm .}}
\put(120,755){\line(-1, 0){ 10}}
\multiput(110,755)(-0.50000,0.25000){21}{\makebox(0.4444,0.6667){\sevrm .}}
\multiput(100,760)(-0.41667,0.41667){13}{\makebox(0.4444,0.6667){\sevrm .}}
\put( 95,765){\line( 0, 1){ 25}}
\put(145,800){\line( 0, 1){ 30}}
\put(145,775){\line( 0,-1){ 30}}
\put( 74,831){\line( 0,-1){ 87}}
\put( 23,754){\makebox(0,0)[lb]{\raisebox{0pt}[0pt][0pt]
{\twlrm $a \in \cal S$}}}
\put( 66,820){\makebox(0,0)[lb]{\raisebox{0pt}[0pt][0pt]{\twlrm $b$}}}
\put(100,820){\makebox(0,0)[lb]{\raisebox{0pt}[0pt][0pt]{\twlrm $a$}}}
\put(149,822){\makebox(0,0)[lb]{\raisebox{0pt}[0pt][0pt]{\twlrm $c$}}}
\put(230,799){\makebox(0,0)[lb]{\raisebox{0pt}[0pt][0pt]{\twlrm $a$}}}
\put(353,799){\line(-1, 0){ 25}}
\put(328,799){\line( 1,-1){ 15}}
\put(353,769){\line(-1, 0){ 25}}
\put(328,769){\line( 1, 1){ 15}}
\put(482,787){\line( 0,-1){ 20}}
\multiput(482,767)(0.25000,-0.50000){21}{\makebox(0.4444,0.6667){\sevrm .}}
\multiput(487,757)(0.50000,-0.25000){21}{\makebox(0.4444,0.6667){\sevrm .}}
\put(497,752){\line( 1, 0){ 10}}
\put(482,779){\line( 0, 1){ 20}}
\multiput(482,799)(0.25000,0.50000){21}{\makebox(0.4444,0.6667){\sevrm .}}
\multiput(487,809)(0.50000,0.25000){21}{\makebox(0.4444,0.6667){\sevrm .}}
\put(497,814){\line( 1, 0){ 10}}
\put(528,779){\line( 0, 1){ 20}}
\multiput(528,799)(-0.25000,0.50000){21}{\makebox(0.4444,0.6667){\sevrm .}}
\multiput(523,809)(-0.50000,0.25000){21}{\makebox(0.4444,0.6667){\sevrm .}}
\put(513,814){\line(-1, 0){ 10}}
\put(528,787){\line( 0,-1){ 20}}
\multiput(528,767)(-0.25000,-0.50000){21}{\makebox(0.4444,0.6667){\sevrm .}}
\multiput(523,757)(-0.50000,-0.25000){21}{\makebox(0.4444,0.6667){\sevrm .}}
\put(513,752){\line(-1, 0){ 10}}
\put(423,774){\framebox(35,25){}}
\put(433,799){\line( 0, 1){  5}}
\multiput(433,804)(-0.25000,0.50000){21}{\makebox(0.4444,0.6667){\sevrm .}}
\multiput(428,814)(-0.41667,0.41667){13}{\makebox(0.4444,0.6667){\sevrm .}}
\put(423,819){\line(-1, 0){ 10}}
\multiput(413,819)(-0.50000,-0.25000){21}{\makebox(0.4444,0.6667){\sevrm .}}
\multiput(403,814)(-0.41667,-0.41667){13}{\makebox(0.4444,0.6667){\sevrm .}}
\put(398,809){\line( 0,-1){ 25}}
\put(433,774){\line( 0,-1){  5}}
\multiput(433,769)(-0.25000,-0.50000){21}{\makebox(0.4444,0.6667){\sevrm .}}
\multiput(428,759)(-0.41667,-0.41667){13}{\makebox(0.4444,0.6667){\sevrm .}}
\put(423,754){\line(-1, 0){ 10}}
\multiput(413,754)(-0.50000,0.25000){21}{\makebox(0.4444,0.6667){\sevrm .}}
\multiput(403,759)(-0.41667,0.41667){13}{\makebox(0.4444,0.6667){\sevrm .}}
\put(398,764){\line( 0, 1){ 25}}
\put(448,799){\line( 0, 1){ 30}}
\put(448,774){\line( 0,-1){ 30}}
\multiput(378,833)(0.11111,-0.55556){19}{\makebox(0.4444,0.6667){\sevrm .}}
\put(380,823){\line( 1, 0){ 42}}
\multiput(422,823)(0.52941,-0.17647){18}{\makebox(0.4444,0.6667){\sevrm .}}
\multiput(431,820)(0.33333,-0.44444){19}{\makebox(0.4444,0.6667){\sevrm .}}
\put(437,812){\line( 0,-1){ 13}}
\multiput(378,739)(0.11111,0.55556){19}{\makebox(0.4444,0.6667){\sevrm .}}
\put(380,749){\line( 1, 0){ 42}}
\multiput(422,749)(0.52941,0.17647){18}{\makebox(0.4444,0.6667){\sevrm .}}
\multiput(431,752)(0.33333,0.44444){19}{\makebox(0.4444,0.6667){\sevrm .}}
\put(437,760){\line( 0, 1){ 13}}
\put(280,782){\makebox(0,0)[lb]{\raisebox{0pt}[0pt][0pt]{\twlrm $=$}}}
\put(369,819){\makebox(0,0)[lb]{\raisebox{0pt}[0pt][0pt]{\twlrm $b$}}}
\put(452,821){\makebox(0,0)[lb]{\raisebox{0pt}[0pt][0pt]{\twlrm $c$}}}
\put(330,756){\makebox(0,0)[lb]{\raisebox{0pt}[0pt][0pt]{\twlrm
$d \in {\cal S}
$}}}
\put(387,781){\makebox(0,0)[lb]{\raisebox{0pt}[0pt][0pt]{\twlrm $d$}}}
\put(388,730){\makebox(0,0)[lb]{\raisebox{0pt}[0pt][0pt]{\twlrm $b$}}}
\put(534,796){\makebox(0,0)[lb]{\raisebox{0pt}[0pt][0pt]{\twlrm $d$}}}
\end{picture}

\end{figure}

\noindent {\em where $b \in \cal S$ and $c$ is any object, and the box
represents any map built out of the structural maps for the tortile
structure (thus representable by a framed tangle).}
\medskip

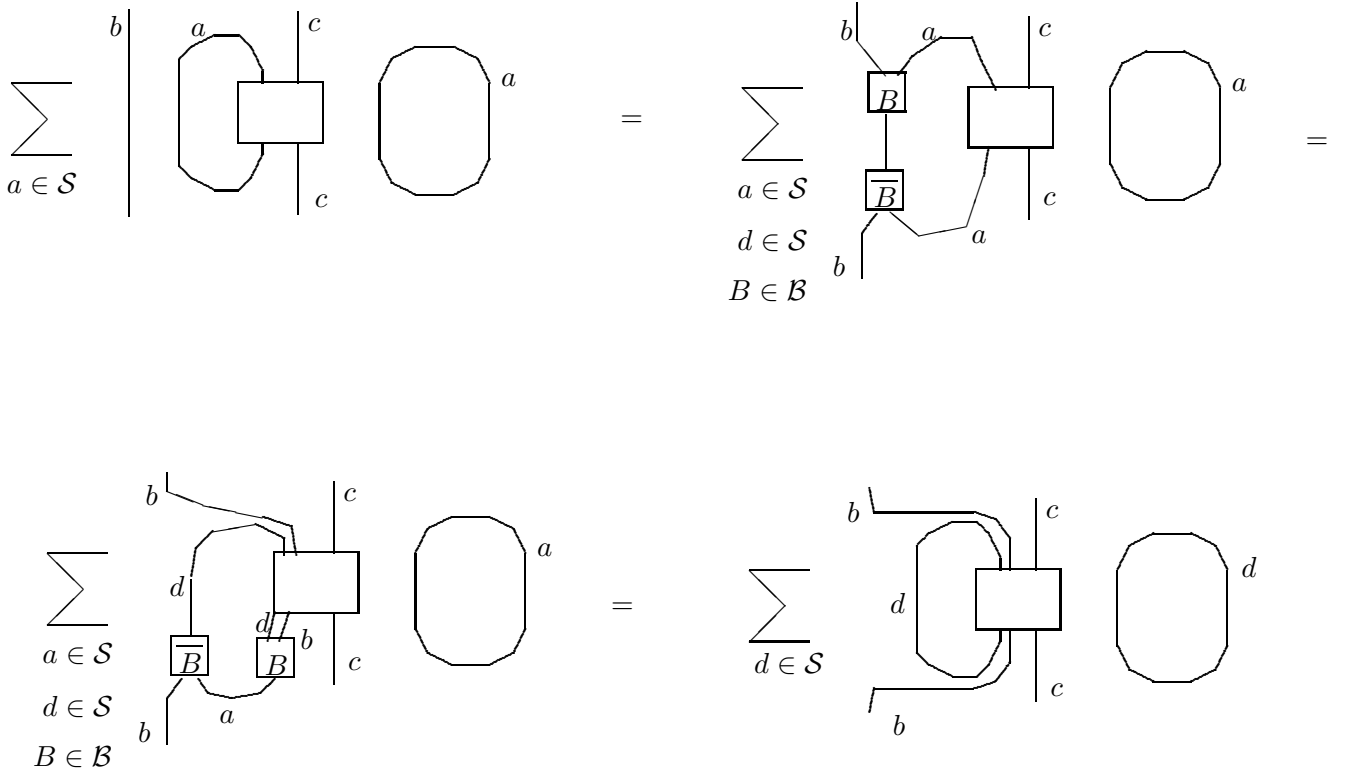
\begin{figure}[ht]
\setlength{\unitlength}{0.0125in}%
\begin{picture}(551,320)(23,514)
\thinlines
\put( 50,800){\line(-1, 0){ 25}}
\put( 25,800){\line( 1,-1){ 15}}
\put( 50,770){\line(-1, 0){ 25}}
\put( 25,770){\line( 1, 1){ 15}}
\put(179,788){\line( 0,-1){ 20}}
\multiput(179,768)(0.25000,-0.50000){21}{\makebox(0.4444,0.6667){\sevrm .}}
\multiput(184,758)(0.50000,-0.25000){21}{\makebox(0.4444,0.6667){\sevrm .}}
\put(194,753){\line( 1, 0){ 10}}
\put(179,780){\line( 0, 1){ 20}}
\multiput(179,800)(0.25000,0.50000){21}{\makebox(0.4444,0.6667){\sevrm .}}
\multiput(184,810)(0.50000,0.25000){21}{\makebox(0.4444,0.6667){\sevrm .}}
\put(194,815){\line( 1, 0){ 10}}
\put(225,780){\line( 0, 1){ 20}}
\multiput(225,800)(-0.25000,0.50000){21}{\makebox(0.4444,0.6667){\sevrm .}}
\multiput(220,810)(-0.50000,0.25000){21}{\makebox(0.4444,0.6667){\sevrm .}}
\put(210,815){\line(-1, 0){ 10}}
\put(225,788){\line( 0,-1){ 20}}
\multiput(225,768)(-0.25000,-0.50000){21}{\makebox(0.4444,0.6667){\sevrm .}}
\multiput(220,758)(-0.50000,-0.25000){21}{\makebox(0.4444,0.6667){\sevrm .}}
\put(210,753){\line(-1, 0){ 10}}
\put(120,775){\framebox(35,25){}}
\put(130,800){\line( 0, 1){  5}}
\multiput(130,805)(-0.25000,0.50000){21}{\makebox(0.4444,0.6667){\sevrm .}}
\multiput(125,815)(-0.41667,0.41667){13}{\makebox(0.4444,0.6667){\sevrm .}}
\put(120,820){\line(-1, 0){ 10}}
\multiput(110,820)(-0.50000,-0.25000){21}{\makebox(0.4444,0.6667){\sevrm .}}
\multiput(100,815)(-0.41667,-0.41667){13}{\makebox(0.4444,0.6667){\sevrm .}}
\put( 95,810){\line( 0,-1){ 25}}
\put(130,775){\line( 0,-1){  5}}
\multiput(130,770)(-0.25000,-0.50000){21}{\makebox(0.4444,0.6667){\sevrm .}}
\multiput(125,760)(-0.41667,-0.41667){13}{\makebox(0.4444,0.6667){\sevrm .}}
\put(120,755){\line(-1, 0){ 10}}
\multiput(110,755)(-0.50000,0.25000){21}{\makebox(0.4444,0.6667){\sevrm .}}
\multiput(100,760)(-0.41667,0.41667){13}{\makebox(0.4444,0.6667){\sevrm .}}
\put( 95,765){\line( 0, 1){ 25}}
\put(145,800){\line( 0, 1){ 30}}
\put(145,775){\line( 0,-1){ 30}}
\put( 74,831){\line( 0,-1){ 87}}
\put( 23,754){\makebox(0,0)[lb]{\raisebox{0pt}[0pt][0pt]{\twlrm
$a \in {\cal S}$}}}
\put( 66,820){\makebox(0,0)[lb]{\raisebox{0pt}[0pt][0pt]{\twlrm $b$}}}
\put(100,820){\makebox(0,0)[lb]{\raisebox{0pt}[0pt][0pt]{\twlrm $a$}}}
\put(149,822){\makebox(0,0)[lb]{\raisebox{0pt}[0pt][0pt]{\twlrm $c$}}}
\put(230,799){\makebox(0,0)[lb]{\raisebox{0pt}[0pt][0pt]{\twlrm $a$}}}
\put(359,596){\line(-1, 0){ 25}}
\put(334,596){\line( 1,-1){ 15}}
\put(359,566){\line(-1, 0){ 25}}
\put(334,566){\line( 1, 1){ 15}}
\put(488,584){\line( 0,-1){ 20}}
\multiput(488,564)(0.25000,-0.50000){21}{\makebox(0.4444,0.6667){\sevrm .}}
\multiput(493,554)(0.50000,-0.25000){21}{\makebox(0.4444,0.6667){\sevrm .}}
\put(503,549){\line( 1, 0){ 10}}
\put(488,576){\line( 0, 1){ 20}}
\multiput(488,596)(0.25000,0.50000){21}{\makebox(0.4444,0.6667){\sevrm .}}
\multiput(493,606)(0.50000,0.25000){21}{\makebox(0.4444,0.6667){\sevrm .}}
\put(503,611){\line( 1, 0){ 10}}
\put(534,576){\line( 0, 1){ 20}}
\multiput(534,596)(-0.25000,0.50000){21}{\makebox(0.4444,0.6667){\sevrm .}}
\multiput(529,606)(-0.50000,0.25000){21}{\makebox(0.4444,0.6667){\sevrm .}}
\put(519,611){\line(-1, 0){ 10}}
\put(534,584){\line( 0,-1){ 20}}
\multiput(534,564)(-0.25000,-0.50000){21}{\makebox(0.4444,0.6667){\sevrm .}}
\multiput(529,554)(-0.50000,-0.25000){21}{\makebox(0.4444,0.6667){\sevrm .}}
\put(519,549){\line(-1, 0){ 10}}
\put(429,571){\framebox(35,25){}}
\put(439,596){\line( 0, 1){  5}}
\multiput(439,601)(-0.25000,0.50000){21}{\makebox(0.4444,0.6667){\sevrm .}}
\multiput(434,611)(-0.41667,0.41667){13}{\makebox(0.4444,0.6667){\sevrm .}}
\put(429,616){\line(-1, 0){ 10}}
\multiput(419,616)(-0.50000,-0.25000){21}{\makebox(0.4444,0.6667){\sevrm .}}
\multiput(409,611)(-0.41667,-0.41667){13}{\makebox(0.4444,0.6667){\sevrm .}}
\put(404,606){\line( 0,-1){ 25}}
\put(439,571){\line( 0,-1){  5}}
\multiput(439,566)(-0.25000,-0.50000){21}{\makebox(0.4444,0.6667){\sevrm .}}
\multiput(434,556)(-0.41667,-0.41667){13}{\makebox(0.4444,0.6667){\sevrm .}}
\put(429,551){\line(-1, 0){ 10}}
\multiput(419,551)(-0.50000,0.25000){21}{\makebox(0.4444,0.6667){\sevrm .}}
\multiput(409,556)(-0.41667,0.41667){13}{\makebox(0.4444,0.6667){\sevrm .}}
\put(404,561){\line( 0, 1){ 25}}
\put(454,596){\line( 0, 1){ 30}}
\put(454,571){\line( 0,-1){ 30}}
\multiput(384,630)(0.11111,-0.55556){19}{\makebox(0.4444,0.6667){\sevrm .}}
\put(386,620){\line( 1, 0){ 42}}
\multiput(428,620)(0.52941,-0.17647){18}{\makebox(0.4444,0.6667){\sevrm .}}
\multiput(437,617)(0.33333,-0.44444){19}{\makebox(0.4444,0.6667){\sevrm .}}
\put(443,609){\line( 0,-1){ 13}}
\multiput(384,536)(0.11111,0.55556){19}{\makebox(0.4444,0.6667){\sevrm .}}
\put(386,546){\line( 1, 0){ 42}}
\multiput(428,546)(0.52941,0.17647){18}{\makebox(0.4444,0.6667){\sevrm .}}
\multiput(437,549)(0.33333,0.44444){19}{\makebox(0.4444,0.6667){\sevrm .}}
\put(443,557){\line( 0, 1){ 13}}
\put(375,616){\makebox(0,0)[lb]{\raisebox{0pt}[0pt][0pt]{\twlrm $b$}}}
\put(458,618){\makebox(0,0)[lb]{\raisebox{0pt}[0pt][0pt]{\twlrm $c$}}}
\put(336,553){\makebox(0,0)[lb]{\raisebox{0pt}[0pt][0pt]{\twlrm
$d \in {\cal S}$}}}
\put(393,578){\makebox(0,0)[lb]{\raisebox{0pt}[0pt][0pt]{\twlrm $d$}}}
\put(394,527){\makebox(0,0)[lb]{\raisebox{0pt}[0pt][0pt]{\twlrm $b$}}}
\put(540,593){\makebox(0,0)[lb]{\raisebox{0pt}[0pt][0pt]{\twlrm $d$}}}
\put(356,798){\line(-1, 0){ 25}}
\put(331,798){\line( 1,-1){ 15}}
\put(356,768){\line(-1, 0){ 25}}
\put(331,768){\line( 1, 1){ 15}}
\put(485,786){\line( 0,-1){ 20}}
\multiput(485,766)(0.25000,-0.50000){21}{\makebox(0.4444,0.6667){\sevrm .}}
\multiput(490,756)(0.50000,-0.25000){21}{\makebox(0.4444,0.6667){\sevrm .}}
\put(500,751){\line( 1, 0){ 10}}
\put(485,778){\line( 0, 1){ 20}}
\multiput(485,798)(0.25000,0.50000){21}{\makebox(0.4444,0.6667){\sevrm .}}
\multiput(490,808)(0.50000,0.25000){21}{\makebox(0.4444,0.6667){\sevrm .}}
\put(500,813){\line( 1, 0){ 10}}
\put(531,778){\line( 0, 1){ 20}}
\multiput(531,798)(-0.25000,0.50000){21}{\makebox(0.4444,0.6667){\sevrm .}}
\multiput(526,808)(-0.50000,0.25000){21}{\makebox(0.4444,0.6667){\sevrm .}}
\put(516,813){\line(-1, 0){ 10}}
\put(531,786){\line( 0,-1){ 20}}
\multiput(531,766)(-0.25000,-0.50000){21}{\makebox(0.4444,0.6667){\sevrm .}}
\multiput(526,756)(-0.50000,-0.25000){21}{\makebox(0.4444,0.6667){\sevrm .}}
\put(516,751){\line(-1, 0){ 10}}
\put(384,788){\framebox(15,16){}}
\put(387,789){\makebox(0,0)[lb]{\raisebox{0pt}[0pt][0pt]{\twlrm $B$}}}
\put(383,747){\framebox(15,16){}}
\put(386,748){\makebox(0,0)[lb]{\raisebox{0pt}[0pt][0pt]{\twlrm $B$}}}
\put(426,773){\framebox(35,25){}}
\put(451,798){\line( 0, 1){ 30}}
\put(451,773){\line( 0,-1){ 30}}
\put(385,759){\line( 1, 0){ 11}}
\put(391,787){\line( 0,-1){ 24}}
\put(379,834){\line( 0,-1){ 16}}
\put(379,818){\line( 4,-5){ 12}}
\multiput(387,745)(-0.33333,-0.44444){19}{\makebox(0.4444,0.6667){\sevrm .}}
\put(381,737){\line( 0,-1){ 19}}
\put(393,746){\line( 6,-5){ 12}}
\put(405,736){\line( 5, 1){ 20}}
\put(425,740){\line( 1, 3){  7}}
\multiput(432,761)(0.09524,0.57143){22}{\makebox(0.4444,0.6667){\sevrm .}}
\multiput(396,803)(0.33333,0.44444){19}{\makebox(0.4444,0.6667){\sevrm .}}
\multiput(402,811)(0.48000,0.32000){26}{\makebox(0.4444,0.6667){\sevrm .}}
\put(414,819){\line( 1, 0){ 13}}
\multiput(427,819)(0.21053,-0.52632){20}{\makebox(0.4444,0.6667){\sevrm .}}
\multiput(431,809)(0.25000,-0.50000){25}{\makebox(0.4444,0.6667){\sevrm .}}
\put(329,752){\makebox(0,0)[lb]{\raisebox{0pt}[0pt][0pt]{\twlrm
$a \in {\cal S}$}}}
\put(372,818){\makebox(0,0)[lb]{\raisebox{0pt}[0pt][0pt]{\twlrm $b$}}}
\put(406,818){\makebox(0,0)[lb]{\raisebox{0pt}[0pt][0pt]{\twlrm $a$}}}
\put(455,820){\makebox(0,0)[lb]{\raisebox{0pt}[0pt][0pt]{\twlrm $c$}}}
\put(536,797){\makebox(0,0)[lb]{\raisebox{0pt}[0pt][0pt]{\twlrm $a$}}}
\put(329,730){\makebox(0,0)[lb]{\raisebox{0pt}[0pt][0pt]{\twlrm
$d \in {\cal S}$}}}
\put(325,709){\makebox(0,0)[lb]{\raisebox{0pt}[0pt][0pt]{\twlrm
$B \in {\cal B}$}}}
\put(457,749){\makebox(0,0)[lb]{\raisebox{0pt}[0pt][0pt]{\twlrm $c$}}}
\put(427,733){\makebox(0,0)[lb]{\raisebox{0pt}[0pt][0pt]{\twlrm $a$}}}
\put(369,719){\makebox(0,0)[lb]{\raisebox{0pt}[0pt][0pt]{\twlrm $b$}}}
\put(567,773){\makebox(0,0)[lb]{\raisebox{0pt}[0pt][0pt]{\twlrm $=$}}}
\put( 65,603){\line(-1, 0){ 25}}
\put( 40,603){\line( 1,-1){ 15}}
\put( 65,573){\line(-1, 0){ 25}}
\put( 40,573){\line( 1, 1){ 15}}
\put(194,591){\line( 0,-1){ 20}}
\multiput(194,571)(0.25000,-0.50000){21}{\makebox(0.4444,0.6667){\sevrm .}}
\multiput(199,561)(0.50000,-0.25000){21}{\makebox(0.4444,0.6667){\sevrm .}}
\put(209,556){\line( 1, 0){ 10}}
\put(194,583){\line( 0, 1){ 20}}
\multiput(194,603)(0.25000,0.50000){21}{\makebox(0.4444,0.6667){\sevrm .}}
\multiput(199,613)(0.50000,0.25000){21}{\makebox(0.4444,0.6667){\sevrm .}}
\put(209,618){\line( 1, 0){ 10}}
\put(240,583){\line( 0, 1){ 20}}
\multiput(240,603)(-0.25000,0.50000){21}{\makebox(0.4444,0.6667){\sevrm .}}
\multiput(235,613)(-0.50000,0.25000){21}{\makebox(0.4444,0.6667){\sevrm .}}
\put(225,618){\line(-1, 0){ 10}}
\put(240,591){\line( 0,-1){ 20}}
\multiput(240,571)(-0.25000,-0.50000){21}{\makebox(0.4444,0.6667){\sevrm .}}
\multiput(235,561)(-0.50000,-0.25000){21}{\makebox(0.4444,0.6667){\sevrm .}}
\put(225,556){\line(-1, 0){ 10}}
\put( 92,552){\framebox(15,16){}}
\put( 95,553){\makebox(0,0)[lb]{\raisebox{0pt}[0pt][0pt]{\twlrm $B$}}}
\put(128,551){\framebox(15,16){}}
\put(131,552){\makebox(0,0)[lb]{\raisebox{0pt}[0pt][0pt]{\twlrm $B$}}}
\put(135,578){\framebox(35,25){}}
\put(160,603){\line( 0, 1){ 30}}
\put(160,578){\line( 0,-1){ 30}}
\put( 94,564){\line( 1, 0){ 11}}
\put(100,592){\line( 0,-1){ 24}}
\multiput( 96,550)(-0.33333,-0.44444){19}{\makebox(0.4444,0.6667){\sevrm .}}
\put( 90,542){\line( 0,-1){ 19}}
\multiput(132,566)(0.13636,0.54545){23}{\makebox(0.4444,0.6667){\sevrm .}}
\multiput(137,566)(0.18182,0.54545){23}{\makebox(0.4444,0.6667){\sevrm .}}
\put(139,602){\line( 0, 1){  7}}
\multiput(139,609)(-0.50000,0.25000){25}{\makebox(0.4444,0.6667){\sevrm .}}
\put(127,615){\line(-6,-1){ 18}}
\multiput(109,612)(-0.41176,-0.41176){18}{\makebox(0.4444,0.6667){\sevrm .}}
\multiput(102,605)(-0.09524,-0.57143){22}{\makebox(0.4444,0.6667){\sevrm .}}
\multiput(144,602)(-0.09524,0.57143){22}{\makebox(0.4444,0.6667){\sevrm .}}
\multiput(142,614)(-0.54545,0.18182){23}{\makebox(0.4444,0.6667){\sevrm .}}
\put(130,618){\line(-5, 1){ 25}}
\put(105,623){\line(-5, 2){ 15}}
\put( 90,629){\line( 0, 1){  8}}
\multiput(103,550)(0.33333,-0.50000){13}{\makebox(0.4444,0.6667){\sevrm .}}
\multiput(107,544)(0.55556,-0.11111){19}{\makebox(0.4444,0.6667){\sevrm .}}
\multiput(117,542)(0.57143,0.09524){22}{\makebox(0.4444,0.6667){\sevrm .}}
\multiput(129,544)(0.40000,0.40000){11}{\makebox(0.4444,0.6667){\sevrm .}}
\multiput(133,548)(0.40000,0.40000){6}{\makebox(0.4444,0.6667){\sevrm .}}
\put(280,782){\makebox(0,0)[lb]{\raisebox{0pt}[0pt][0pt]{\twlrm $=$}}}
\put(152,748){\makebox(0,0)[lb]{\raisebox{0pt}[0pt][0pt]{\twlrm $c$}}}
\put(460,543){\makebox(0,0)[lb]{\raisebox{0pt}[0pt][0pt]{\twlrm $c$}}}
\put( 38,557){\makebox(0,0)[lb]{\raisebox{0pt}[0pt][0pt]{\twlrm
$a \in {\cal S}$}}}
\put( 81,623){\makebox(0,0)[lb]{\raisebox{0pt}[0pt][0pt]{\twlrm $b$}}}
\put(164,625){\makebox(0,0)[lb]{\raisebox{0pt}[0pt][0pt]{\twlrm $c$}}}
\put(245,602){\makebox(0,0)[lb]{\raisebox{0pt}[0pt][0pt]{\twlrm $a$}}}
\put( 38,535){\makebox(0,0)[lb]{\raisebox{0pt}[0pt][0pt]{\twlrm
$d \in {\cal S}$}}}
\put( 34,514){\makebox(0,0)[lb]{\raisebox{0pt}[0pt][0pt]{\twlrm
$B \in {\cal B}$}}}
\put(166,554){\makebox(0,0)[lb]{\raisebox{0pt}[0pt][0pt]{\twlrm $c$}}}
\put( 78,524){\makebox(0,0)[lb]{\raisebox{0pt}[0pt][0pt]{\twlrm $b$}}}
\put(276,578){\makebox(0,0)[lb]{\raisebox{0pt}[0pt][0pt]{\twlrm $=$}}}
\put(112,532){\makebox(0,0)[lb]{\raisebox{0pt}[0pt][0pt]{\twlrm $a$}}}
\put( 91,585){\makebox(0,0)[lb]{\raisebox{0pt}[0pt][0pt]{\twlrm $d$}}}
\put(127,570){\makebox(0,0)[lb]{\raisebox{0pt}[0pt][0pt]{\twlrm $d$}}}
\put(146,563){\makebox(0,0)[lb]{\raisebox{0pt}[0pt][0pt]{\twlrm $b$}}}
\end{picture}

\caption{\label{hsp} Proof of Proposition \protect\ref{handleslide} }
\end{figure}

\noindent {\bf proof:} The proof is the diagrammatic calculation
given in Figure \ref{hsp}.
The first step is an application of Lemma \ref{sumlemma},
the second uses the naturality (and dinaturality) properties of the
structure maps, and the third is an application of Lemma \ref{pdb}.

Thus, the framed link invariant arising by labelling each strand of the link
with the linear combination

\[ \omega_{\cal C} = \sum_{s \in \cal S} dim(s) s \]

is invariant under handle-sliding, regardless of what semisimple tortile
category $\cal C$ we use. (It is a trivial to see that the invariant
does not depend on the choice of $\cal S$.)

Now, if we let $a_+$ (resp. $a_-$) be the value of the $+1$-
(resp. $-1$-)framed $\omega_{\cal C}$-labelled unknot,
and assume that $\cal C$ satisfies

\begin{defin} A semisimple tortile category over a field $k$
is {\em 3-conformed} if the values $a_+$ and $a_-$ are non-zero.
\end{defin}

\noindent then letting  $x = (a_+ a_-)^\frac{1}{2}$ and
$y = (a_+/a_-)^\frac{1}{2}$, it follows immediately from the same
sort of argument given in [RT] that if $M$ is the 3-manifold obtained
as the boundary of the 4-dimensional handle-body with one 0-handle, and
2-handles attached using $L$, then

\[ {\cal I_C}(M) = \omega_{\cal C}(L) x^{-|L|} y^{-\sigma(L)} \]

\noindent depends only on the diffeomorphism type of $M$, where $|L|$ is
the number of components of $L$ and $\sigma(L)$ is the signature of the
linking matrix. We shall call ${\cal I_C}(M)$ a generalized Reshetikhin/Tureav
invariant of $M$. Note that we have used a different non-degeneracy
condition than that used by Turaev [T].

In a similar way, if we let $b$ be the evaluation of an
$\omega_{\cal C}$-labelled Hopf link, then if $W$ is the 4-manifold
obtained by attaching 2-handles to undotted components of $L$, and
``hollowing out 2-handles'' along dotted components of $L$ as in Kirby [Ki],
then

\[ {\cal B_C}(W) = \omega_{\cal C}(L) b^\frac{\nu(L)-|L|}{2} N^{-\nu(L)} \]

\noindent depends only on the diffeomorphism type of $W$, where $|L|$ is
as above and $\nu(L)$ is the nullity of the linking matrix. We will call
${\cal B_C}(W)$ a generalized Broda invariant.  Note first that for
${\cal B_C}(W)$ to be defined, we need a different non-degeneracy condition:

\begin{defin} A semi-simple tortile category is {\em 4-conformed} if $b$
and $N$ are both non-zero.
\end{defin}

We will see in Section \ref{centerm} that under hypotheses satisfied by
the TL and KR categories at principal $4r^{th}$-roots of unity,
$N = b = x^2$ and $a_+$ and $a_-$ are both non-zero provided $N$ is.

In the next section, we present Roberts' analysis [Ro1] (cf. also [CKY])
of the relation between
the original Crane-Yetter invariant [CY], the original Broda invariant [B], and
the corresponding 3-manifold invariant, the Reshetikhin/Turaev invariant [RT],
all in the TL formulation.

\clearpage \section{TL translation of $CY(W)$ and Roberts' Chainmail
Method}

\subsection{\label{translation} Translation}

The purpose of this section is to give a sketch of Justin Roberts' beautiful
method of understanding the Crane-Yetter 4-manifold invariant in the
case of $U_q(sl_2)$.  In order to accomplish this connection we need to
translate the original formulation of the Crane-Yetter invariant in
terms of Kirillov-Reshetikhin recoupling into Temperley-Lieb recoupling,
in that Roberts' method is cleanest in the TL theory. This
translation has already been done in [CKY]. For completeness, we repeat this
construction here.

First recall the general definition for a Crane-Yetter invariant:

\[ CY_{\cal C}(M) = \sum_{\lambda \in \Lambda_{\cal CSB}({\bf T})}
 \ll \lambda \gg \]

\noindent where

\[
 \ll \lambda \gg  =  N^{n_0 - n_1} \prod_{\parbox{.6in}{\small faces
$\sigma$}}
dim(\lambda(\sigma)) \prod_{\parbox{.7in}{\small tetrahedra $\tau $}}
dim(\lambda
(\tau))^{-1}
\prod_{\parbox{.8in}{\small 4-simplices $\xi$}}
\| \lambda, \xi \|
\]

\noindent and $\| \lambda, \xi \|$ is the 15j-network appropriate to
the 4-simplex $\xi$ and the coloring $\lambda$. Recall $n_0$ (resp. $n_1$)
is the number of vertices (resp. edges) in the triangulation and
$N$ is the sum of the squares of the quantum dimensions.

In the case of where our category is the truncation of $Rep(U_q(sl_2))$ at
a root of unity, we choose as $\cal S$, the irreducible
representations of $U_q(sl_2)$ at $A = \sqrt(q)$ a $4r^{th}$-root of unity,
labelling them $\{ 0, 1, 2, ... , r-2\}$;
The bases $\cal B$ for the hom-spaces  consist of the projections and
inclusions given by the KR 3-vertices; and $N$, the sum of the squares of
the quantum dimensions, has the specific value $N = -\frac{2r}{(q-q^{-1})^2}$.
($q = \exp(i\pi/r)$).

Because the labels at the node of the 15j-symbol are uniquely determined
by the labels of the arcs incident, we can regard the labelling as
a coloring of faces and tetrahedra by integers (twice spins). That is,
the 15j-symbol becomes in this case simply a particular KR network.

Now, we have remarked in the first part of Section 2 that there is a simple
translation from KR theory to TL theory effected by the formula of
Figure \ref{KRtoTL}.

Recall that in Figure \ref{KRtoTL} $\theta (a,b,c)$ is the evaluation of the
TL theta net with lables $a, b,$ and $c$. The result of applying this
translation to the 15j-symbols results in a formula for the
($U_q(sl_2)$) Crane-Yetter invariant $CY(W)$ in the TL theory:

\[ CY(M) = \sum_{\lambda \in \Lambda({\bf T})}
 \ll \lambda \gg \]

\noindent where the sum runs over all labellings of faces and tetrahedra
by elements of $\{0,1,...,r-2\}$ and

\begin{eqnarray*}
 \ll \lambda \gg & = & N^{n_0 - n_1} \prod_{\parbox{.6in}{\small faces
$\sigma$}}
dim(\lambda(\sigma)) \prod_{\parbox{.7in}{\small tetrahedra $\tau $}}
dim(\lambda(\tau)) \theta(\lambda(\tau),\lambda(\tau_0),\lambda(\tau_2))
\theta(\lambda(\tau),\lambda(\tau_1),\lambda(\tau_3)) \\
 & & \times \prod_{\parbox{.8in}{\small 4-simplices $\xi$}}
\| \lambda, \xi \|_TL
\end{eqnarray*}

	Here $\| \lambda, \xi \|_TL$ denotes the TL
15j-symbol associated to $\xi$ and the coloring $\lambda$, that is
the network given by the diagram of Figure \ref{gen15j}, with TL 3-vertices
in place of the ends of every ``dumbbell'', and $\tau_0, \tau_1, \tau_2,
\tau_3$ are the faces of the tetrahedron $\tau$ obtained by omitting
the lowest numbered,...,highest numbered vertex of $\tau$ in the ordering
on $\bf T$.

	This completes the translation of $CY(W)$ into the TL recoupling
theory.

\subsection{Roberts' Chain Mail}

	Roberts considers a triangulated 4-manifold $W$ and its dual
handlebody decomposition $D^\ast$. The 0-handles of $D^\ast$ correspond to
the 4-simplices of $W$; the 1-handles of $D^\ast$ correspond to the
tetrahedra of $W$; the 2-handles of $D^\ast$ have framed attaching curves
 on the boundary of $M = \partial N$, where $N$ is the union of the 0- and
1-handles of $D^\ast$---these 2-handles correspond to the faces of the
triangulation of $W$.

	Letting $N^\prime$ denote $N$ with the 2-handles attached and
$M^\prime = \partial N^\prime$, note that both $M$ and $M^\prime$ are
connected sums of $S^1 \times S^2$'s.  By adding more $S^1 \times S^2$'s
(by adding 1-handles which cancel all but one of the 0-handles of $D^\ast$),
Roberts produces a surgery description of $N^{\prime \prime}$ by a
labelled link in $S^3$ (cf. Kirby [Ki]), where closing up $N^{\prime \prime}$
by 3- and 4-handles give $W^\prime = W \# d(S^1 \times S^3)$ for
$d = n_4-1$ ($n_4$ is the number of 4-simplexes in the original triangulation
on $W$).

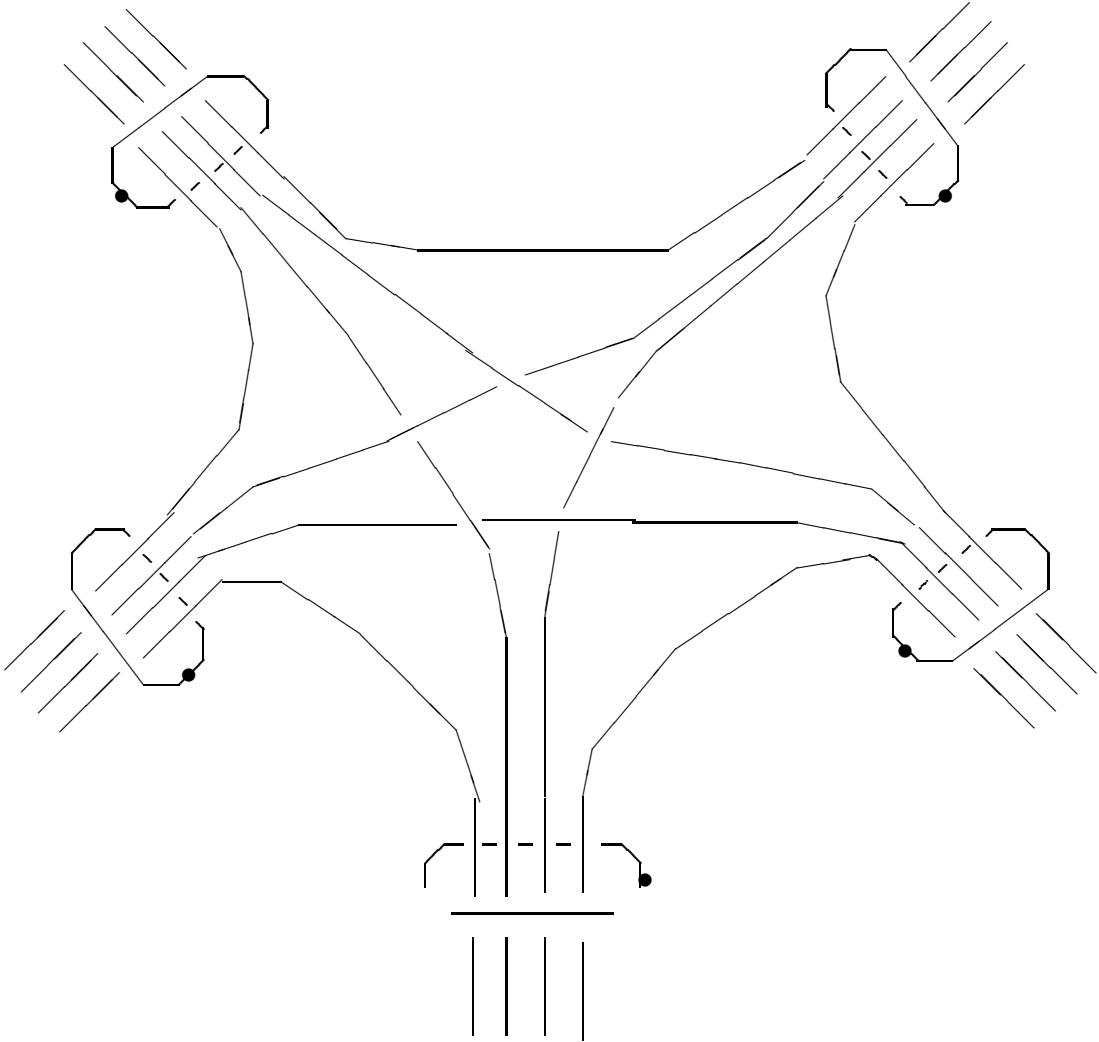
\begin{figure}[htb]
\setlength{\unitlength}{0.0125in}%
\begin{picture}(457,435)(44,358)
\thinlines
\put(388,753){\line( 0, 1){ 10}}
\multiput(388,763)(0.40000,0.40000){26}{\makebox(0.4444,0.6667){\sevrm .}}
\put(398,773){\line( 1, 0){ 15}}
\put(413,773){\line( 3,-4){ 30}}
\put(443,733){\line( 0,-1){ 15}}
\multiput(443,718)(-0.40000,-0.40000){26}{\makebox(0.4444,0.6667){\sevrm .}}
\put(433,708){\line(-1, 0){ 10}}
\put(448,793){\line(-1,-1){ 25}}
\put(457,785){\line(-1,-1){ 25}}
\put(464,776){\line(-1,-1){ 25}}
\put(471,767){\line(-1,-1){ 25}}
\put(413,762){\line(-1,-1){ 33}}
\put(420,752){\line(-1,-1){ 33}}
\put(426,744){\line(-1,-1){ 33}}
\put(433,734){\line(-1,-1){ 33}}
\multiput(419,711)(0.42857,-0.42857){8}{\makebox(0.4444,0.6667){\sevrm .}}
\multiput(388,750)(0.42857,-0.42857){8}{\makebox(0.4444,0.6667){\sevrm .}}
\multiput(395,740)(0.42857,-0.42857){8}{\makebox(0.4444,0.6667){\sevrm .}}
\multiput(403,730)(0.42857,-0.42857){8}{\makebox(0.4444,0.6667){\sevrm .}}
\multiput(410,722)(0.42857,-0.42857){8}{\makebox(0.4444,0.6667){\sevrm .}}
\put(388,749){\line( 0, 1){  6}}
\put(421,708){\line( 1, 0){  5}}
\put(109,707){\line(-1, 0){ 10}}
\multiput( 99,707)(-0.40000,0.40000){26}{\makebox(0.4444,0.6667){\sevrm .}}
\put( 89,717){\line( 0, 1){ 15}}
\put( 89,732){\line( 4, 3){ 40}}
\put(129,762){\line( 1, 0){ 15}}
\multiput(144,762)(0.40000,-0.40000){26}{\makebox(0.4444,0.6667){\sevrm .}}
\put(154,752){\line( 0,-1){ 10}}
\put( 69,767){\line( 1,-1){ 25}}
\put( 77,776){\line( 1,-1){ 25}}
\put( 86,783){\line( 1,-1){ 25}}
\put( 95,790){\line( 1,-1){ 25}}
\put(100,732){\line( 1,-1){ 33}}
\put(110,739){\line( 1,-1){ 33}}
\put(118,745){\line( 1,-1){ 33}}
\put(128,752){\line( 1,-1){ 33}}
\multiput(151,738)(0.42857,0.42857){8}{\makebox(0.4444,0.6667){\sevrm .}}
\multiput(112,707)(0.42857,0.42857){8}{\makebox(0.4444,0.6667){\sevrm .}}
\multiput(122,714)(0.42857,0.42857){8}{\makebox(0.4444,0.6667){\sevrm .}}
\multiput(132,722)(0.42857,0.42857){8}{\makebox(0.4444,0.6667){\sevrm .}}
\multiput(140,729)(0.42857,0.42857){8}{\makebox(0.4444,0.6667){\sevrm .}}
\put(113,707){\line(-1, 0){  6}}
\put(154,740){\line( 0, 1){  5}}
\put(461,572){\line( 1, 0){ 10}}
\multiput(471,572)(0.40000,-0.40000){26}{\makebox(0.4444,0.6667){\sevrm .}}
\put(481,562){\line( 0,-1){ 15}}
\put(481,547){\line(-4,-3){ 40}}
\put(441,517){\line(-1, 0){ 15}}
\multiput(426,517)(-0.40000,0.40000){26}{\makebox(0.4444,0.6667){\sevrm .}}
\put(416,527){\line( 0, 1){ 10}}
\put(501,512){\line(-1, 1){ 25}}
\put(493,503){\line(-1, 1){ 25}}
\put(484,496){\line(-1, 1){ 25}}
\put(475,489){\line(-1, 1){ 25}}
\put(470,547){\line(-1, 1){ 33}}
\put(460,540){\line(-1, 1){ 33}}
\put(452,534){\line(-1, 1){ 33}}
\put(442,527){\line(-1, 1){ 33}}
\multiput(419,541)(-0.42857,-0.42857){8}{\makebox(0.4444,0.6667){\sevrm .}}
\multiput(458,572)(-0.42857,-0.42857){8}{\makebox(0.4444,0.6667){\sevrm .}}
\multiput(448,565)(-0.42857,-0.42857){8}{\makebox(0.4444,0.6667){\sevrm .}}
\multiput(438,557)(-0.42857,-0.42857){8}{\makebox(0.4444,0.6667){\sevrm .}}
\multiput(430,550)(-0.42857,-0.42857){8}{\makebox(0.4444,0.6667){\sevrm .}}
\put(457,572){\line( 1, 0){  6}}
\put(416,539){\line( 0,-1){  5}}
\put(127,527){\line( 0,-1){ 10}}
\multiput(127,517)(-0.40000,-0.40000){26}{\makebox(0.4444,0.6667){\sevrm .}}
\put(117,507){\line(-1, 0){ 15}}
\put(102,507){\line(-3, 4){ 30}}
\put( 72,547){\line( 0, 1){ 15}}
\multiput( 72,562)(0.40000,0.40000){26}{\makebox(0.4444,0.6667){\sevrm .}}
\put( 82,572){\line( 1, 0){ 10}}
\put( 67,487){\line( 1, 1){ 25}}
\put( 58,495){\line( 1, 1){ 25}}
\put( 51,504){\line( 1, 1){ 25}}
\put( 44,513){\line( 1, 1){ 25}}
\put(102,518){\line( 1, 1){ 33}}
\put( 95,528){\line( 1, 1){ 33}}
\put( 89,536){\line( 1, 1){ 33}}
\put( 82,546){\line( 1, 1){ 33}}
\multiput( 96,569)(-0.42857,0.42857){8}{\makebox(0.4444,0.6667){\sevrm .}}
\multiput(127,530)(-0.42857,0.42857){8}{\makebox(0.4444,0.6667){\sevrm .}}
\multiput(120,540)(-0.42857,0.42857){8}{\makebox(0.4444,0.6667){\sevrm .}}
\multiput(112,550)(-0.42857,0.42857){8}{\makebox(0.4444,0.6667){\sevrm .}}
\multiput(105,558)(-0.42857,0.42857){8}{\makebox(0.4444,0.6667){\sevrm .}}
\put(127,531){\line( 0,-1){  6}}
\put( 94,572){\line(-1, 0){  5}}
\put(121,511){\circle*{6}}
\put( 93,712){\circle*{6}}
\put(438,712){\circle*{6}}
\put(421,521){\circle*{6}}
\put(312,425){\circle*{6}}
\put(241,459){\line( 0,-1){ 41}}
\put(254,460){\line( 0,-1){ 42}}
\put(270,459){\line( 0,-1){ 39}}
\put(286,460){\line( 0,-1){ 40}}
\put(236,440){\line(-1, 0){  8}}
\multiput(228,440)(-0.40000,-0.40000){21}{\makebox(0.4444,0.6667){\sevrm .}}
\put(220,432){\line( 0,-1){ 10}}
\put(220,422){\line( 1,-1){ 11}}
\put(231,411){\line( 1, 0){ 34}}
\put(294,440){\line( 1, 0){  8}}
\multiput(302,440)(0.40000,-0.40000){21}{\makebox(0.4444,0.6667){\sevrm .}}
\put(310,432){\line( 0,-1){ 10}}
\put(310,422){\line(-1,-1){ 11}}
\put(299,411){\line(-1, 0){ 34}}
\put(240,401){\line( 0,-1){ 41}}
\put(254,401){\line( 0,-1){ 41}}
\put(270,401){\line( 0,-1){ 41}}
\put(286,399){\line( 0,-1){ 41}}
\put(275,440){\line( 1, 0){  6}}
\put(244,440){\line( 1, 0){  6}}
\put(259,440){\line( 1, 0){  6}}
\put(161,720){\line( 1,-1){ 26}}
\put(187,694){\line( 6,-1){ 30}}
\put(217,689){\line( 1, 0){ 58}}
\put(275,689){\line( 1, 0){ 47}}
\put(322,689){\line( 3, 2){ 57}}
\put(400,700){\line(-2,-5){ 12}}
\put(388,670){\line( 1,-6){  6}}
\put(394,634){\line( 4,-5){ 44}}
\put(134,698){\line( 1,-2){  9}}
\put(143,680){\line( 1,-6){  5}}
\put(148,650){\line(-1,-6){  6}}
\put(142,614){\line(-5,-6){ 30}}
\put(135,550){\line( 1, 0){ 25}}
\put(160,550){\line( 3,-2){ 33}}
\put(193,528){\line( 1,-1){ 40}}
\put(233,488){\line( 1,-3){ 10}}
\put(286,460){\line( 1, 5){  4}}
\put(290,480){\line( 5, 6){ 35}}
\put(325,522){\line( 3, 2){ 51}}
\put(376,556){\line( 6, 1){ 30}}
\multiput(406,561)(0.50000,-0.33333){7}{\makebox(0.4444,0.6667){\sevrm .}}
\put(250,632){\line(-2,-1){ 46}}
\put(217,609){\line( 2,-3){ 30}}
\put(244,576){\line( 1, 0){ 64}}
\put(237,647){\line( 3,-2){ 51}}
\put(278,581){\line( 1, 2){ 21}}
\put(262,637){\line( 3, 1){ 45}}
\put(307,652){\line( 4, 3){ 56}}
\put(363,694){\line( 1, 1){ 24}}
\put(143,707){\line( 5,-6){ 45}}
\put(188,653){\line( 2,-3){ 22}}
\put(298,609){\line( 6,-1){ 54}}
\put(352,600){\line( 5,-1){ 55}}
\put(407,589){\line( 6,-5){ 18}}
\put(276,571){\line(-1,-6){  6}}
\put(270,535){\line( 0,-1){ 75}}
\put(233,574){\line(-1, 0){ 66}}
\put(167,574){\line(-3,-1){ 42}}
\put(205,609){\line(-3,-1){ 57}}
\put(148,590){\line(-5,-4){ 25}}
\put(247,562){\line( 1,-5){  7}}
\put(254,527){\line( 0,-1){ 70}}
\put(307,575){\line( 1, 0){ 69}}
\put(376,575){\line( 5,-1){ 45}}
\put(395,712){\line(-6,-5){ 78}}
\put(317,647){\line(-4,-5){ 16}}
\put(152,712){\line( 4,-3){ 88}}
\end{picture}
\caption{\label{chainmail} The Portion of Chainmail Corresponding to a
4-simplex}
\end{figure}

	The surgery curves of the presentation of $N^{\prime \prime}$ then
take the local form shown in Figure \ref{chainmail} or its mirror image,
with one such region occuring for each 4-simplex of $W$.

	In Figure \ref{chainmail}
the dotted curves are meridians corresponding to the 1-handles
(recall from Kirby [Ki] that ``hollowing out'' a 2-handle along an
unknot is equivalent to attaching a 1-handle), the other curves are
(parts of) attaching curves for the 2-handles.

	One can then define the (original) Broda invariant [B]
 of $W^\prime $ by
labelling every curve with $\omega = \sum_{i=0}^{r-2} \Delta_i \cdot i$.
Here $i$ denotes the parallel cabling of $i$ arcs (in blackboard framing)
with a q-symmetrizer attached as in Section 2. One then normalizations
of the bracket evaluation of the resulting sum by multiplying by
$N^\frac{|L|+\nu(L)}/2$, where $|L|$ is the number of components of
the ``chainmail link'' $L$, and $\nu(L)$ is the nullity of its linking
matrix. The resulting invariant is then seen to
be $I(W) = \kappa^{\sigma(W^\prime)}$ where

\[ \kappa = \exp(\frac{i\pi(-3-r^2)}{2r} - \frac{i\pi}{4}) \]

\noindent and $\sigma(W)$ is the signature of the 4-manifold $W$, but since
$W^\prime $ is a connected sum of $W$ with a manifold with trivial signature,
$\sigma(W) = \sigma(W^\prime )$.

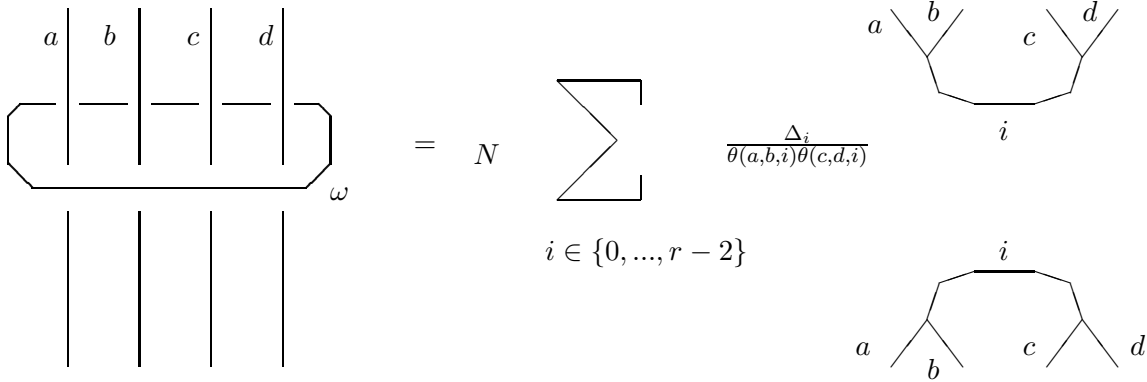
\begin{figure}[htb]
\setlength{\unitlength}{0.0125in}%
\begin{picture}(470,159)(65,635)
\thinlines
\put(330,750){\line( 0, 1){ 10}}
\put(330,760){\line(-1, 0){ 35}}
\put(295,760){\line( 1,-1){ 25}}
\put(330,720){\line( 0,-1){ 10}}
\put(330,710){\line(-1, 0){ 35}}
\put(295,710){\line( 1, 1){ 25}}
\put(435,790){\line( 3,-4){ 15}}
\put(465,790){\line(-3,-4){ 15}}
\put(500,790){\line( 3,-4){ 15}}
\put(530,790){\line(-3,-4){ 15}}
\put(450,770){\line( 1,-3){  5}}
\put(455,755){\line( 3,-1){ 15}}
\put(470,750){\line( 1, 0){ 25}}
\put(515,770){\line(-1,-3){  5}}
\put(510,755){\line(-3,-1){ 15}}
\put(495,750){\line(-1, 0){ 25}}
\put(435,640){\line( 3, 4){ 15}}
\put(465,640){\line(-3, 4){ 15}}
\put(500,640){\line( 3, 4){ 15}}
\put(530,640){\line(-3, 4){ 15}}
\put(450,660){\line( 1, 3){  5}}
\put(455,675){\line( 3, 1){ 15}}
\put(470,680){\line( 1, 0){ 25}}
\put(515,660){\line(-1, 3){  5}}
\put(510,675){\line(-3, 1){ 15}}
\put(495,680){\line(-1, 0){ 25}}
\put( 90,790){\line( 0,-1){ 65}}
\put(120,790){\line( 0,-1){ 65}}
\put(150,790){\line( 0,-1){ 65}}
\put(180,790){\line( 0,-1){ 65}}
\put( 90,705){\line( 0,-1){ 65}}
\put(120,705){\line( 0,-1){ 65}}
\put(150,705){\line( 0,-1){ 65}}
\put(180,705){\line( 0,-1){ 65}}
\put( 75,715){\line( 1, 0){115}}
\multiput(190,715)(0.40000,0.40000){26}{\makebox(0.4444,0.6667){\sevrm .}}
\put(200,725){\line( 0, 1){ 20}}
\multiput(200,745)(-0.41667,0.41667){13}{\makebox(0.4444,0.6667){\sevrm .}}
\put(195,750){\line(-1, 0){ 10}}
\multiput( 75,715)(-0.40000,0.40000){26}{\makebox(0.4444,0.6667){\sevrm .}}
\put( 65,725){\line( 0, 1){ 20}}
\multiput( 65,745)(0.41667,0.41667){13}{\makebox(0.4444,0.6667){\sevrm .}}
\put( 70,750){\line( 1, 0){ 15}}
\put( 95,750){\line( 1, 0){ 20}}
\put(125,750){\line( 1, 0){ 20}}
\put(155,750){\line( 1, 0){ 20}}
\put( 80,775){\makebox(0,0)[lb]{\raisebox{0pt}[0pt][0pt]{\twlrm $a$}}}
\put(105,775){\makebox(0,0)[lb]{\raisebox{0pt}[0pt][0pt]{\twlrm $b$}}}
\put(140,775){\makebox(0,0)[lb]{\raisebox{0pt}[0pt][0pt]{\twlrm $c$}}}
\put(170,775){\makebox(0,0)[lb]{\raisebox{0pt}[0pt][0pt]{\twlrm $d$}}}
\put(200,710){\makebox(0,0)[lb]{\raisebox{0pt}[0pt][0pt]{\twlrm $\omega$}}}
\put(235,730){\makebox(0,0)[lb]{\raisebox{0pt}[0pt][0pt]{\twlrm $=$}}}
\put(260,725){\makebox(0,0)[lb]{\raisebox{0pt}[0pt][0pt]{\twlrm $N$}}}
\put(290,685){\makebox(0,0)[lb]{\raisebox{0pt}[0pt][0pt]
{\twlrm $i \in \{0,...,r-2\}$}}}
\put(365,730){\makebox(0,0)[lb]{\raisebox{0pt}[0pt][0pt]
{\twlrm $\frac{\Delta_i}{\theta (a,b,i) \theta (c,d,i)}$}}}
\put(425,780){\makebox(0,0)[lb]{\raisebox{0pt}[0pt][0pt]{\twlrm $a$}}}
\put(450,785){\makebox(0,0)[lb]{\raisebox{0pt}[0pt][0pt]{\twlrm $b$}}}
\put(490,775){\makebox(0,0)[lb]{\raisebox{0pt}[0pt][0pt]{\twlrm $c$}}}
\put(515,785){\makebox(0,0)[lb]{\raisebox{0pt}[0pt][0pt]{\twlrm $d$}}}
\put(420,645){\makebox(0,0)[lb]{\raisebox{0pt}[0pt][0pt]{\twlrm $a$}}}
\put(450,635){\makebox(0,0)[lb]{\raisebox{0pt}[0pt][0pt]{\twlrm $b$}}}
\put(490,645){\makebox(0,0)[lb]{\raisebox{0pt}[0pt][0pt]{\twlrm $c$}}}
\put(535,645){\makebox(0,0)[lb]{\raisebox{0pt}[0pt][0pt]{\twlrm $d$}}}
\put(480,685){\makebox(0,0)[lb]{\raisebox{0pt}[0pt][0pt]{\twlrm $i$}}}
\put(480,735){\makebox(0,0)[lb]{\raisebox{0pt}[0pt][0pt]{\twlrm $i$}}}
\end{picture}

\caption{\label{encircut} Using Encirclement to Cut 4 Strands}
\end{figure}

	This would be uninteresting if it were not for the formula
of Figure \ref{encircut} which follows directly from
Lickorish's encirclement lemma [L],

(Of course, in Figure \ref{encircut} $(a,b,i)$ and $(c,d,i)$ must be
admissible triples for a 3-vertex.  This includes the condition
(imposed by $A$ being a $4r^{th}$ root of unity) that $a+b+i \leq 2r-4$.)
$N$ is as usual the sum of the squares of the quantum dimensions.

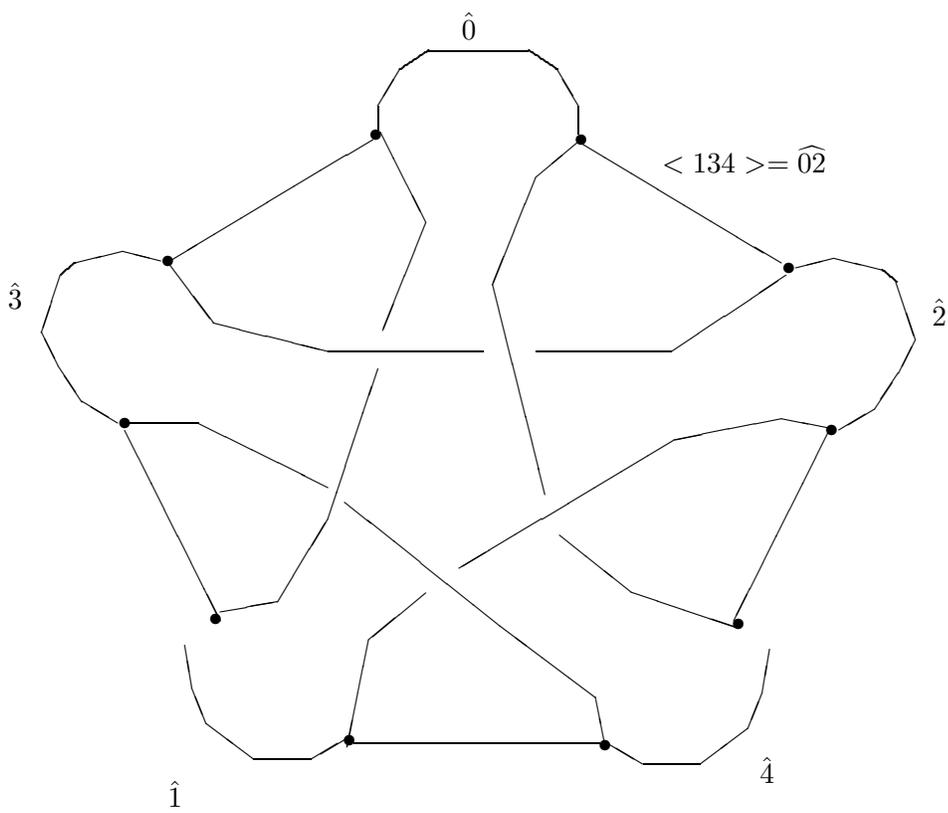
\begin{figure}
\setlength{\unitlength}{0.0125in}%
\begin{picture}(387,331)(70,417)
\thinlines
\put(137,646){\circle*{4}}
\put(119,578){\circle*{4}}
\put(134,646){\line(-4, 1){ 16}}
\put(118,650){\line(-4,-1){ 20}}
\multiput( 98,645)(-0.42857,-0.35714){15}{\makebox(0.4444,0.6667){\sevrm .}}
\put( 92,640){\line(-1,-3){  8}}
\put( 84,616){\line( 1,-2){  7}}
\put( 91,602){\line( 2,-3){ 10}}
\put(101,587){\line( 5,-3){ 15}}
\put(157,496){\circle*{4}}
\put(213,445){\circle*{4}}
\put(155,496){\line(-1,-1){ 11}}
\put(144,485){\line( 1,-6){  3}}
\put(147,467){\line( 2,-5){  6}}
\put(153,452){\line( 4,-3){ 20}}
\put(173,437){\line( 1, 0){ 24}}
\put(197,437){\line( 5, 3){ 15}}
\put(397,643){\circle*{4}}
\put(415,575){\circle*{4}}
\put(400,643){\line( 4, 1){ 16}}
\put(416,647){\line( 4,-1){ 20}}
\multiput(436,642)(0.42857,-0.35714){15}{\makebox(0.4444,0.6667){\sevrm .}}
\put(442,637){\line( 1,-3){  8}}
\put(450,613){\line(-1,-2){  7}}
\put(443,599){\line(-2,-3){ 10}}
\put(433,584){\line(-5,-3){ 15}}
\put(376,494){\circle*{4}}
\put(320,443){\circle*{4}}
\put(378,494){\line( 1,-1){ 11}}
\put(389,483){\line(-1,-6){  3}}
\put(386,465){\line(-2,-5){  6}}
\put(380,450){\line(-4,-3){ 20}}
\put(360,435){\line(-1, 0){ 24}}
\put(336,435){\line(-5, 3){ 15}}
\put(414,575){\line(-1,-2){ 40}}
\put(224,699){\circle*{4}}
\put(310,697){\circle*{4}}
\put(225,699){\line( 0, 1){ 12}}
\put(225,711){\line( 3, 5){  9}}
\multiput(234,726)(0.48000,0.32000){26}{\makebox(0.4444,0.6667){\sevrm .}}
\put(246,734){\line( 1, 0){ 21}}
\put(309,699){\line( 0, 1){ 12}}
\put(309,711){\line(-3, 5){  9}}
\multiput(300,726)(-0.48000,0.32000){26}{\makebox(0.4444,0.6667){\sevrm .}}
\put(288,734){\line(-1, 0){ 21}}
\put(119,575){\line( 1,-2){ 40}}
\put(215,444){\line( 1, 0){103}}
\put(223,697){\line(-5,-3){ 85}}
\put(309,696){\line( 5,-3){ 85}}
\put(138,644){\line( 3,-4){ 18}}
\put(156,620){\line( 4,-1){ 48}}
\put(204,608){\line( 1, 0){ 65}}
\put(309,696){\line(-6,-5){ 18}}
\put(291,681){\line(-2,-5){ 18}}
\put(273,636){\line( 1,-4){ 22}}
\put(414,576){\line(-5, 1){ 20}}
\put(394,580){\line(-5,-1){ 45}}
\put(349,571){\line(-5,-3){ 90}}
\put(320,443){\line(-1, 5){  4}}
\put(316,463){\line(-4, 3){ 40}}
\put(276,493){\line(-5, 4){ 65}}
\put(159,499){\line( 6, 1){ 24}}
\put(183,503){\line( 3, 5){ 21}}
\put(204,538){\line( 1, 3){ 21}}
\put(245,507){\line(-6,-5){ 24}}
\put(221,487){\line(-1,-5){  9}}
\put(301,531){\line( 5,-4){ 30}}
\put(331,507){\line( 3,-1){ 45}}
\put(291,608){\line( 1, 0){ 57}}
\put(348,608){\line( 3, 2){ 48}}
\put(226,700){\line( 1,-2){ 19}}
\put(245,662){\line(-2,-5){ 18}}
\put(119,578){\line( 1, 0){ 31}}
\put(150,578){\line( 2,-1){ 54}}
\put(260,739){\makebox(0,0)[lb]{\raisebox{0pt}[0pt][0pt]{\twlrm $\hat{0}$}}}
\put(137,417){\makebox(0,0)[lb]{\raisebox{0pt}[0pt][0pt]{\twlrm $\hat{1}$}}}
\put( 70,626){\makebox(0,0)[lb]{\raisebox{0pt}[0pt][0pt]{\twlrm $\hat{3}$}}}
\put(457,619){\makebox(0,0)[lb]{\raisebox{0pt}[0pt][0pt]{\twlrm $\hat{2}$}}}
\put(385,427){\makebox(0,0)[lb]{\raisebox{0pt}[0pt][0pt]{\twlrm $\hat{4}$}}}
\put(344,683){\makebox(0,0)[lb]{\raisebox{0pt}[0pt][0pt]
{\twlrm $<134> = \widehat{02}$}}}
\end{picture}

\caption{\label{15jbycutting} 15j-networks obtained by cutting chainmail}
\end{figure}

Application of this formula to $I(W)$ rewrites it as a sum of products
involving $\Delta_i$'s corresponding to labelled tetrahedra;
reciprocals of $\theta$-net evaluations corresponding to the
``even'' and ``odd'' faces of the tetrahedra,  $\Delta_i$'s on
faces, and evaluations of networks shown in Figure \ref{15jbycutting}.

The labels on the network in Figure \ref{15jbycutting} are given in
terms of the ordered 4-simplex $<01234>$ and will allow
the reader to compare the combinatorics of its structure to that of
the network in Figure \ref{gen15j} to see that it is indeed just
the quantum 15j-symbol in the TL formulation. (As usual $\hat{a}$ means
the face obtained by omitting $a$).

Putting all this together with the normalization factors for the 4-manifold
invariant $I(W)$, we find that

\[ I(W) = N^{-\frac{\chi(W)}{2}} CY(W), \]

where $\chi(W)$ is the Euler characteristic of $W$.

Thus, equivalently

\[ CY(W) = N^{\frac{\chi(W)}{2}} \kappa^{\sigma(W)}. \]

\clearpage \section{\label{centerm} The Center and Roberts' Chainmail Method}

	The keys to Roberts' approach [Ro1] to interpreting the original
Crane-Yetter invariant [CY] (cf. also [CKY1]), was the Lickorish Encirclement
Lemma (see [L]). The notion of the center of a braided monoidal category
introduced in [CKY2] was motivated by the desire to understand this result
in more generality.

	The analog of Lickorish's lemma at the appropriate level
of generality is

\begin{lemma}
\,?
\end{lemma}

\begin{figure}[h]
\begin{centering}
\setlength{\unitlength}{0.0125in}%
\begin{picture}(280,110)(35,600)
\thinlines
\put(180,675){\line(-1, 0){ 10}}
\put(170,635){\line( 1, 0){ 15}}
\put(200,675){\line(-1, 0){ 10}}
\multiput(170,675)(-0.40000,-0.40000){26}{\makebox(0.4444,0.6667){\sevrm .}}
\put(160,665){\line( 0,-1){ 20}}
\multiput(160,645)(0.40000,-0.40000){26}{\makebox(0.4444,0.6667){\sevrm .}}
\put(170,635){\line( 1, 0){ 15}}
\multiput(200,675)(0.40000,-0.40000){26}{\makebox(0.4444,0.6667){\sevrm .}}
\put(210,665){\line( 0,-1){ 20}}
\multiput(210,645)(-0.40000,-0.40000){26}{\makebox(0.4444,0.6667){\sevrm .}}
\put(200,635){\line(-1, 0){ 15}}
\put(115,675){\line(-1, 0){ 10}}
\put(105,635){\line( 1, 0){ 15}}
\put(135,675){\line(-1, 0){ 10}}
\multiput(105,675)(-0.40000,-0.40000){26}{\makebox(0.4444,0.6667){\sevrm .}}
\put( 95,665){\line( 0,-1){ 20}}
\multiput( 95,645)(0.40000,-0.40000){26}{\makebox(0.4444,0.6667){\sevrm .}}
\put(105,635){\line( 1, 0){ 15}}
\multiput(135,675)(0.40000,-0.40000){26}{\makebox(0.4444,0.6667){\sevrm .}}
\put(145,665){\line( 0,-1){ 20}}
\multiput(145,645)(-0.40000,-0.40000){26}{\makebox(0.4444,0.6667){\sevrm .}}
\put(135,635){\line(-1, 0){ 15}}
\put( 70,670){\line( 0, 1){  5}}
\put( 35,675){\line( 1,-1){ 20}}
\put( 70,675){\line(-1, 0){ 30}}
\put( 70,640){\line( 0,-1){  5}}
\put( 35,635){\line( 1, 1){ 20}}
\put( 70,635){\line(-1, 0){ 30}}
\put( 35,675){\line( 1, 0){  5}}
\put(115,675){\line( 1, 0){ 10}}
\put(185,710){\line( 0,-1){ 70}}
\put(185,630){\line( 0,-1){ 30}}
\put( 35,635){\line( 1, 0){  5}}
\put(115,620){\makebox(0,0)[lb]{\raisebox{0pt}[0pt][0pt]{\twlrm $j$}}}
\put(215,650){\makebox(0,0)[lb]{\raisebox{0pt}[0pt][0pt]{\twlrm $j$}}}
\put(190,690){\makebox(0,0)[lb]{\raisebox{0pt}[0pt][0pt]{\twlrm $n$}}}
\put(265,650){\makebox(0,0)[lb]{\raisebox{0pt}[0pt][0pt]{\twlrm =}}}
\put( 40,615){\makebox(0,0)[lb]{\raisebox{0pt}[0pt][0pt]
{\twlrm $j \in \cal S$}}}
\put(315,645){\makebox(0,0)[lb]{\raisebox{0pt}[0pt][0pt]
{\twlrm $N \chi_{Z({cal C})}(n) Id_n$}}}
\end{picture}
\caption{}
\end{centering}
\end{figure}

\noindent {\em where, as usual $N$ is the sum of the squares of the
dimensions of the objects in $\cal S$ and $\chi_{Z({\cal C})}(n) = 1$ if
$n$ is an object in the center, and $0$ otherwise.}
\medskip

	The proof of this Lemma is analogous to that given in [KaLi] or [L]:
If $n$ is in the center, we can unbraid it from the loops labelled $j$,
and obtain $N Id_n$.  Otherwise, select an object with which $n$ does
not braid trivially, handle-slide this over the $\omega_{\cal C}$-labled
loop, and observe that the non-triviality of the braiding implies that the
endomorphism of $n$ depicted on the right must be 0, since otherwise
it cannot compose with two different maps to give the same result (by
simplicity of $n$).


	Observe that in the case where the category has trivial center, this
lemma allows us to simply erase (!) components passing through unknotted loops
labelled $\omega_{\cal C}$ just as Lickorish's original formulation did
in the case of TL diagrams.

	Moreover, in the notation of Section \ref{3mfsec}, we have

\begin{propo} \label{sigonly}
If $\cal C$ is a semi-simple tortile category with trivial center, then
$\cal C$ is 3- and 4-conformed, with $N = b = x^2$.  Moreover, the
generalized Broda invariant $\cal B_C$ associated to $\cal C$ satisfies

\[ {\cal B_C}(W) = y^{\sigma(W)} \]

and the generalized Crane-Yetter invariant $CY_{\cal C}$ satisfies

\[ CY_{\cal C}(W) = N^\frac{\chi(W)}{2} y^{\sigma(W)}. \]
\end{propo}

\noindent{\bf proof:} Once we establish the first statement, the second
follows by an analysis essentially identical to that in the previous section.
The first statement follows by applying the generalized Lickorish encirclement
lemma and triviality of the center to the $\omega_{\cal C}$-labelled
0-framed Hopf link, and Lemma \ref{handleslide} followed by the generalized
Lickorish encirclement lemma and triviality of the center to the
$\omega_{\cal C}$ labelled disjoint union of a $+1$-framed unknot and a
$-1$-framed unknot. $\Box$
\smallskip

	Proposition \ref{sigonly} is a generalization of the result of
[CKY1] (cf. also [Ro1]) expressing the original Crane-Yetter invariant in
terms of the signature and Euler character of the manifold.
As noted in [CKY1] and in the introduction, this result should be regarded
as giving a purely combinatorial
expression for the signature of a 4-manifold in terms of a triangulation.

	At first it may not be clear that this result can be read in this
way since $y$ is (for known examples) a root of unity.  However, in the
case of the TL formulation of $Rep_!(U_q(sl_2))$ with $A$ chosen to be the
principal $4r^{th}$ root of unity, by explicit calculations of Roberts [Ro1],
we have that

\[ y =  e^{\frac{-i\pi (3+r^2)}{2r} - \frac{i\pi}{4}} \]

	It is not hard to show that if $r$ is chosen to be a multiple of
4 and relatively prime to 3, then $y$ will be a primitive $2r^{th}$ root
of unity.  Thus choosing such an $r$ greater than the rank of
the second homology of the manifold (or for simplicity, greater than the
number of 2-simplices in the triangulation used), it is then possible to
extract the signature of the manifold from the original Crane-Yetter
invariant [CY].

	Explicitly, if we let $r = 4n$ ($ (n,3) = 1 $) for $n$ sufficiently
large, then the signature of $W$ is the unique solution between $-r$ and $r$ to
the equation

\[ \sigma \frac{-i\pi (3 + 2n)}{8n} = \log (CY(W) N^\frac{-\chi(W)}{2})
\pmod{2\pi}\]
\noindent where $\log $ is the principal branch of the complex natural
logarithm.

	A curious (though not particularly important) question is whether
there are any artinian semisimple tortile categories which are
sufficiently ``non-unitary'' that $y$ is not a root of unity. If so, the
generalized Crane-Yetter invariant for those categories would allow
us to compute the signature directly as a logarithm of a state-sum.

	An important point for further research is a comparison of this
combinatorial expression for the signature with that given by
Gelfand and Macpherson [GM].

	Another aspect of this formulation of the signature of a 4-manifold
bears consideration: The generalized Crane-Yetter invariants are all
invariants associated to a TQFT. This fact follows from general principles
set down in [Y6].  The transition amplitudes of an generalized Crane-Yetter
TQFT, can thus be regarded as giving ``relative signatures'' for
cobordisms. Put another way, Crane-Yetter theory allows us to ``factor''
4-manifold signatures along any 3-manifold.

	Still, from another point of view, Proposition \ref{sigonly}
 is rather disappointing:  the Crane-Yetter and Broda
constructions for any semi-simple tortile category with trivial center
merely give rise to various encodings of the signature and Euler character
of the 4-manifold.  The question of whether this construction can give
yield other information, say about homotopy type or smooth structure,
turns crucially on the properties of the center of the category used.

	Two particular points in the construction suggest lines of
further research:

	The first is the demonstration in the case of a trivial center
that $x^2 = N$.  In general, the handle-sliding followed by the encirclement
lemma shows that

\[ x^2 = N z_+ = N z_- \]

\noindent where $z_\pm$ is the value of a $\pm 1$-framed unknot labelled
with $\sum_{i\in Z({\cal C})} dim(i) i $.

	Thus, our category will fail to be 3-conformed if $z_\pm = 0$.
In this case, the reduction of the generalized Broda invariant to signature
will break down.

	Similarly there is a property which semi-simple tortile categories
could possess which would destroy the reduction of the generalized
Crane-Yetter invariant to a generalized Broda invariant:  observe that
in Roberts chain-mail argument, the encirclement lemma is used to
cut an edge labelled with a summand of some $i\otimes i^\ast$ (for $i$
an object in our set of simple objects).  The reduction of the generalized
Crane-Yetter invariant to a generalized Broda invariant is thus depends on
the category satifying:

\begin{defin}
A semi-simple tortile category has {\em centrally trivial duality} if
for every simple object $i$, the only direct summand of $i\otimes i^\ast$
to lie in the center is the monoidal identity $I$.
\end{defin}

	It is a question for future research whether there are any
(interesting) semi-simple tortile categories which fail to have
centrally trivial duality or have $z_\pm = 0$.  If the answer is yes,
the corresponding Crane-Yetter and Broda invariants will have to be
analysed.  We will not conjecture whether they will be more interesting
or less interesting than the known examples, though we hope the former and
suspect the latter.

\section{Extensions to Manifolds with (Co)homology Classes}

	We conclude by reviewing the extensions of the Crane-Yetter type
invariants to 4-manifolds equipped with 2-dimensional (co)homology classes
introduced in [Y2] and [Ro2].

	In the case of the $U_q(sl_2)$ theory, both constructions are
quite easy to state:

Consider a 2-dimensional $Z/2$-homology class $\alpha$ on $W$,
and choose a representation of
it as a sum of distinct 2-simplices of a triangulation.

Roberts' construction gives an invariant of the pair $(W,\alpha)$
by restricting the labels in the Crane-Yetter state-sum to odd labels on
the 2-simplices representing it and even labels elsewhere.

Yetter's construction modifies the Crane-Yetter state-sum by multiplying
each term by
\[ (-1)^{ \left\{ \parbox[b]{2in}{\scriptsize \rm  number of odd labels on
2-simplices
in the
representative 2-cycle} \right\} } . \]

	Roberts [Ro2] has provided interpretations for his version of
this construction, and (private communication) given a ``Fourier transform''
formula relating the two approaches.

	The corresponding constructions for general artinian semisimple
tortile categories depend on the notion of a grading of a
semisimple monoidal category:

\begin{defin}
If $\cal C$ is a semisimple monoidal category (over some field), a
{\em grading} of $\cal C$ over an abelian group $A$ is a map $|\, |$ from the
set (or class) of simple object of $\cal C$ to $A$ satisfying

\begin{enumerate}
\item If $s \cong t$ then $|s| = |t|$.
\item If $s\otimes t \cong \oplus_{i\in I} u_i$ then $|u_i| = |s| |t|$ for all
$i\in I$
\end{enumerate}
\end{defin}

	 It is immediate from general principles that if $\cal C$ is
essentially small, then there exists an abelian group $gr({\cal C})$, equipped
with a grading of $\cal C$, and univeral among such. We call this group
the {\em universal grading group} of $\cal C$. As an example, observe that
the universal grading group for $Rep(U_q(sl_n)$ is $Z/n$, the grade
being given by the number of blocks in a Young diagram for the irrep (taken
mod n).

	Note, that for any abelian group $A$, there is an abelian group of
characters $\tilde{A}$. It is shown in Yetter [Y2] that the group of characters
of the universal grading group is isomorphic to the group of functorial
monoidal automorphisms (monoidal natural automorphisms of the identity
functor). Yetter's invariant of a 4-manifold equipped with a 2-dimensional
homology class $[\alpha]$ over $\widetilde{gr({\cal G})}$ is then given by

\[ Y_{\cal C}(M, [\alpha]) = \sum_{\lambda \in \Lambda_{\cal CSB}({\bf T})}
 \ll \lambda, [\alpha] \gg \]

\noindent where

\[
 \ll \lambda, [\alpha] \gg  =
N^{n_0 - n_1} \prod_{\parbox{.6in}{\small faces $\sigma$}}
tr(\alpha^\sigma_{\lambda(\sigma )})
\prod_{\parbox{.7in}{\small tetrahedra $\tau $}}
dim(\lambda
(\tau))^{-1}
\prod_{\parbox{.8in}{\small 4-simplices $\xi$}}
\| \lambda, \xi \|
\]

\noindent $\alpha$ is a representative cycle for $[\alpha]$ subordinate to
the triangulation, $\alpha^\sigma$ is the coefficent of $\sigma$ (which is
a natural transformation),
and $\| \lambda, \xi \|$ is the 15j-network appropriate to
the 4-simplex $\xi$ and the coloring $\lambda$.

Roberts invariant of a 4-manifold equipped with a cohomology class $[\beta]$
in $gr({\cal C})$ is then given by the Crane-Yetter formula, but with the
assignments of spins on the faces restricted by
$|\lambda(\sigma )| = \beta(\sigma)$, where $\beta$ is a representative
cocycle subordinate to the triangulation.

Proof of the invariance properties of these state-sums, and a discussion
of results related to them will be given in the sequel.

\clearpage \section{Appendix on Diagrammatic Notation}

	The following brief outline of diagrammatic notation for maps in
tortile categories is adapted from [Y3].  It is one of the surprising facts,
common
in quantum topology, that the diagrammatics given here for
general tortile categories is, in the case of $Rep(U_q(sl_2)$, in exact
coincidence with the
diagrammatics already developped by Kauffman [Ka1,Ka2,Ka3]
in his application of
q-deformed versions of Penrose spin-networks [P] to the construction of
statistical mechanical models for knot invariants.
The reader will already be familiar
with the most refined version of Kauffman's diagrammatics given in
Kauffman/Lins [KaLi], as this is the ``TL theory'' in the main exposition.
We concern ourselves here with the general case:

	Diagrammatic notation is best adapted to ``arrows-only''
descriptions of categories, so we make no distinction between an object and
the identity map on the object.
Identity maps are denoted by labelled curves descending the page; the tensor
product $\otimes $ is denoted by setting side-by-side.

\begin{figure}[htbp]
\centering
\setlength{\unitlength}{0.01in}%
\begin{picture}(130,80)(95,705)
\thinlines
\put( 95,785){\line( 0,-1){ 80}}
\put(190,785){\line( 0,-1){ 80}}
\put(215,785){\line( 0,-1){ 80}}
\put(105,770){\makebox(0,0)[lb]{\raisebox{0pt}[0pt][0pt]
{\twlrm X $\otimes $ Y}}}
\put(195,760){\makebox(0,0)[lb]{\raisebox{0pt}[0pt][0pt]{\twlrm X}}}
\put(225,760){\makebox(0,0)[lb]{\raisebox{0pt}[0pt][0pt]{\twlrm Y}}}
\end{picture}
\caption{Two ways to denote $X \otimes Y$}
\end{figure}
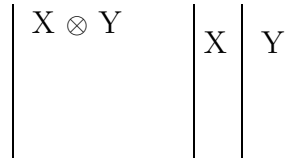

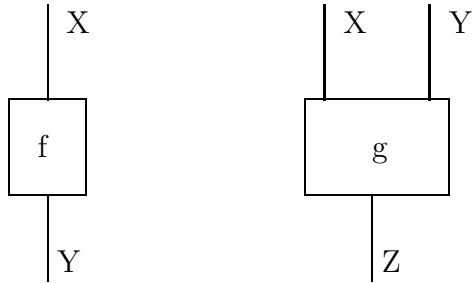
\begin{figure}[htbp]
\centering
\setlength{\unitlength}{0.01in}%
\begin{picture}(230,145)(105,635)
\thinlines
\put(105,680){\framebox(40,50){}}
\put(260,680){\framebox(75,50){}}
\put(125,730){\line( 0, 1){ 50}}
\put(125,680){\line( 0,-1){ 45}}
\put(125,635){\line( 0, 1){  0}}
\put(270,730){\line( 0, 1){ 50}}
\put(325,730){\line( 0, 1){ 50}}
\put(295,680){\line( 0,-1){ 45}}
\put(120,700){\makebox(0,0)[lb]{\raisebox{0pt}[0pt][0pt]{\twlrm f}}}
\put(135,765){\makebox(0,0)[lb]{\raisebox{0pt}[0pt][0pt]{\twlrm X}}}
\put(130,640){\makebox(0,0)[lb]{\raisebox{0pt}[0pt][0pt]{\twlrm Y}}}
\put(280,765){\makebox(0,0)[lb]{\raisebox{0pt}[0pt][0pt]{\twlrm X}}}
\put(335,765){\makebox(0,0)[lb]{\raisebox{0pt}[0pt][0pt]{\twlrm Y}}}
\put(300,640){\makebox(0,0)[lb]{\raisebox{0pt}[0pt][0pt]{\twlrm Z}}}
\put(295,700){\makebox(0,0)[lb]{\raisebox{0pt}[0pt][0pt]{\twlrm g}}}
\end{picture}
\caption{Maps $f:X\rightarrow Y$ and $g:X\otimes Y\rightarrow Z$
\label{samplemaps}}
\end{figure}

	In general, maps are denoted by boxes with incoming edges above
denoting the source of the map, and outgoing edges below denoting the target
of the map, as for example in
Figure \ref{samplemaps}.

	Some maps and objects, however, have special notations whose use is
justified by the coherence theorems of [S] and [FY2]. In particular, curves
labelled with the monoidal
identity object $I$ may be omitted.

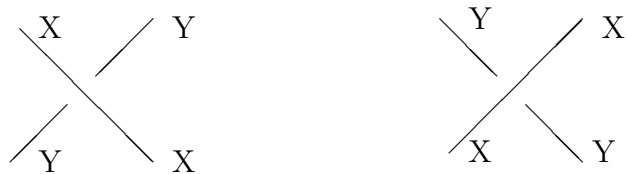
\begin{figure}[htbp]
\centering
\setlength{\unitlength}{0.01in}%
\begin{picture}(310,84)(75,695)
\thinlines
\put( 80,770){\line( 1,-1){ 70}}
\put(150,775){\line(-1,-1){ 30}}
\put(105,730){\line(-1,-1){ 30}}
\put(305,705){\line( 1, 1){ 70}}
\put(300,775){\line( 1,-1){ 30}}
\put(345,730){\line( 1,-1){ 30}}
\put( 90,765){\makebox(0,0)[lb]{\raisebox{0pt}[0pt][0pt]{\twlrm X}}}
\put(160,765){\makebox(0,0)[lb]{\raisebox{0pt}[0pt][0pt]{\twlrm Y}}}
\put( 90,695){\makebox(0,0)[lb]{\raisebox{0pt}[0pt][0pt]{\twlrm Y}}}
\put(160,695){\makebox(0,0)[lb]{\raisebox{0pt}[0pt][0pt]{\twlrm X}}}
\put(315,770){\makebox(0,0)[lb]{\raisebox{0pt}[0pt][0pt]{\twlrm Y}}}
\put(385,765){\makebox(0,0)[lb]{\raisebox{0pt}[0pt][0pt]{\twlrm X}}}
\put(315,700){\makebox(0,0)[lb]{\raisebox{0pt}[0pt][0pt]{\twlrm X}}}
\put(380,700){\makebox(0,0)[lb]{\raisebox{0pt}[0pt][0pt]{\twlrm Y}}}
\end{picture}
\caption{The braiding and its inverse}
\end{figure}

\begin{figure}[htbp]
\centering
\setlength{\unitlength}{0.01in}%
\begin{picture}(280,80)(155,715)
\thinlines
\put(155,795){\line( 0,-1){ 40}}
\put(155,755){\line( 1,-4){  5}}
\multiput(160,735)(0.25000,-0.50000){21}{\makebox(0.4444,0.6667){\sevrm .}}
\multiput(165,725)(0.40000,-0.40000){26}{\makebox(0.4444,0.6667){\sevrm .}}
\put(175,715){\line( 1, 0){ 10}}
\put(215,795){\line( 0,-1){ 40}}
\put(215,755){\line(-1,-4){  5}}
\multiput(210,735)(-0.25000,-0.50000){21}{\makebox(0.4444,0.6667){\sevrm .}}
\multiput(205,725)(-0.40000,-0.40000){26}{\makebox(0.4444,0.6667){\sevrm .}}
\put(195,715){\line(-1, 0){ 10}}
\put(370,715){\line( 0, 1){ 40}}
\put(370,755){\line( 1, 4){  5}}
\multiput(375,775)(0.25000,0.50000){21}{\makebox(0.4444,0.6667){\sevrm .}}
\multiput(380,785)(0.40000,0.40000){26}{\makebox(0.4444,0.6667){\sevrm .}}
\put(390,795){\line( 1, 0){ 10}}
\put(430,715){\line( 0, 1){ 40}}
\put(430,755){\line(-1, 4){  5}}
\multiput(425,775)(-0.25000,0.50000){21}{\makebox(0.4444,0.6667){\sevrm .}}
\multiput(420,785)(-0.40000,0.40000){26}{\makebox(0.4444,0.6667){\sevrm .}}
\put(410,795){\line(-1, 0){ 10}}
\put(160,780){\makebox(0,0)[lb]{\raisebox{0pt}[0pt][0pt]{\twlrm X}}}
\put(220,775){\makebox(0,0)[lb]{\raisebox{0pt}[0pt][0pt]{\twlrm X$^\ast $}}}
\put(375,715){\makebox(0,0)[lb]{\raisebox{0pt}[0pt][0pt]{\twlrm X$^\ast $}}}
\put(435,715){\makebox(0,0)[lb]{\raisebox{0pt}[0pt][0pt]{\twlrm X}}}
\end{picture}
\caption{The evaluation and coevaluation maps for a right dual}
\end{figure}
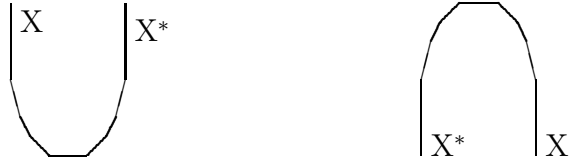

	The coherence theorems of [S] and [FY2] then insure that
manipulating the diagrams by ``generalized framed Reidemeister moves''
results in a different notation for the
same map.

	Further suppression of labelling is possible by orienting the
curves and considering downward oriented curves as indicating the labelling
object, and upward oriented
curves as indicating its right dual.

\begin{figure}[htbp]
\centering
\setlength{\unitlength}{0.01in}%
\begin{picture}(247,232)(145,573)
\thinlines
\put(325,595){\framebox(25,30){}}
\put(337,594){\line( 0,-1){ 17}}
\put(337,624){\line( 1, 1){ 34}}
\put(371,658){\line( 0, 1){  8}}
\multiput(371,666)(-0.40000,0.40000){11}{\makebox(0.4444,0.6667){\sevrm .}}
\put(367,670){\line(-1, 0){  9}}
\multiput(358,670)(-0.40000,-0.40000){21}{\makebox(0.4444,0.6667){\sevrm .}}
\multiput(350,662)(0.40000,-0.40000){21}{\makebox(0.4444,0.6667){\sevrm .}}
\put(368,644){\line( 1,-1){ 16}}
\put(384,628){\line( 0,-1){ 48}}
\put(325,757){\framebox(25,30){}}
\put(337,788){\line( 0, 1){ 17}}
\put(337,758){\line( 1,-1){ 34}}
\put(371,724){\line( 0,-1){  8}}
\multiput(371,716)(-0.40000,-0.40000){11}{\makebox(0.4444,0.6667){\sevrm .}}
\put(367,712){\line(-1, 0){  9}}
\multiput(358,712)(-0.40000,0.40000){21}{\makebox(0.4444,0.6667){\sevrm .}}
\multiput(350,720)(0.40000,0.40000){21}{\makebox(0.4444,0.6667){\sevrm .}}
\put(368,738){\line( 1, 1){ 16}}
\put(384,754){\line( 0, 1){ 48}}
\put(155,795){\line( 0,-1){ 40}}
\put(155,755){\line( 1,-4){  5}}
\multiput(160,735)(0.25000,-0.50000){21}{\makebox(0.4444,0.6667){\sevrm .}}
\multiput(165,725)(0.40000,-0.40000){26}{\makebox(0.4444,0.6667){\sevrm .}}
\put(175,715){\line( 1, 0){ 10}}
\put(215,795){\line( 0,-1){ 40}}
\put(215,755){\line(-1,-4){  5}}
\multiput(210,735)(-0.25000,-0.50000){21}{\makebox(0.4444,0.6667){\sevrm .}}
\multiput(205,725)(-0.40000,-0.40000){26}{\makebox(0.4444,0.6667){\sevrm .}}
\put(195,715){\line(-1, 0){ 10}}
\put(145,575){\line( 0, 1){ 40}}
\put(145,615){\line( 1, 4){  5}}
\multiput(150,635)(0.25000,0.50000){21}{\makebox(0.4444,0.6667){\sevrm .}}
\multiput(155,645)(0.40000,0.40000){26}{\makebox(0.4444,0.6667){\sevrm .}}
\put(165,655){\line( 1, 0){ 10}}
\put(205,575){\line( 0, 1){ 40}}
\put(205,615){\line(-1, 4){  5}}
\multiput(200,635)(-0.25000,0.50000){21}{\makebox(0.4444,0.6667){\sevrm .}}
\multiput(195,645)(-0.40000,0.40000){26}{\makebox(0.4444,0.6667){\sevrm .}}
\put(185,655){\line(-1, 0){ 10}}
\put(150,575){\makebox(0,0)[lb]{\raisebox{0pt}[0pt][0pt]{\twlrm X}}}
\put(215,575){\makebox(0,0)[lb]{\raisebox{0pt}[0pt][0pt]{\twlrm X$^\ast$}}}
\put(220,780){\makebox(0,0)[lb]{\raisebox{0pt}[0pt][0pt]{\twlrm X}}}
\put(160,785){\makebox(0,0)[lb]{\raisebox{0pt}[0pt][0pt]{\twlrm X$^\ast$}}}
\put(342,573){\makebox(0,0)[lb]{\raisebox{0pt}[0pt][0pt]{\twlrm X}}}
\put(392,791){\makebox(0,0)[lb]{\raisebox{0pt}[0pt][0pt]{\twlrm X}}}
\put(340,794){\makebox(0,0)[lb]{\raisebox{0pt}[0pt][0pt]{\twlrm X$^\ast$}}}
\put(388,577){\makebox(0,0)[lb]{\raisebox{0pt}[0pt][0pt]{\twlrm X$^\ast$}}}
\put(327,766){\makebox(0,0)[lb]{\raisebox{0pt}[0pt][0pt]{\twlrm $\theta $}}}
\put(328,602){\makebox(0,0)[lb]{\raisebox{0pt}[0pt][0pt]
{\twlrm $\theta ^{-1}$}}}
\put(269,752){\makebox(0,0)[lb]{\raisebox{0pt}[0pt][0pt]{\twlrm =}}}
\put(272,604){\makebox(0,0)[lb]{\raisebox{0pt}[0pt][0pt]{\twlrm =}}}
\end{picture}
\caption{\label{leftdual} Left duality in terms of right duality, braiding
and $\theta $}
\end{figure}
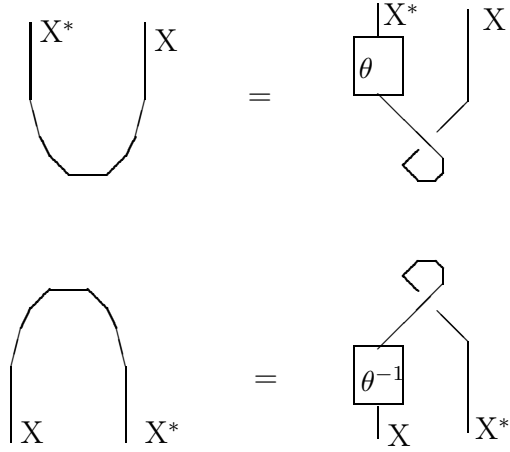

	Finally, we must note one other subtlety of diagrammatic notation as
used here: there are two types of maxima and minima in the diagrammatic
notation, those defining
the right dual, and those in which the labels $X$ and $X^\ast $ have been
exchanged. Maxima and minima of the second type are the structure maps for
$X^\ast $ as a left
dual, and may be expressed in terms of the right duality, the braiding and
the ``twist'' map $\theta $ as in Figure \ref{leftdual}.

	Since the categories involved are $K$-linear, we can also denote maps
by $K$-linear combinations of diagrams with the same source and target.
Similarly, the
``biproduct'' condition on the projections and inclusions of direct sum
decompositions into simple objects have diagrammatic expressions of the form
given in Figure
\ref{biprod}, where $A, A^\prime$ are elements of a basis for the relevant
hom-space, $X$ denotes an arbitrary object, and $S$ a simple object.

\begin{figure}[htbp]
\centering
\setlength{\unitlength}{0.01in}%
\begin{picture}(264,538)(111,246)
\thinlines
\put(181,398){\framebox(40,36){}}
\put(202,434){\line( 0, 1){ 38}}
\put(202,359){\line( 0, 1){ 38}}
\put(181,285){\framebox(40,36){}}
\put(202,321){\line( 0, 1){ 38}}
\put(202,246){\line( 0, 1){ 38}}
\put(201,710){\framebox(40,36){}}
\put(222,746){\line( 0, 1){ 38}}
\put(222,671){\line( 0, 1){ 38}}
\put(201,597){\framebox(40,36){}}
\put(222,633){\line( 0, 1){ 38}}
\put(222,558){\line( 0, 1){ 38}}
\put(266,654){\makebox(0,0)[lb]{\raisebox{0pt}[0pt][0pt]
{\twlrm = $\ \delta_{A A^\prime} 1_S$}}}
\put(206,768){\makebox(0,0)[lb]{\raisebox{0pt}[0pt][0pt]{\twlrm $S$}}}
\put(206,660){\makebox(0,0)[lb]{\raisebox{0pt}[0pt][0pt]{\twlrm $X$}}}
\put(203,564){\makebox(0,0)[lb]{\raisebox{0pt}[0pt][0pt]{\twlrm $S^\prime$}}}
\put(214,722){\makebox(0,0)[lb]{\raisebox{0pt}[0pt][0pt]
{\twlrm $\overline{A}$}}}
\put(214,609){\makebox(0,0)[lb]{\raisebox{0pt}[0pt][0pt]{\twlrm $A^\prime$}}}
\put(364,471){\line( 0,-1){225}}
\multiput(137,400)(0.20000,0.60000){6}{\makebox(0.4444,0.6667){\sevrm .}}
\put(138,403){\line(-1, 0){ 27}}
\put(111,403){\line( 1,-1){ 14}}
\multiput(137,378)(0.20000,-0.60000){6}{\makebox(0.4444,0.6667){\sevrm .}}
\put(138,375){\line(-1, 0){ 27}}
\put(111,375){\line( 1, 1){ 14}}
\put(210,458){\makebox(0,0)[lb]{\raisebox{0pt}[0pt][0pt]{\twlrm $X$}}}
\put(373,456){\makebox(0,0)[lb]{\raisebox{0pt}[0pt][0pt]{\twlrm $X$}}}
\put(207,350){\makebox(0,0)[lb]{\raisebox{0pt}[0pt][0pt]{\twlrm $S$}}}
\put(198,410){\makebox(0,0)[lb]{\raisebox{0pt}[0pt][0pt]
{\twlrm $\overline{A}$}}}
\put(198,296){\makebox(0,0)[lb]{\raisebox{0pt}[0pt][0pt]{\twlrm $A$}}}
\put(209,246){\makebox(0,0)[lb]{\raisebox{0pt}[0pt][0pt]{\twlrm $X$}}}
\put(110,357){\makebox(0,0)[lb]{\raisebox{0pt}[0pt][0pt]
{\twlrm $S \in {\cal S}$}}}
\put(114,336){\makebox(0,0)[lb]{\raisebox{0pt}[0pt][0pt]{\twlrm $A$}}}
\put(276,352){\makebox(0,0)[lb]{\raisebox{0pt}[0pt][0pt]{\twlrm =}}}
\end{picture}
\caption{\label{biprod}}
\end{figure}

	The relation expressed in Figure \ref{biprod} is used in tandem with
Lemma \ref{schur} in our diagrammatic calculations whenever we remove edges
whose
labels were summed over (the pictorial version of contracting indices).

\clearpage
\section{References}
\bigskip

\noindent [A] Atiyah, M., ``Topological Quantum Field Theories,''
{\em Publ. Math. I.H.E.S.} {\bf 68} (1989) 175-186.

\noindent [B] Broda, B., `` Surgical Invariants of 4-Manifolds,'' preprint
(1993).

\noindent [Ca] Casler, B.G., ``An Embedding Theorem for Connected 3-Manifolds
with Boundary,'' {\em Proc. AMS} {\bf 16} (1965) 559-566.

\noindent [CFS] Chung, S., Fukama, M. and Shapere, A., ``The Structure of
Topological Field Theories in Three Dimensions,'' {\em Int. J. Mod Phys A}
(1994) 1305-1360.

\noindent [Cr1] Crane, L., ``2D-Physics and 3D-Topology,'' {\em Comm. Math.
Phys.} {\bf 135} (1991) 615-640.

\noindent [Cr2] Crane, L., ``Four Dimensional TQFT; a Triptych,'' in
{\em Quantum Topology}, Kauffman,
L.H. and Baadhio, R.A. eds., World Scientific Press (1993) 116-119.

\noindent [CF] Crane, L., and Frenkel, I., ``Four Dimensional Topological
Field Theory, Hopf Categories and the Canonical Basis,'' (1994) preprint.

\noindent [CKY1] Crane, L., Kauffman, L.H. and Yetter, D.N., ``Evaluating the
Crane-Yetter Invariant,'' in {\em Quantum Topology}, Kauffman,
L.H. and Baadhio, R.A. eds., World Scientific Press (1993) 131-138.

\noindent [CKY2] Crane, L., Kauffman, L.H. and Yetter, D.N., ``On the Failure
of the Lickorish Encirclement Lemma for Temperley-Lieb Recoupling Theory
at Certain Roots of Unity,'' Proceedings of the XXII Conference on
Differential Geometric Methods in Theoretical Physics (to appear).

\noindent [CY] Crane, L. and Yetter, D.N., ``A Categorical Construction of
4D Topological Quantum Field Theories,'' in {\em Quantum Topology}, Kauffman,
L.H. and Baadhio, R.A. eds., World Scientific Press (1993) 120-130.

\noindent [D] Donaldson, S.K., ``An Application of Gauge Theory to Four
Dimensional Topology,'' {\em J. Diff. Geom.} {\bf 18} 269-316.

\noindent [DK] Donaldson, S.K. and Kronheimer, P.B., {\em The Geometry of
Four-Manifolds}, Oxford Mathematical Monographs, Oxford Univ. Press (1990).

\noindent [FY1] Freyd, P. J. and Yetter, D. N., ``Braided Compact Closed
Categories with Applications to Low-Dimensional Topology,''
{\em Adv. in Math.} {\bf 77} (2) (1989) 156-182.

\noindent [FY2] Freyd, P. J. and Yetter, D. N., ``Coherence Theorems via
Knot Theory,'' {\em JPAA} {\bf 78} (1992) 49-76.

\noindent [GM] Gelfand, I.M and Macpherson, R.D., ``A Combinatorial Formula for
the Pontrjagin Classes,'' {\em Bull. AMS} {\bf 26} (2) (1992) 304-308.

\noindent [GK] Gelfand, S. and Kazhdan, D., ``Examples of Tensor Categories,''
{\em Invent. Math.} {\bf 109} (1992) 595-617.

\noindent [JS1] Joyal, A. and Street, R., ``Braided Monoidal Categories,''
Macquarie Mathematics Reports No. 860081 (preprint) (1986).

\noindent [Ka1] Kauffman, L.H. ``Spin Networks and Knot Polynomials,'' {\em
Intl. J. Mod. Phys. A} {\bf 5} (1)  (1990) 93-115.

\noindent [Ka2] Kauffman, L.H. {\em Knots and Physics}, World. Sci. Press,
1991.

\noindent [Ka3] Kauffman, L.H. ``Map Coloring, q-Deformed Spin Networks and
Turaev-Viro Invariants for 3-Manifolds,'' in {\em The Proceedings for the
Conference on Quantum Groups -- Como, Italy, June 1991}, M. Rasetti, ed.,
World Sci. Press, {\em Intl. J. of Mod. Phys. B} {\bf 6} (11), (12) (1992)
1765-1794.

\noindent [KaLi] Kauffman, L.H. and Lins, S.L. {\em Temperley-Lieb Recoupling
Theory and Invariants of 3-Manifolds}, Princeton University Press (1994).

\noindent [KeLa] Kelly, G. M. and Laplaza, M. L., ``Coherence for Compact
Closed Categories,'' {\em JPAA} {\bf 19} (1980) 193-213.

\noindent [KM] Kronheimer, P.B. and Mrowka, T.S., ``Recurrence Relations and
Asymptotics for Four-Manifold Invariants,'' {\em Bull. AMS} {\bf 30} (2) (1994)
215-221.

\noindent [Ku] Kuperberg, G., ``Involutory Hopf Algebras and 3-Manifold
Invariants,'' {\em Int. J. of Math.} {\bf 2} (1) (1991) 41-66.

\noindent [JS2] Joyal, A. and Street, R., ``The Geometry of Tensor
Calculus, I,'' {\em Adv. in Math.} {\bf 88} (1) (1991) 55-112.

\noindent [KR] Kirillov, A. N. and Reshetikhin, N. Yu., ``Representations of
the Algebra $U_q(sl(2))$, $q$-Orthogonal Polynomials and Invariants of Links,''
in {\em Infinite Dimensional Lie Algebras and Groups}, V. G. Kac, ed., World
Scientific Adv. Series in Math. Phys., vol. 7 (1989).

\noindent [CWM] Mac Lane, S., {\em Categories for the Working Mathematician},
Springer-Verlag (1971).

\noindent [MS] Moore, G. and Seiberg, N., ``Classical and Quantum Conformal
Field Theory,'' {\em Comm. Math. Phys} {\bf 123} (1989) 177-254.

\noindent [P] Penrose, R., ``Applications of Negative Dimensional Tensors,''
in {\em Combinatorial Mathematics and its Applications} (D. J. A. Welsh, ed)
Acad. Press (1971).

\noindent [RP] Ponzano, G. and Regge, T., ``The Semi-classical Limit of Racah
Coefficients'' in {\em Spectoscopic and Group-Theoretical Methods in Physics}
F. Bloch ed., North-Holland (1968).

\noindent [RT] Reshetikhin, N. Yu. and Turaev, V. G., ``Ribbon Graphs and
Their Invariants Derived from Quantum Groups,'' {\em Comm. Math. Phys.}
{\bf 127} (1) (1990) 1-26.

\noindent [Ro1] Roberts, J. ``Skein Theory and Turaev-Viro Invariants'' (1993)
preprint.

\noindent [Ro2] Roberts, J., ``Refined State-Sum Invariants of 3- and
4-Manifolds,'' (1993) preprint.

\noindent [S] Shum, M.-C.,  ``Tortile Tensor Categories'', {\em JPAA} {\bf 93}
(1994) 57-110 .

\noindent [T] Turaev, V. G., ``Quantum Invariants of 3-Manifolds'',
{\em Publication de l'Institutue de Recherche Math\'{e}matique Avanc\'{e}e}
{\bf 509/P-295} (1992).

\noindent [TV] Turaev, V. G. and Viro O. Y., ``State Sum Invariants of
3-Manifolds and Quantum 6j-Symbols,'' {\em Topology} {\bf 31} (4) (1992)
865-902 .

\noindent [W1] Witten, E., ``Topological Quantum Field Theory,'' {\em Comm.
Math. Phys.} {\bf 117} (1988) 353-386.

\noindent [W2] Witten, E., ``Quantum Field Theory and the Jones Polynomial,''
{\em Comm. Math. Phys.} {\bf 121} (1989) 351-399.

\noindent [W3] Witten, E., ``Supersymmetric Yang-Mills Theory on a
4-Manifold,'' IASSNS preprint (1994).

\noindent [Y1] Yetter, D. N., ``Framed Tangles and a Theorem of Deligne on
Braided Deformations of Tannakian Categories,'' in {\em Deformation Theory
and Quantum Groups with
Applications to Mathematical Physics}, M. Gerstenhaber and
J. D. Stasheff, eds., AMS Contemp. Math. vol. 134 (1992).

\noindent [Y2] Yetter, D.N., ``Homologically Twisted Invariants Related to
(2+1)- and (3+1)-Dimensional State-Sum Topological Field Theories,''
(1993) e-preprint hep-th/9311082.

\noindent [Y3] Yetter, D.N., ``State-Sum Invariants of 3-Manifolds Associated
to Artinian Semisimple Tortile Categories,'' {\em Topology and
Its Applications} {\bf 58} (1) (1994) 47-80.

\noindent [Y4] Yetter, D. N., ``Topological Quantum Field Theories Associated
to Finite Groups and Crossed $G$-Sets,'' {\em JKTR} {\bf 1} (1) (1992) 1-20.

\noindent [Y5] Yetter, D. N., ``TQFT's from Homotopy 2-Types,'' {\em JKTR}
{\bf 2} (1) (1993) 113-123.

\noindent [Y6] Yetter, D.N., ``Triangulations and TQFT's'' in
{\em Quantum Topology}, Kauffman,
L.H. and Baadhio, R.A. eds., World Scientific Press (1993) 354-370.

\end{document}